# Conservative adventurers have more future academic impact

Xingsheng Yang[1,2†], Zhaoru Ke[1,3†], Xiangyi Xiao[1], Qing Ke[2], Fengnan Gao[4*], Haipeng Zhang[1*]

[1]School of Information Science and Technology, ShanghaiTech University, Shanghai, 201210, PR China.
[2]School of Data Science, City University of Hong Kong, Hong Kong, 999077, PR China.
[3]Yangtze River Waterway Transportation Monitoring and Emergency Center, Changjiang River Administration of Navigational Affairs, Wuhan, 430014, Hubei, PR China.
[4]School of Mathematics and Statistics, University College Dublin, Dublin D04 C1P1, Ireland.

*Corresponding author(s). E-mail(s): fengnan.gao@ucd.ie; zhanghp@shanghaitech.edu.cn;
Contributing authors: xingsheng.yang@my.cityu.edu.hk; 1164523499@qq.com; xiaoxy2025@shanghaitech.edu.cn; q.ke@cityu.edu.hk;
†These authors contributed equally to this work.

**Abstract**

Some scientists explore unfamiliar topics, while others exploit existing ones, yet the link between these choices and academic performance remains unclear. Prior studies offer conflicting evidence, often relying on single metrics and overlooking confounding factors. To address this, we complement the traditional switch frequency metric with switch distances, control for confounders, and establish a clear connection between past switching behaviors and future performance.
We identify a group, 'conservative adventurers', who frequently switch topics within close domains, excelling in future performance compared to others (up to 19% more citations per future paper). This rare behavioral pattern suggests an effective balance between exploration and exploitation. Beyond correlations, we question whether conservative adventuring can be intentionally adopted as a strategy. While proving intentionality is challenging, individuals who drastically transition to conservative adventurers likely do so purposefully, achieving significant performance gains.
Our findings, based on three datasets covering 31,780,857 papers in physics, biomedicine, and chemistry primarily since the 20th century (1976–2015 for physics; 1900–2021 for biomedicine and chemistry), provide insights for scientific career understanding and planning, particularly for junior scientists.

# Introduction

While advancing their academic pursuits, scientists choose their research topics, and these choices shape their individual careers (Jia, Wang, & Szymanski, 2017; Kleinberg & Oren, 2011; Zeng et al., 2019) as well as the evolution of science (Gates, Ke, Varol, & Barabási, 2019; Shi, Foster, & Evans, 2015; Sinatra, Deville, & Szell, 2015). Like phenomena seen in various human activities and the broader natural world (J.D. Cohen, McClure, & Yu, 2007; March, 1991; Mehlhorn et al., 2015), some scientists choose to exploit their current domains, following a conservative strategy, while the more risk-taking ones prefer to explore domains that are less familiar to them (Foster, Rzhetsky, & Evans, 2015; Topper & Kuhn, 1977), leading to the classic discussion on the trade-off between exploration and exploitation (J.D. Cohen et al., 2007; March, 1991; Mehlhorn et al., 2015).

Studies in sociology of science and philosophy of science have long investigated research topic choices (Topper & Kuhn, 1977; Zuckerman, 1978). Besides the work on generating mechanisms (Rzhetsky, Foster, Foster, & Evans, 2015; Yin, Wang, Evans, & Wang, 2019; Zeng et al., 2019), there has been plenty of recent attention paid to research topic switching, which depicts how frequently a scientist changes research topics, reflecting their exploitation-exploration preferences. Previous studies have discovered apparent correlations between switching and academic performance (Amjad, Daud, & Song, 2018; Hill et al., 2025; Huang, Lu, Bu, & Huang, 2022; Pramanik, Gora, Sundaram, Ganguly, & Mitra, 2019; Yu, Szymanski, & Jia, 2021; Zeng et al., 2019). Peculiarly, the found correlations, at both the paper level (Foster et al., 2015; Hill et al., 2025) and the individual level (Amjad et al., 2018; Huang et al., 2022; Pramanik et al., 2019; Yu et al., 2021; Zeng et al., 2019), go opposite directions, suggesting that the intricate interplay behind the curtain of correlations is yet to be unveiled. Some studies on individual behaviors reported negative correlations between switch tendencies and scholarly performances (Amjad et al., 2018; Zeng et al., 2019), while others found positive connections (Huang et al., 2022; Pramanik et al., 2019; Yu et al., 2021), resulting in conflicting suggestions for scientists. One educated guess is that there may be a Simpson's paradox, where confounding variables have not been appropriately addressed. Besides the common factors such as the beginning year of career, potential mechanisms in this scenario include switching to hot areas (Wei et al., 2013), utilizing specific knowledge combinations (Uzzi, Mukherjee, Stringer, & Jones, 2013), seeking advantageous collaborations (R. Wu & Shi, 2025; Wuchty, Jones, & Uzzi, 2007; F. Xu, Wu, & Evans, 2022; H. Xu et al., 2024; Zhang, 2009), or accessing institutional resources (Zhai, Xia, Chi, & Li, 2024) that may yield more academic impact. In the analysis presented in the Section 4, we show how the seemingly negative correlation between topic switch frequencies and academic performance in our data disappears, when we further consider the fact that many scientists with past successes are less likely to switch topics in the future.

The incompleteness of previous studies could also stem from the reliance on single metrics of the topic switch behaviors. A one-dimensional indicator, as opposed to a combined one, may not be sufficient in encapsulating the complexity of these behaviors. One common set of metrics captures the tendencies of topic switching by measuring the frequencies of switches (Zeng et al., 2019) or the degree of topic distribution concentration (Amjad et al., 2018; Pramanik et al., 2019). Some other metrics focus on switch distances, typically measured by how much the topic distributions of new papers deviate from those of past papers (Hill et al., 2025; Yang, 2025; Yu et al., 2021). The switch frequency metric would fail to differentiate a far switch from acoustics to lasers and a much closer switch from statistics to machine learning. Someone who frequently makes larger jumps and someone who makes smaller jumps at the same frequency have intuitively different behavior patterns while under the switch frequency metric, they may be the same. However, if we solely consider switch distances, an individual who rarely switches topics but takes a significant leap would be treated the same as someone who frequently makes smaller topic moves. By complementing the switch frequency metric with switch distances and extensively controlling for confounding factors, we have teased out the predictive power of the 'conservative adventuring' behavioral pattern in relation to future academic performance from previous inconsistencies. We find that 'conservative adventurers' who frequently switch topics (with high switch frequencies) but do so within closely related domains (with low switch distances), have significantly better future performance, even when they are still junior scientists. Their advantages against others are then precisely quantified. Along the temporal dimension, we further investigate how individuals' switching behaviors under the two metrics change throughout their careers and how these behaviors evolve across generations of scientists.

On top of the correlation, we further ask the question from a practical point of view — can 'conservative adventuring' serve as an effective strategy for scientists to achieve greater future impacts? Or equivalently, if someone intentionally adopts a 'conservative adventuring' approach, will she have more future impact? We do not know the exact reasons behind the behaviors reflected in our data. However, it is a plausible guess those who change their behaviors drastically may have intentionally altered their research strategies, since most scientists maintain stable behavioral patterns under our metrics. We correspondingly conduct quasi-experiments on these 'drastic changers'.

To unravel the complexity of switching and its impact on academic performance, we utilize three large datasets for analysis: a PubMed dataset formulated in the Microsoft Academic Graph (Sinha et al., 2015) including 29,776,639 papers in biomedicine, the American Physical Society (APS) dataset (APS, 2022) that covers 678,961 physics papers, and the American Chemical Society (ACS) dataset with 1,325,257 chemistry papers (Section 4). The main text focuses on the APS dataset, while all the reported results also hold for the PubMed dataset (section A4.5) and the ACS dataset (section A4.6). To ensure the conservation of effect, we perform regressions and conduct propensity score weighting with various alternatives in measurement computations (e.g., topic graph overlap vs. node2vec for computing paper distance) and sample selections (e.g., different minimum number of publications requirements).

# Background

## Exploration vs. Exploitation in Topic-Switching Behavior

Topic switch behavior constitutes a central concern in the career development of scientists. The academic community has extensively examined both its theoretical underpinnings and its consequential impacts. The 'Essential Tension' hypothesis (Topper & Kuhn, 1977), emphasizes the conflict that scientists encounter when selecting research topics. This tension arises from the need to balance tradition and innovation, forcing scientists to navigate between exploration and exploitation. Exploration of new topics represents an open and innovative strategy, albeit with inherent risks, whereas exploitation of existing knowledge is a more conservative approach aimed at efficiency. This behavior is not only shaped by economic and social capital but also constrained by cultural factors (Bourdieu, 1975).

Different patterns of topic-switching behaviors directly impact an individual's career trajectory (Rzhetsky et al., 2015; Venturini, Sikdar, Rinaldi, Tudisco, & Fortunato, 2024; Zeng et al., 2019). Specifically, scientists tend to adopt a more conservative strategy when choosing topics, prioritizing well-established domains (Rzhetsky et al., 2015), which enhances their chances of securing collaboration (Venturini et al., 2024), obtaining research grants (Hoppe et al., 2019) and producing academic output (Zeng et al., 2019).

## Metrics of Topic-Switching Behaviors

Early work characterizes topic switching by how often authors transition across topics within their publication histories. Zeng et al. (2019) construct co-citing networks for each scientist, detect topic communities, and define switches as transitions between communities over time. Amjad et al. (2018) apply the author–conference–topic model to infer author-level topic distributions and operationalize topic drift as changes in these distributions across periods. Pramanik et al. (2019) develop an algorithm on large-scale bibliometric data to identify migrating researchers by mapping papers to fields and counting inter-field transitions. Earlier qualitative work also tracks topic mobility by analyzing titles to code topics longitudinally and record transitions (Franzoni, Simpkins, Li, & Ram, 2010). While these studies capture how often scientists move across topics, they cannot distinguish how far those moves extend, treating distant and proximate shifts equivalently.

Another stream of work focuses on switch distances, capturing how far new research departs from prior work. Yu et al. (2021) represent each scientist's topic profile as a topic vector space and compute cosine distances between consecutive publication vectors to quantify the extent of direction change. Hill et al. (2025) introduce a pivot metric, measuring vector similarity between journal distributions of consecutive papers. Yang (2025) compute knowledge distances based on overlaps in cited references between new and past papers, using this to detect topic-shift magnitude. By measuring distance alone and omitting switching frequency, these approaches obscure the cadence of topic boundary crossings. To overcome these limitations, we integrate switch frequency and switch distance into a two-dimensional characterization of topic switching, allowing simultaneous assessment of how often and how far scientists move.

## Contradictory Correlations

Existing studies have examined the relationship between scientists' topic switch behavior and their academic performance, yet the findings are contradictory. Some papers suggest that frequent topic switching negatively impacts academic performance (Amjad et al., 2018; Hill et al., 2025; Rzhetsky et al., 2015; Sun, Livan, Ma, & Latora, 2021; Zeng et al., 2019). Zeng et al. (2019) analyze academic publication data from the field of physics and find a negative correlation between topic switching and the author's citations. Moreover, Amjad et al. (2018) argue that scientists who persist in a single research theme tend to achieve higher academic influence and attract greater attention compared to those who frequently shift topics. Similarly, Rzhetsky et al. (2015) identify that switching increases the risk of publication failure.

Conversely, other studies adopt an opposing perspective (Huang et al., 2022; Pramanik et al., 2019; Yu et al., 2021). Yu et al. (2021) assert that a strong positive correlation exists between topic switching and increases in academic impact, measured by average citation counts.

Adding further nuance, a recent study has explored the relationship between topic switching and various dimensions of scientific innovation beyond traditional citation-based impact metrics (Yang, 2025). This study highlights positive links between topic switching and indicators of innovative research, including disruption, novelty, and interdisciplinarity, while also recognizing potential trade-offs with traditional metrics.

These contradictions come not only from limitations in how studies measure topic switching but also from experimental design flaws. Specifically, in examining the relationship between scientists' topic switch behavior and academic performance, it is crucial to consider the potential influence of confounding factors. The existing literature identifies several factors that affect academic performance, such as age (Jones & Weinberg, 2011), academic ability (Acuna, Allesina, & Kording, 2012; Hirsch, 2007), career length (Huang et al., 2022; Zeng et al., 2019), and research topics (Polyakov, Polyakov, & Iftekhar, 2017; Wei et al., 2013). For example, Huang et al. (2022) demonstrate that scientists' topic switch behaviors vary across career stages and are associated with distinct academic outcomes. Notably, high-performing scientists, especially those in the early stages of their careers, tend to explore emerging and interdisciplinary topics, rather than concentrating on established topics. Additionally, Polyakov et al. (2017) find that the research topics of a paper largely influence the number of citations and identify the hot topics in economic research. Therefore accounting for these confounding factors is necessary to draw clear conclusions.

## Research Objectives

Our study investigates the following research objectives:

- **Reconcile Contradictory Findings on Topic Switching and Academic Performance:** Systematically resolve the competing conclusions in prior literature by investigating how topic-switching behaviors impact long-term academic success, while strictly controlling for previously overlooked confounding factors using advanced regression and causal inference methods.
- **Develop Integrated Metrics for Topic-Switching Behaviors:** Build switch frequency and distance into a unified framework to capture the trade-off between risk-taking for impact and risk-aversion for productivity—perspectives typically analyzed in isolation by prior research.
- **Explore the Potential of "Conservative Adventuring" as a Strategy:** Examine whether "conservative adventuring"—frequent exploration within closely related domains—can be intentionally adopted as a practical strategy to enhance future academic performance.

## Materials and Methods

### Data

We have studied three large-scale datasets, covering disciplines of physics, chemistry, and biomedicine, respectively. In the main text, we present the analysis of the APS (American Physical Society) dataset and in the Appendix, the comparable chemistry and biomedicine datasets are analyzed, respectively. The APS dataset contains all papers from the journals in the APS from 1976 to 2015. Each paper is associated with up to five 6-digit PACS codes, and the PACS codes have a hierarchical structure, where the first digit represents the top level of classifications (Jia et al., 2017; Zeng et al., 2019). The chemistry dataset includes papers published in the ACS (American Chemical Society) journals and collected by the Microsoft Academic Graph (MAG), which assigns each paper several field codes under a hierarchical classification scheme (Hao, Xu, Li, & Evans, 2026; Park, Maity, Wuchty, & Wang, 2026; Peng, Teplitskiy, Romero, & Horvát, 2025). The biomedicine dataset is constructed from the PubMed data collected by MAG. Besides the widely-used MAG codes, papers in this PubMed dataset can be mapped to the hierarchically organized Medical Subject Headings (MeSH), as an alternative source of classification codes (Kang, Danziger, Rehman, & Evans, 2025; Sourati & Evans, 2023). For the APS dataset, we perform author name disambiguation using the approaches of other studies of the APS data (Martin, Ball, Karrer, & Newman, 2013; Sinatra, Wang, Deville, Song, & Barabási, 2016) and obtain 395,678 unique authors. The authorships in the ACS and PubMed datasets have already been disambiguated by MAG. In all three datasets, we keep authors with at least 10 papers. These steps result in 25,245 authors with 281,958 papers in the APS dataset, 62,316 authors with 747,061 papers in the ACS dataset, and 1,218,355 authors with 13,072,174 papers in the PubMed dataset (1,260,125 authors with 13,649,286 papers, if MeSH is used for PubMed.). Details on the datasets and processing procedures are presented in the Appendix (section A1).

## Switching Metrics: EP and ED

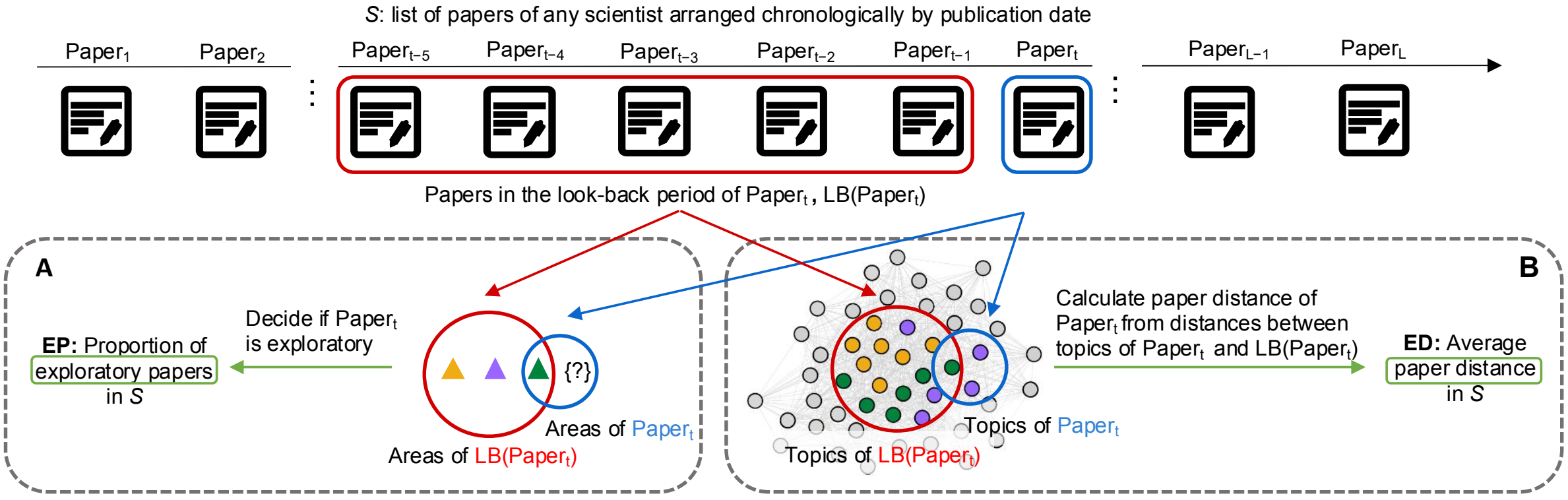

**Fig. 1 Computing exploration propensity (EP) and exploration distance (ED).** Suppose a scientist has a set of papers, denoted by $S$. (A) EP is calculated as the proportion of exploratory papers in $S$. A paper $Paper_t$ is exploratory if it covers at least one research area that has not been studied by the scientist in a past period, i.e., if it has at least one area not covered by LB($Paper_t$)–the papers by the scientist in a look-back period (the past five papers in this example). (B) ED measures how different (in terms of topic similarities) each paper of the scientist is from his or her previous paper, on average. Specifically, $Paper_t$'s distance is the distance between its topic set and that of LB($Paper_t$), where we average the pairwise distances of all possible topic pairs between these two topic sets.

To characterize scientists' topic switch behaviors, we introduce two metrics: exploration propensity (EP) and exploration distance (ED).

EP is the frequency that a scientist switches to unfamiliar areas (Fig. 1A). These switches are represented by the number of exploratory papers published by the scientist. Specifically, 'areas' of a paper are represented by the first two digits of the paper's PACS codes in the APS dataset (section A1.1) and an exploratory paper covers at least one area that is different from these of the scientist's past papers in the look-back period, which can either be defined as her past $J$ papers or the papers in the last $K$ years. The EP is then calculated as the fraction of the exploratory papers among all the scientist's papers. In the study presented in the main text, we compare each paper to the five papers before it to decide whether it is exploratory and study the outcomes. In the Appendix (section A7.5), we test 1 to 15 papers and 1 to 15 years as alternative look-back periods for the calculation of EP, and we also test infinity as the look-back period, which means we include all past papers in comparisons.

We complement switching frequency with distance. ED is the average distance of a scientist's papers to her previous papers, measured by research topic similarities, in the same look-back period as in the definition of the EP (Fig. 1B). For each paper, we calculate its distance to the scientist's last $J$ papers or papers within the last $K$ years (consistent with the choices in EP) and take the average over all papers. Like the calculation of EP, we calculate the distance of each paper to the five papers before and vary the look-back periods in the Appendix (section A7.5). Specifically, to compute the distance between a paper and the set of past papers, we measure the distance between two sets of topics, one from the focal paper and the other from the past papers in the look-back period, and the paper distance is the average distance of all possible topic pairs between these two topic sets. To calculate the distance between any given pair of topics, we construct a network where nodes represent said topics, and weighted edges indicate the frequency with which the two topics occur together in papers. There are two primary methods for determining this distance. The first involves analyzing the degree to which the neighbors of the nodes overlap after appropriate normalizations (Pan, Sinha, Kaski, & Saramäki, 2012). The second forms low-dimensional vector representations of the nodes in the network through a graph embedding technique known as node2vec (Grover & Leskovec, 2016), and calculates the cosine distance between the two nodes in question. Recall that each unique six-digit PACS code is a topic in the APS dataset. More details about the calculations and alternative definitions of EP and ED are in the Appendix (sections A2 and A7.3).

As a side note, while large language models may conceptually evaluate the relative closeness of topics, they lack the precision required for these quantitative calculations, which rely on structured methods and rigorously defined mathematical metrics.

## Performance Metrics: log-citations

To evaluate a scientist's academic impact, we employ citation-based performance metrics. We define a paper's 5 (or 10)-year log-citations (Mukherjee, Romero, Jones, & Uzzi, 2017; Poege, Harhoff, Gaessler, & Baruffaldi, 2019; Sinatra et al., 2016), denoted by log-c5 (or log-c10), by the logarithm of the number of citations it receives within five (or ten) years after publication (sections A1.1.3 and A7.4.1), and we typically measure a scientist's performance by averaging her papers' log-c5. Citations of scientific publications are known to follow heavy-tailed distributions (Brzezinski, 2015) and taking logarithms on said citations drastically reduces the undesirable variability in their numeric values (section A1.1.3).

## Regression Analyses

We design regressions to study the relationship between the proposed switching metrics (EP/ED) and the future performance (log-citations). The data of any scientist is split into two parts, where the first part before the split point corresponds to her 'past' for computing independent variables, and the second consists of the 'future' for an assessment of her performance in the rest of her career as the dependent variable.

### Regression with EP

To examine the impact of scientists' exploration propensity (EP) on future academic performance, we first construct a regression model that includes EP and multiple control variables.

The independent variables include the EP, the number of papers, and the scientist's past performance represented by the log-c5 per paper, and the future performance is evaluated by the average log-c5 of her future papers. We also control for the year and area of the scientist's first paper, since papers of different research areas and times often display distinct citation patterns (Radicchi, Fortunato, & Castellano, 2008). All data points used in the same regression share a split point, which is either the first $N$ career years (years that any individual has spent in her scientific career), or the first $M$ papers. Throughout the main text, the results are presented under the split point of ten career years, unless specified otherwise. The following regression (1) studies the relation between the scientists' past EP and the future performances:

$$\begin{aligned}\text{LogCit}_{\text{future}} = \beta_0 + \beta_1 \text{LogCit}_{\text{past}} + \beta_2 P_{\text{past}} \\ + \beta_3 \text{year}_{\text{first}} + \beta_4 \text{area}_{\text{first}} + \beta_5 \text{EP}_{\text{past}} + \text{Noise},\end{aligned} \tag{1}$$

For each choice of $N$ or $M$ for the split point, we process the data of any scientist with proper splitting and run a regression. Particularly, $\text{LogCit}_{\text{past}}$ is the average logarithm of the citations that the scientist's past papers receive, $P_{\text{past}}$ is the number of her published papers in the past, $\text{year}_{\text{first}}$ is the year of her first paper understood as a category variable, and $\text{area}_{\text{first}}$ is the area of her first paper, again understood as a category variable. The dependent variable, $\text{LogCit}_{\text{future}}$ is computed using the data dated after the split point by averaging the logarithms of the number of citations per paper in the scientist's future career.

### Regression with EP and ED

To further investigate the joint influence of EP and ED on future performance, we extend regression model (1) by constructing an augmented model (Eq. 2) that includes both EP and ED as key predictors:

$$\begin{aligned}\text{LogCit}_{\text{future}} = \beta_0 + \beta_1 \text{LogCit}_{\text{past}} + \beta_2 P_{\text{past}} + \beta_3 \text{year}_{\text{first}} \\ + \beta_4 \text{area}_{\text{first}} + \beta_5 \text{EP}_{\text{past}} + \beta_6 \text{ED}_{\text{past}} + \text{Noise}.\end{aligned} \tag{2}$$

where the newly added $\text{ED}_{\text{past}}$ is computed from the scientist's past papers. Similarly, we run one regression for each choice of split point. All the regression results are reported in the Appendix (section A4.1). Note that the alternative ways of measuring future performance as robustness checks are investigated in the Appendix (section A7.4).

### Individual Fixed-effect Analysis

To strengthen the robustness of our findings, we construct dynamic panel data and employ individual and time fixed-effect regressions. This approach controls for unobserved, time-invariant individual heterogeneity—such as inherent personal traits or baseline academic abilities—that might confound the relationship between topic-switching behaviors and future performance.

Specifically, we divide the timeline into evenly spaced periods. Within each period, we calculate a scientist's average log-citations, EP, ED, and other covariates based exclusively on their publications in that timeframe. This compartmentalization alleviates complex time dependencies present in the static regression framework. We apply the widely adopted Arellano-Bond estimator (system GMM) (Blundell & Bond, 1998) to solve the following model:

$$\begin{aligned}\text{LogCit}_{i,t} =& \beta_0 + \beta_1 \text{LogCit}_{i,t-1} + \beta_2 \text{EP}_{i,t-1} + \beta_3 \text{ED}_{i,t-1} + \beta_4 P_{i,t-1} \\ &+ \beta_5 \text{Career_year}_{i,t-1} + u_i + \eta_t + \text{Noise}_{i,t},\end{aligned} \tag{3}$$

where $\text{LogCit}_{i,t}$ is the average logarithm of the citations of scientist $i$ in period $t$, $\text{EP}_{i,t-1}$ and $\text{ED}_{i,t-1}$ measure scientist $i$'s EP and ED in period $t-1$, $P_{i,t-1}$ and $\text{Career_year}_{i,t-1}$ are the number of papers and last career year of scientist $i$ in period $t-1$, and $u_i$ and $\eta_t$ denote the individual and time fixed effects, respectively. Detailed validation tests and results for this dynamic panel approach are documented in Appendix section A4.3 and Table A10.

## Propensity Score Weighting

To systematically quantify differences in scientific careers across topic switch behaviors, we classify scientists into four groups based on whether their EP/ED metrics fall within the top or bottom half of the distribution. Because a scientist's group assignment is intertwined with other baseline characteristics, we designate one group as the baseline and apply the propensity score weighting (PSW) scheme. This approach assigns weights based on scientists' estimated odds of being in their respective groups, effectively controlling for inter-group discrepancies and rendering the groups statistically comparable.

Specifically, we use the highly versatile nonparametric generalized boosted model (GBM) (Friedman, 2001) to estimate these propensity scores. We then conduct a weighted regression where each scientist's weight is the reciprocal of their propensity score. The derived regression coefficients naturally provide the Average Treatment Effect (ATE) relative to the baseline. The ATE quantifies the counterfactual difference in future log-citations between hypothetical scenarios where all scientists adopted the treatment versus the baseline strategy.

## Ruling out Other Factors

To assess the robustness of our results against potential unobserved confounding, we calculate E-values (VanderWeele & Ding, 2017) for the statistical significance of EP and ED in our regression models. Large E-values indicate that substantial unmeasured confounding would be required to explain away our observed effects, thereby increasing confidence in the reliability of our findings.

While we have controlled for many easily conceivable factors and a quick calculation in the above regression analysis on the E-values (table A5) indicates that our results have taken enough confounding factors into account, it is still conceptually possible that other less visible factors may have more than negligible roles. For instance, after the split point, (A) perhaps the conservative adventurers are more likely to switch to hot areas, which often results in more citations (Wei et al., 2013)? Or (B) are they more likely to publish papers with high novelty and high conventionality, which were shown to be superior (Lin, Li, Ji, Xie, & Chen, 2025; Uzzi et al., 2013)? We have formulated more such hypotheses to propose factors that may explain away the discovered phenomenon as follows: (C) Do conservative adventurers work with larger teams such that more citations follow (R. Wu & Shi, 2025; Wuchty et al., 2007), or do their scholarly impacts mainly come from research in which they take secondary roles (Zhang, 2009)? (D) Do they change institutions more frequently (Allison & Long, 1990) or do they work in Ivy League schools such that they enjoy more resources and opportunities to produce high-impact work? (E) Is conservative adventuring the behavior of the focal scientist, or is it contributed by co-authors who bring in their preferences of research topics (F. Xu et al., 2022)?

To clarify such questions, we conduct several regression analyses where the independent variables further include the variables indicating the above listed factors of concern along with the usual ones used in the regression analysis. When such additional variables are considered or even if we include all the variables of these hypotheses in one of the regression studies, our results (section A6) are robust as the regression coefficients of EP and ED have the same signs and are always statistically significant, despite that the explanatory power of EP and ED combined may be weakened slightly. This implies that although more factors may be at play, the differences of scientists indicated quantitatively by EP and ED metrics are sound.

# Results

In this section, we begin by unpacking the reasons behind the conflicting links between topic switching and performance. We then employ the EP and ED metrics in regressions while accounting for confounding factors. Next, we group scientists based on their switching patterns to compare subsequent outcomes across groups. We further analyze the temporal dynamics of these behaviors across career stages and ages, and finally examine whether transitions into specific EP/ED profiles represent potentially strategic shifts.

## The Observed Correlation and its Disappearance

We present an example, using our data, to illustrate how the aforementioned contradictions in observed correlations may have arisen. As defined in sections 4 and A2.1, we use the exploration propensity (EP) of a scientist to measure her likelihood of switching to an unfamiliar area, estimated by the proportion of papers that explore unfamiliar areas among all her papers. We measure a scientist's performance by averaging her papers' log-c5 (see sections 4, A1.1.3, and A7.4.1).

Figure 2A presents an apparent negative correlation between the EP and overall career performance, which seemingly implies that someone who stays focused will have better performance. However, this implication can be easily challenged. For instance, the negative correlation can be explained the other way around—scientists who enjoy past successes are more likely to stick with their existing areas (Fig. 2B). When we look at a specific group of scientists with similar past performance (and thus the past performance is controlled for), the correlation between their past EP and future performance is almost non-existent (Fig. 2C).

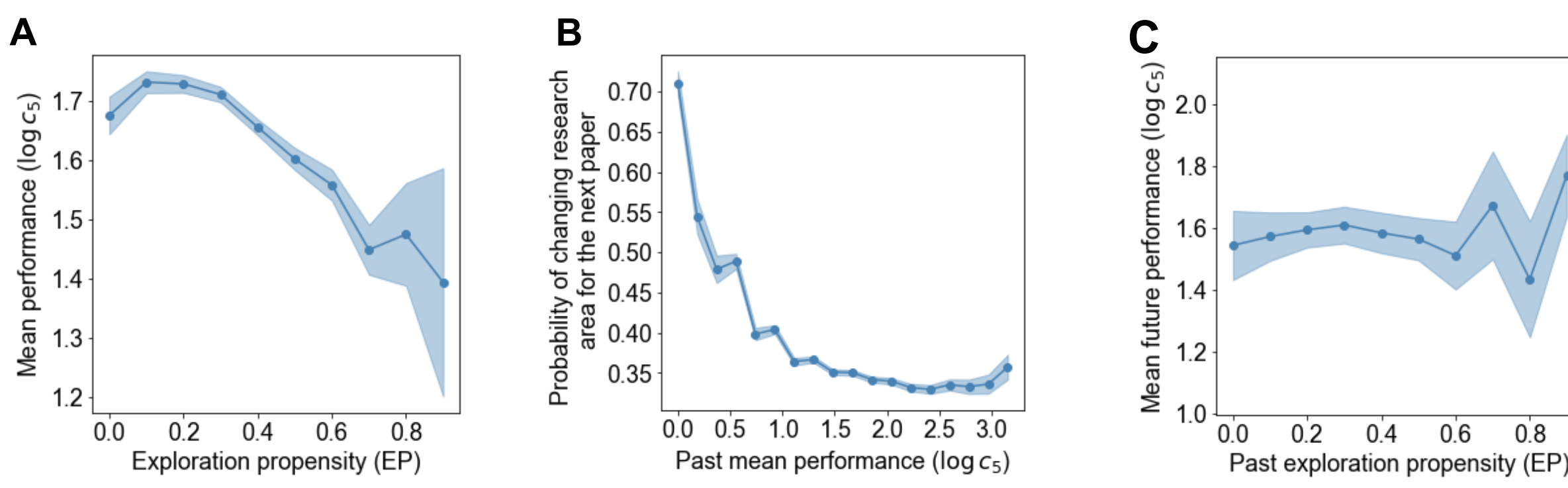

**Fig. 2 Interplay between switching and career performance.** (A) EPs and career performance. We calculate the EP and log-c5 per paper for each scientist's entire career and divide the scientists into groups by aligning their EPs to the nearest integer multiples of 0.1. Each blue point marks the mean of the log-c5 per paper of the scientists in that group, with 95% confidence bands. The career performance shows a decreasing trend as the EP increases. (B) Scientists with higher past performance (log-c5 per past paper) have a lower probability of changing research areas for their next paper. Each time a scientist publishes a new paper, we compute her past mean log-c5 to put the publication into a group of publications with similar past performance among all authors. For each publication in a particular group, we decide whether it is exploratory and calculate the portion of exploratory papers in the group as the group's probability of changing research area for the next paper (section A3.2). (C) Past EPs and future performance of a group of scientists whose past performance is controlled for. We regard the first ten years of each scientist's career as 'past', select the scientists whose past performance is in the small range of [1.74,1.90], and plot future performance against past EP, in a fashion like (A). The disappeared correlation suggests that past performance can be one confounding factor not to be ignored.

## Conservative Adventurers

We are interested in explaining the differences in scientists' future performance by the variations in their past research area switching behaviors measured by EP. As detailed in section 4, we conduct regression analysis correlating scientists' 'past' EP up to a split point with their 'future' performance, where we control for 'past' performance (log-citations), 'past' number of papers, and the area and year of the first paper. The regression results (table A2, column (S3) on the left) confirm the highly-intuitive relationship between future performance and past performance (Clauset, Larremore,

& Sinatra, 2017; Kong et al., 2020; Merton, 1968; H. Wang, Huang, & Bu, 2025), and indicate that EP has a statistically significant positive effect on scientists' future performance ($P = 2.2 \times 10^{-13}$). For those more likely to explore new areas (i.e., with higher EPs), their future papers have significantly higher average log-citations, as shown in the marginal effect plot in Fig. 3A, where future performance increases with respect to increasing EP after all other variables are fixed at their respective averages. Further analyses under varying temporal split points suggest a robust relationship between switching and performance (table A3).

Adding exploration distance (ED) to our regression setup as one more independent variable (section 4), we find that it has statistically significant negative effects on the future performance ($P = 3.7 \times 10^{-9}$), while the positive effects of the EP remain statistically significant (table A2, column (S4) on the left). The results are still true across comprehensive choices of split points with respect to both career length (2-15 years) and the number of papers (2-15) (table A4). These indicate that for someone with high EP but low ED, whom we categorize as a 'conservative adventurer', her future papers have higher average citations (Figs. 3B and 3C). Since it is generally expected that those with higher EP have higher ED and vice versa (Pearson correlation between EP and ED: 0.6039), individuals with high EP and yet low ED deviate from the norm and achieve a balance.

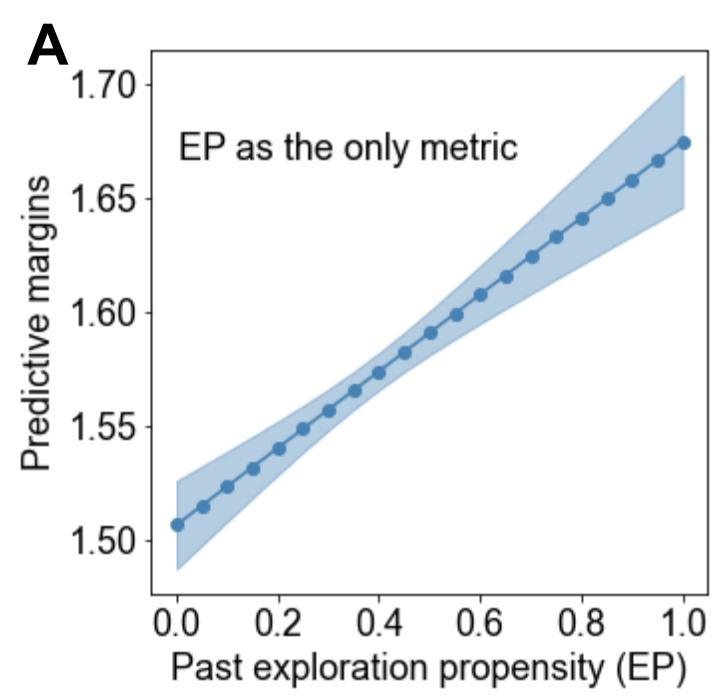

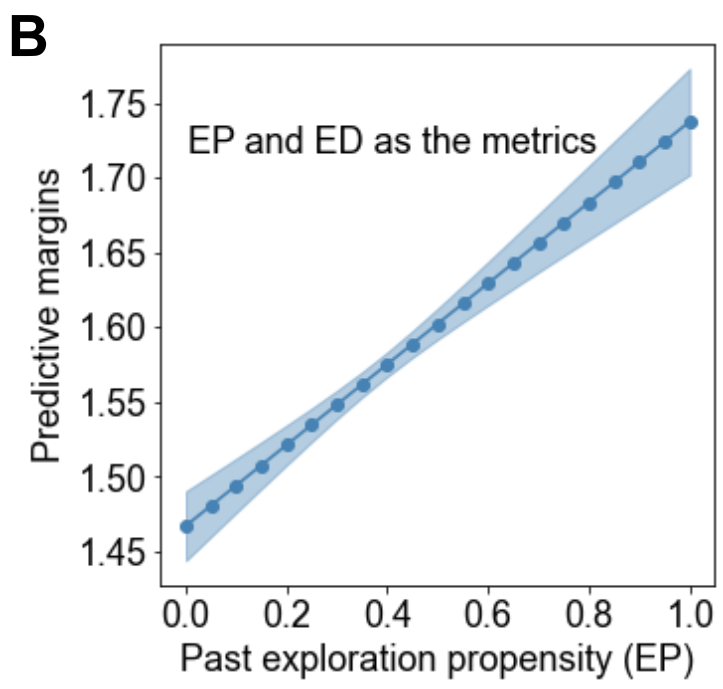

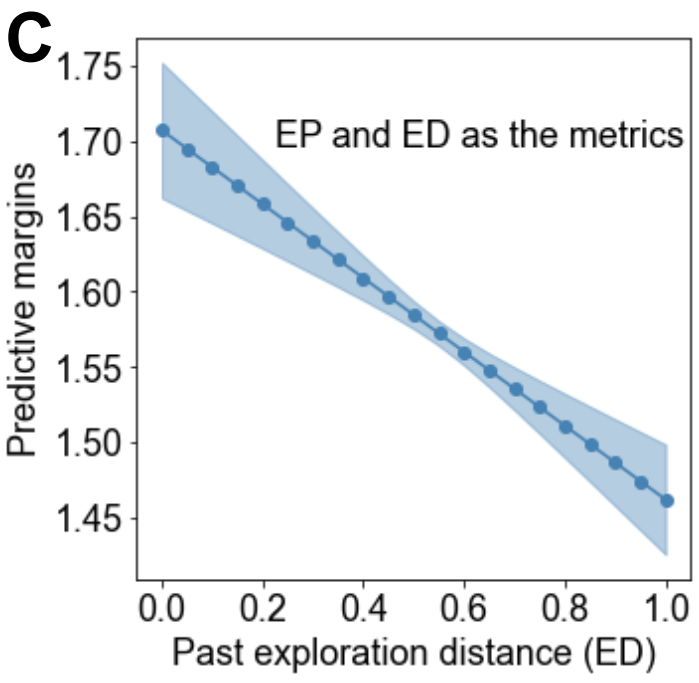

**Fig. 3 Marginal effect of the exploration metrics.** (A) We regress the future performance against the EP as the only metric while controlling for the usual variables. The marginal effect of the EP on the future performance is graphed. (B) Similar to (A), we run the regression with the EP and ED as the metrics and plot the marginal effect of the EP on the future performance. (C) The same as (B), except that the marginal effect of the ED on the future performance is shown.

Both the EP and ED give E-values (VanderWeele & Ding, 2017) (table A5) much larger than ones that scientists commonly consider safe to report their findings (larger E-values suggest more reliable findings, see 4). Besides, the regression results are consistent when we perturb the EP and ED with Gaussian noises (section A7.8). Note that similar conclusions hold consistently if we change the measurement of academic performance to the log-c10 per future paper or the log-citations of a scientist's most impactful future paper (Sinatra et al., 2016) (section A7.4). We also validate the robustness of our findings across career start periods and demographic subgroups (section A7.1): the core effect of the high-EP-low-ED pattern remains statistically significant for scientists with different career initiation periods, and our key conclusions hold consistently regardless of author name characteristics (i.e., scientists with Asian versus non-Asian last names) in the APS dataset. Furthermore, to eliminate potential temporal biases arising from the different starting years across datasets, we restricted the PubMed and ACS samples to post-1976 cohorts to match the APS timeframe. Our core conclusions regarding EP and ED remain highly consistent under this strict timeframe alignment (Appendix A7.5). We control for possible confounding factors, e.g., topic contribution from co-authors (Mariani et al., 2024; F. Xu et al., 2022), and high-EP-low-ED scientists preferring hot areas (Wei et al., 2013), combining novelty and conventionality in papers (Lin et al., 2025; Uzzi et al., 2013), seeking advantageous collaborations (R. Wu & Shi, 2025; Wuchty et al., 2007; Zhang, 2009), or utilizing institutional resources (Zhai et al., 2024) that may yield more academic impact. On the basis of this, we additionally conduct mediation analysis (section A6). To further consolidate the discovery, we convert the original data into dynamic panel data and perform author and time fixed-effect regressions (see section 4, and section A4.3). This approach allows us to account for the unobserved stable differences between individual scientists in the regression model. Passing the above robustness checks, our findings on EP and ED in the regression analyses remain (statistically) striking.

To better understand these high-EP-low-ED scientists, as well as other behavioral patterns quantified by EP and ED, we divide the scientists into four groups along two dimensions, one dividing EP into upper or lower halves of all EPs, and the other dividing ED into upper and lower halves. As anticipated, the high-EP-low-ED group is relatively small, consisting of only 14% of the population. Figure 4 visualizes one scientist for each group (chosen such that the four scientists started their careers in the same area for illustration purposes, see section A5). Here each node represents a topic, and any two topics that have co-occurred in at least one paper are connected by an edge, the length of which is the topic distance between the two. Nodes sharing the same color are in the same area defined by the APS and if a scientist has multiple papers with a particular topic, the corresponding node is proportionally larger. In Fig. 4, the high-ED representatives (right column) have visibly more spread topics and those in the high-EP groups (top row) have more colors, which indicates that they explore more areas. For the high-EP-low-ED group of conservative adventurers, the representative scientist has colorful nodes of moderate sizes that are clustered together, which implies that this scientist makes steady moves when choosing topics, yet these topics span a variety of areas.

## Quantifying the Effects via Propensity Score Weighting

The above analysis revealed a clear relationship between future performance and EP/ED, but we are further interested in precisely quantifying the difference between the four groups defined above. We first compute the citations per paper for each scientist's future career and obtain the group averages. In Fig. 5A, we see that 'conservative adventurers' have the highest future performance across split points from 2 to 15 career years; when the split point is two career years, for example, the advantage over the second place is almost 20%. However, this line of thinking does not take the inter-group differences in other independent variables into account. To remedy this, we borrow the standard method of Propensity Score Weighting (PSW) that handles multiple treatments (B.K. Lee, Lessler, & Stuart, 2010), which in our scenario correspond to the four high/low EP/ED groups. The method weights each scientist according to her estimated likelihood to be in her own group to make the groups comparable and calculates the pairwise ATE between groups (section 4 and section A4.2.3).

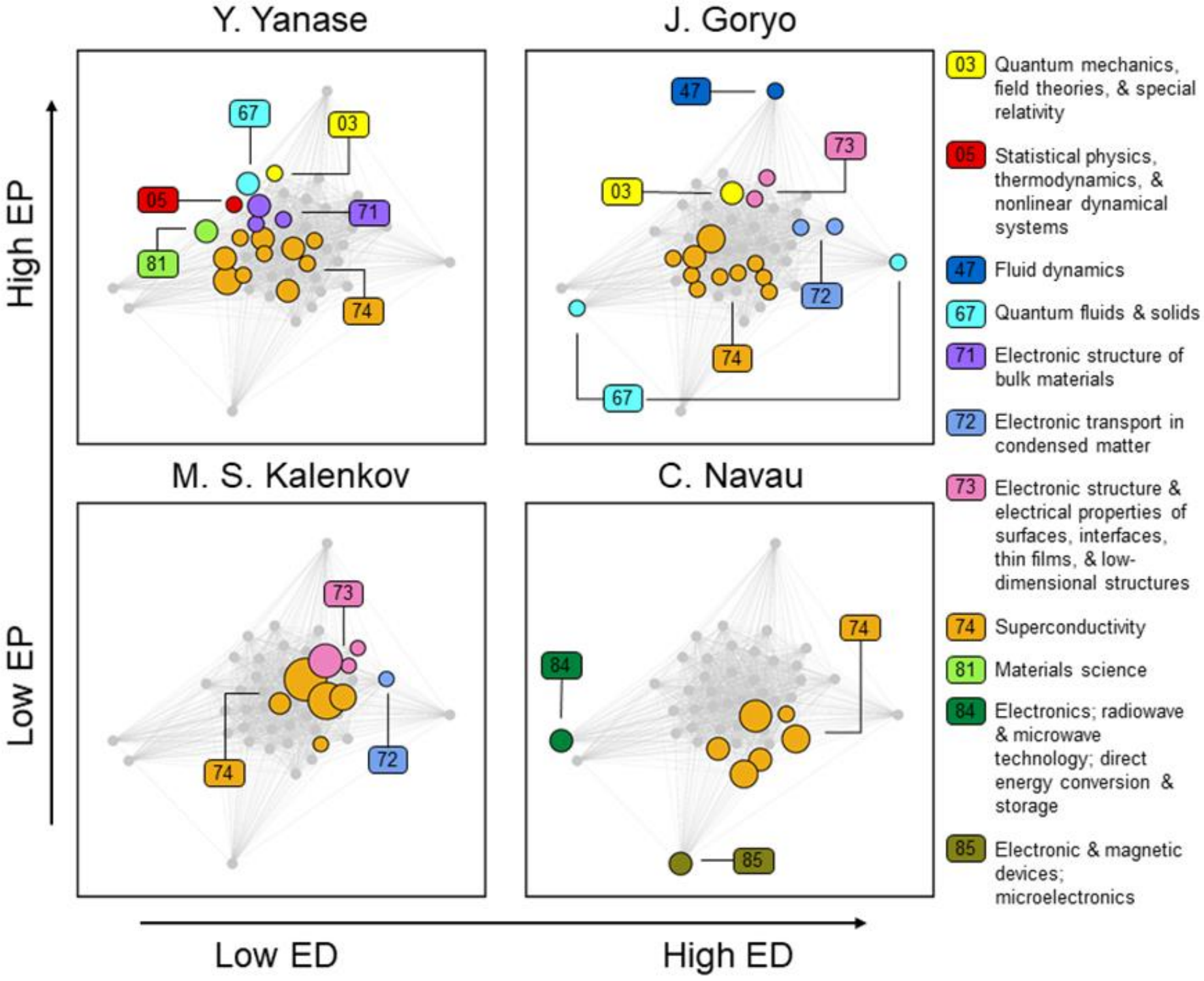

**Fig. 4 Sample scientist from each of the four high/low EP/ED groups.** Each node, colored or grey, represents a topic, and each edge linking two nodes indicates that the corresponding topics have occurred together in at least one paper, with its length showing the distance between the topics (section A2.2.2). In each subplot, only the topics covered by the corresponding scientist's papers are colored and expanded, where the colors correspond to the area containing the topic/node. Any topics/nodes of the same color (excluding grey in the background) are under the same area. The node size reflects how many of the scientist's papers are on that topic. Among the four groups, the high-EP-low-ED (conservative adventurers) and low-EP-high-ED ones are relatively rare combinations, each taking up about 14% of the entire population, while the other two combinations each represent roughly 36%.

We carry out our study in diverse split points consistent with those in the previous regression analyses. Figure 5B illustrates the pairwise ATEs across different split points on career years, against the baseline high-EP-high-ED groups. We see that the high-EP-low-ED groups (conservative adventurers) are always significantly better than the others. This implies that combining both serves as a powerful indicator of more successful scientific careers ahead. We also observe that the low-EP-high-ED groups have negative ATEs compared with the baseline and the low-EP-low-ED groups do not have a clear advantage/disadvantage over the baseline, with the ATEs being around zero. We further compare the conservative adventurers with the apparently opposite low-EP-high-ED groups by changing the baseline (Fig. 5C). The 'conservative adventurers' have statistically significant positive ATEs against the new baseline since the corresponding confidence intervals only contain positive elements. If the split point is set to four career years, the ATE of being in the high-EP-low-ED group is as much as 19% more citations per future paper (percentage translated from 0.1738 in the logarithmic scale, 95% CI = [0.1252, 0.2225], table A7). In a simplified setup where conservative adventurers are compared against the rest three groups combined, the ATE of being conservative adventurers is about 10% more citations per future paper (Fig. 5C). These findings pass thorough robustness checks and are consistent on the PubMed dataset (section A4.5) and the ACS dataset (section A4.6).

As the regressions suggest that a higher EP and lower ED are associated with better future performance, we next study how the inter-group differences change, if we adjust the thresholds to obtain more disparate group divisions. Instead of using the previous 'upper and lower halves' threshold, we vary it to 'upper and lower Q%' to obtain the high/low groups, with Q being 40, 30, and 25, respectively. We see in Fig. 5D that when we form the groups with more extreme EPs and EDs, the 'conservative adventurer' group displays increasing advantage over other groups. We conduct further analyses almost parallel to the above via propensity score matching (Ho, Imai, King, & Stuart, 2007)—a common (and less refined) causal inference technique, and they give consistent findings (sections A4.2 and A7.6).

## Behavior Evolution

It is natural to ask how the EP and ED evolve through scientists' careers and how scientists of different generations behave under these two metrics. Figure 6A displays the average EP and ED of scientists in their careers. The EP shows a decreasing trend while the ED grows over time, despite the early-stage sharp drop of the EP that may be partially associated with the way it is calculated (when there are relatively fewer past papers in the early stage, the EP can be high). Curiously, this suggests that junior scientists are more conservative adventurers than the future senior scientists that they become.

In Fig. 6B, we compare the EPs and EDs of three groups of scientists who started their careers in different periods of time (i.e., 1985-1989, 1990-1994, and 1995-1999). Kolmogorov–Smirnov (K–S) tests on the distributions of the EP and ED, respectively, suggest that scientists from these three periods have similar distributions of EPs, while scientists of newer generations become more conservative as indicated by having smaller EDs, which is consistent with existing findings (Rzhetsky et al., 2015).

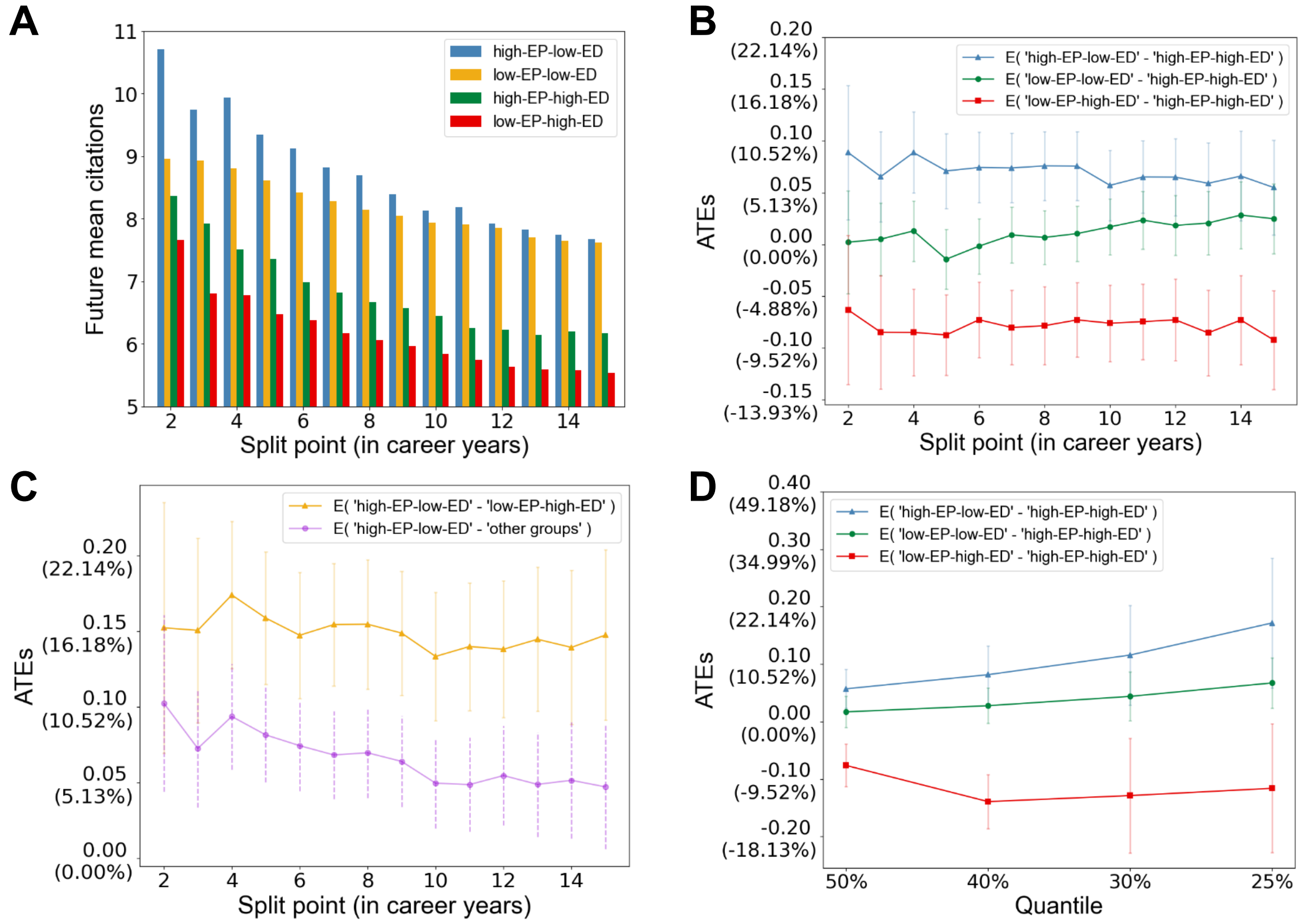

**Fig. 5 Better future performance of conservative adventurers.** (A) We compute the average future citations for each of the four high/low EP/ED groups with varying career years as split points. The conservative adventurer (high-EP-low-ED) groups have consistently the highest future performance, while the opposite (low-EP-high-ED) groups have the lowest. (B) We select the high-EP-high-ED groups as the baseline and calculate the ATEs between the baseline and the other three groups. The y-axis marks the ATEs of the three groups over the baseline with 95% confidence intervals, corresponding to the future citations per paper in the logarithmic scale with translated percentages in parentheses. The 'conservative adventurers' have the largest ATEs across all groups. (C) We compare the 'conservative adventurers' against the low-EP-high-ED baseline; the ATEs reported in orange are significantly larger than zero, with the largest ATE being 0.1738 (19%) at the split point of four career years. The violet line shows the (statistically significant) ATEs with conservative adventurers as the treatment group and everyone else as the control. When the split point is four years, conservative adventurers compared to everyone else have an ATE of about 0.094, which translates to 10% more future citations per paper. (D) We try different thresholds on EPs and EDs to form the four groups with the split point of 10 years. There is an increasing trend in ATEs when larger EPs and smaller EDs are required to form the conservative adventurer group.

How do memberships of the four high/low EP/ED groups change over time? In Fig. 6C, we track the scientists who started their careers before 2000 and remained active beyond 2009. We compute the group memberships at four time points—the ends of 2000, 2003, 2006 and 2009—using all data prior to the respective time point and observe how the memberships evolved. We see that over 60% of the 'conservative adventurers' in 2000 had remained in the same group in 2003 and the ratio was more than 72%, when we look at the transition from 2006 to 2009. The two groups in the middle are relatively stable across the years, with the ratios of staying being over 83%. Meanwhile, members rarely flow between completely opposite groups. For example, only 1.0% of the scientists in the low-EP-high-ED group turns into conservative adventurers from 2000 to 2003.

## Conservative Adventuring as a Strategy?

While our empirical analysis establishes a robust association between the conservative-adventurer profile and future academic impact, a critical question remains: is this merely a retrospectively observed pattern, or can it be consciously adopted as an actionable research strategy?

To address this, we draw a clear conceptual distinction between behavioral patterns and research strategies. In our framework, a behavioral pattern is purely descriptive and retrospective. It represents the quantitative measurement of a scientist's topic-switching record over a specific historical time window (e.g., their first ten career years). For instance, documenting that a researcher fell into the low-EP–high-ED quadrant during this past period is a measurement of their historical behavior.

Conversely, we conceptualize a strategy as a proactive, intentional transition in one's research approach. Because human behavior is fundamentally dynamic, researchers can and frequently do alter their topic-selection habits over time. When a scientist undergoes a drastic and purposeful shift from one established behavioral pattern to another in subsequent years—such as transitioning from a low-EP–high-ED profile to become a conservative adventurer (high-EP–low-ED) aimed at pursuing enhanced academic performance—we identify this deliberate transition as adopting a strategy.

In an ideal experimental setting, we would randomly assign scientists to adopt this specific strategy to observe the outcomes. Lacking such feasibility, we rely on observational datasets, acknowledging that the exact motivations behind any observed switching behaviors are multifaceted. Personal traits, such as conscientiousness and openness (Feist, 1998), or early-career learning curves common to knowledge production (Yelle, 1979), undoubtedly influence scientific behaviors.

If these unobserved factors remained entirely constant over time within individuals, the individual and time fixed-effect regressions (sections 4 and A4.3) would capture pure within-person effects—meaning an individual's future performance changes strictly according to alterations in their EP and ED. However, because traits and environments can evolve, we employ an alternative approach focusing on "drastic changers." Given our earlier observation (Fig. 6C) that most scientists maintain consistent research agendas over long periods, those who drastically shift from an opposite behavioral quadrant are likely executing a purposeful strategic change. These individuals serve as our observational proxy for a treatment group. By using a relatively late split point (10 career years) to identify these drastic changers, we further mitigate the confounding influence of early-stage learning effects.

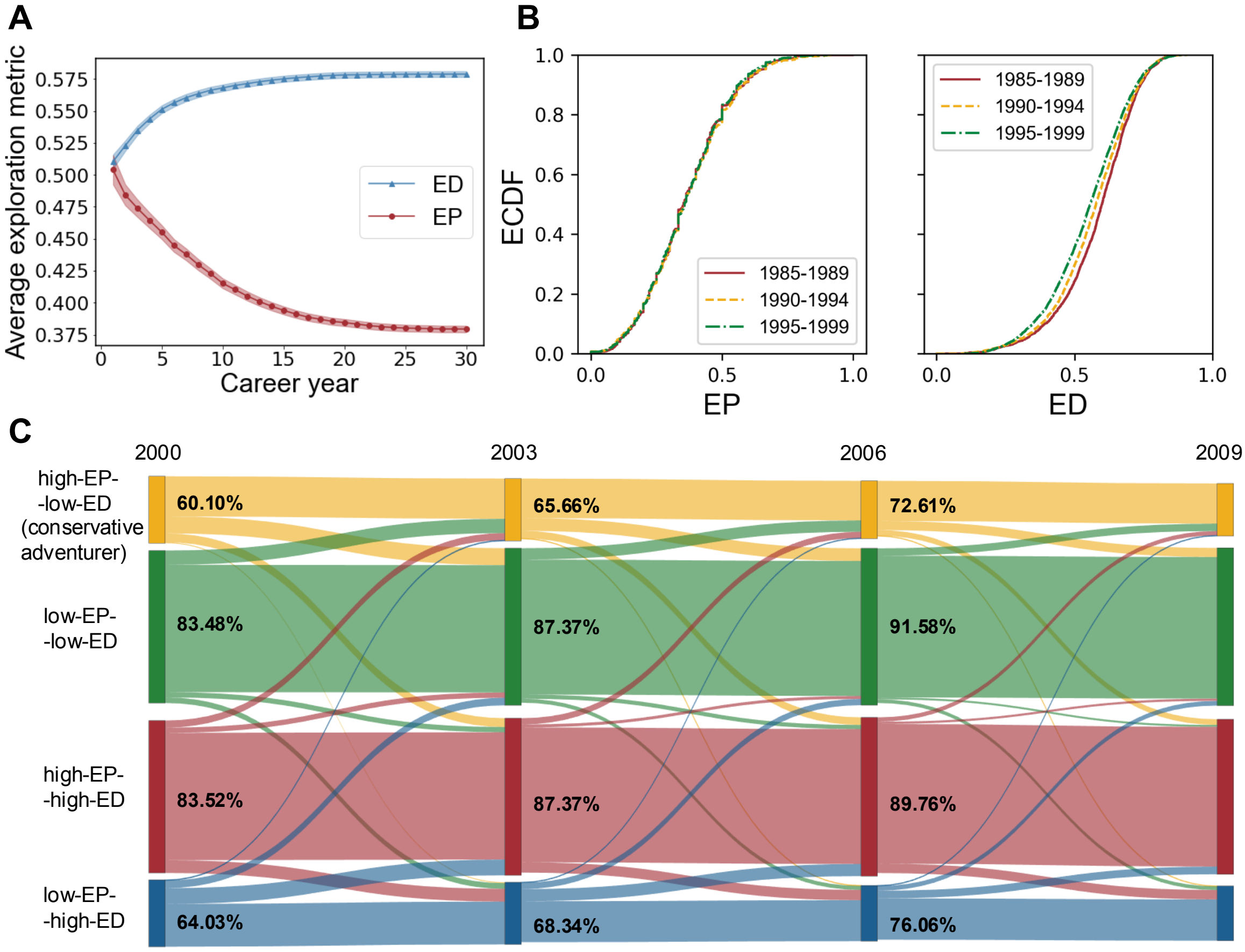

**Fig. 6 Temporal perspectives of EP and ED.** (A) The average EP and ED of scientists in each year of their careers with 95% confidence bands. We select scientists with careers of at least 15 years (58% of the population). For each year, we calculate each scientist's career EP and ED until the end of that year and compute the average. Notably, junior scientists have higher EPs and lower EDs than they do in later career stages. (B) The EP and ED distributions of scientists who are selected from those specified in (A) and whose careers began in 1985-1989, 1990-1994, and 1995-1999, respectively. The EP and ED are calculated within their first 15 career years. The three distributions of EP are similar (K–S tests, $P > 0.48$), while the distributions of ED are different ($P < 10^{-5}$). This suggests that scientists tend to become more conservative in choosing research topics in the past decades. (C) Member exchanges among the four groups in the 2000s. We select the 11,791 scientists who have at least two papers before 2000 and at least one paper after 2009, and divide them into groups in 2000, 2003, 2006, and 2009, respectively, where we set the threshold for having a high EP (ED) at the 50th percentile. For the 'conservative adventurers' in 2000, 37.21% of them were still 'conservative adventurers' by 2009, and 29.48% of them are in this group in all four snapshots.

Similar to Fig. 6C, we categorize scientists into four high/low EP/ED groups before and after the 10-year split. Strikingly, 8.6% of scientists with an initial low-EP-high-ED pattern drastically transitioned into conservative adventurers after this split, subsequently experiencing a 4.8% increase in citations per paper. Conversely, 6.9% of initial conservative adventurers drastically changed their behaviors in the opposite direction, suffering a 19.0% decrease in performance. Beyond these descriptive calculations, our PSW analysis compares these drastic changers against matched counterparts who maintained their initial groupings post-split. Scientists who actively transitioned into conservative adventurers experienced a 34.1% increase in citations per paper, whereas those who moved in the opposite direction saw a 26.2% decline.

These results, detailed in section A4.4, suggest that conservative adventuring can indeed operate as an effective strategy associated with higher research impact (excluding survival bias, see section A7.9), though it is currently rarely adopted. Nevertheless, we stop short of claiming definitive causality, as other exogenous factors—such as shifting societal demands (Hill et al., 2025; Omenn, 2006) and policy interventions (Merigó, Cancino, Coronado, & Urbano, 2016; Noorden, 2018)—might also prompt these substantial behavioral shifts.

# Discussion

By jointly quantifying EP and ED, we take a first step toward disentangling how topic switching relates to academic performance and toward reconciling the mixed evidence reported by prior studies. We discover that conservative adventurers have better future academic impact, a finding that remains consistent whether split points are defined by career years or publication volume. By categorizing scientists into four groups based on their switching behaviors, we identify a relatively rare but highly effective pattern of frequent exploration within closely related areas. Our unified EP–ED framework effectively captures the "essential tension" between risk-taking for impact and risk-aversion for productivity (Foster et al., 2015; Topper & Kuhn, 1977), which were often examined in isolation. Unlike previous correlational studies, our approach extensively controls for confounding factors and reveals the quantifiable predictive power of early-career switching strategies. Our study indicates that scientists with higher exploration propensity tend to achieve greater future impact, but those balancing risk-taking with shorter topic shifts gain even greater advantages. These scientists actively choose to explore diverse yet closely related topics, allowing them to expand their research horizons while maintaining continuity.

Crucially, our temporal robustness checks reveal a striking pattern: the performance advantage of the conservative-adventurer profile persists from the early twentieth to the twenty-first century. A fundamental question arises: how can a specific behavioral pattern remain consistently optimal despite profound shifts in scientific macro-environments, institutional structures, and publishing technologies over a century? The answer lies in the distinction between evolving micro-behaviors and the persistent macro-correlation driven by the scientific reward system.

While the specific topics scientists explore and the methodologies they employ have radically transformed, the foundational mechanisms of knowledge assimilation and peer evaluation have endured. Science is perpetually characterized by the tension between tradition and innovation (Topper & Kuhn, 1977). Regardless of the historical era, the scientific community's evaluation systems inherently favor novel ideas that are firmly anchored in established knowledge, while heavily penalizing radical leaps into overly distant cognitive domains (Boudreau, Guinan, Lakhani, & Riedl, 2016). This structural persistence is corroborated by prior longitudinal studies showing that atypical knowledge combinations grounded in high conventionality consistently yield the highest impact (Uzzi et al., 2013), and that "small pivots" universally lead to higher impact over the past five decades (Hill et al., 2025; Zeng et al., 2019). Thus, the enduring success of the conservative adventurer reflects science's immutable, structural demand for bounded novelty.

## Theoretical Implications

Our findings contribute to the longstanding discussion on the trade-off between exploration and exploitation (J.D. Cohen et al., 2007; March, 1991; Mehlhorn et al., 2015). While prior studies often emphasize either extreme exploration (Hill et al., 2025) or risk-averse strategies for productivity (Rzhetsky et al., 2015), we demonstrate that balancing these opposing strategies through deliberate, moderate exploration leads to higher academic impact. By introducing EP and ED as quantifiable metrics, we provide a unified framework to analyze behaviors previously studied separately, such as risk-taking for impact and risk-aversion for productivity. This balance not only enhances individual performance but also contributes to a broader understanding of how scientists navigate the tension between innovation and continuity.

The robust performance of the "conservative adventurer"—characterized by frequent topic switching (high EP) within proximal semantic distances (low ED)—can be deeply understood through the lens of recombinant innovation (Fleming, 2001; Weitzman, 1998). Scientific breakthroughs are fundamentally driven by novel combinations of existing knowledge components. A high exploration propensity ensures a continuous, proactive search for new combinatorial possibilities, allowing scientists to act as knowledge brokers across sub-disciplinary boundaries. Crucially, maintaining a low exploration distance ensures that newly explored topics remain conceptually adjacent to the scientist's existing knowledge base. This proximity drastically increases the likelihood that the recombinant components are intellectually compatible, yielding coherent and impactful innovations rather than disjointed conceptual failures.

Furthermore, the success of this strategy highlights the profound cognitive limits and the "burden of knowledge" (Jones, 2009) inherent in contemporary scientific careers. Transitioning into entirely new domains (high ED) imposes exorbitant cognitive learning costs, requiring scientists to master fundamentally different paradigms and literatures from scratch, which can temporarily diminish productivity. By contrast, conservative adventurers maximize interdisciplinary knowledge spillovers while minimizing these learning costs. Operating within proximal semantic distances allows researchers to leverage their existing absorptive capacity (W.M. Cohen, Levinthal, et al., 1990) and core methodologies. They frequently refresh their research agenda with adjacent novelties without paying the prohibitive cognitive penalty of extreme conceptual leaps. Additionally, our study resolves inconsistencies in prior research on topic switching and academic performance. While some studies reported negative correlations (Amjad et al., 2018; Zeng et al., 2019), others found positive ones (Huang et al., 2022; Yu et al., 2021). By validating our findings across datasets from physics, biomedicine, and chemistry, we demonstrate that moderate exploration paired with shorter topic shifts represents an optimal strategy. These results suggest that prior literature has provided, at best, an incomplete picture, which our study helps to clarify.

## Practical Implications

Our findings also offer actionable insights for young scientists and their mentors. Young scientists are behaviorally more conservative adventurers compared to their senior counterparts, who tend to focus more narrowly but make bolder switches when they do change topics. This generational difference may stem from increasingly specialized training driven by the expanding knowledge base in most fields (Jones, 2009; Park, Leahey, & Funk, 2005). While conservatism may provide early-career stability, our results suggest that adopting a balanced approach—combining risk-taking with continuity—can enhance future academic impact.

Among studies predicting future scientific impact, our metrics EP and ED stand out as practical and actionable early indicators. In contrast, many established predictors, such as past productivity (Allison & Stewart, 1974; Kong et al., 2020; D.H. Lee, 2019), mentorship (Long, Allison, & McGinnis, 1979; Ma, Mukherjee, & Uzzi, 2020; Williamson & Cable, 2003), or research grants (Bornmann, Wallon, & Ledin, 2008; Horta, Cattaneo, & Meoli, 2018), are less flexible or actionable for early-career researchers. By highlighting the potential of conservative adventuring as a behavioral strategy, we provide a path for young scientists to proactively shape their careers. However, the effectiveness of this strategy remains to be conclusively demonstrated and warrants further investigation.

## Comparison with Existing Work

As discussed in the 'Theoretical Implications' and 'Practical Implications' sections, our study has highlighted differences from prior research. We further compare with a recent study by Yang (2025), which examines the correlation between topic switching (measured by switch distance) and innovation metrics such as disruption, novelty, and interdisciplinarity.

While their work emphasizes the role of switching in fostering innovation, our study incorporates switch frequency to provide a broader perspective on its relationship with academic performance. Additionally, our findings suggest that investigating 'drastic changers' as quasi-experimental cases could extend their work, offering further insights into the impact of specific switching patterns. By focusing on the balance between risk-taking and continuity, our study complements their findings and provides actionable implications for academic impact.

## Limitations and Future Directions

Our quasi-experimental evidence suggests conservative adventuring may lead to greater academic impact, yet our study is limited by potential endogeneity arising from inherent characteristics and external demands (Hill et al., 2025; Merigó et al., 2016; Noorden, 2018; Omenn, 2006). Future research could address these limitations by examining exogenous societal or policy shocks.

Furthermore, our study is limited in fully capturing the precise cultural or institutional contexts of researchers. While we controlled for demographic variations in our supplementary robustness checks, we acknowledge the complex effects of global academic mobility and institutional socialization. For instance, researchers of Asian descent may be deeply socialized within Western academic institutions, profoundly shaping their professional strategies. Future studies utilizing longitudinal institutional affiliation histories are needed to properly assess how specific geographical and cultural environments shape scientific topic-switching behaviors.

Additionally, our dependent variable focuses on general academic impact measured by citations, a metric that fundamentally rewards cumulative knowledge-building and "bounded novelty." It is important to note that the conservative-adventurer strategy may not necessarily generalize to the attainment of major scientific prizes (e.g., Nobel Prizes). Prior literature indicates that elite awards often recognize singular, extreme-risk disruptive leaps rather than bounded exploration (J. Wang, Veugelers, & Stephan, 2017; L. Wu, Wang, & Evans, 2019), and award-winning scientists frequently exhibit distinct, highly radical topic-switching trajectories (Foster et al., 2015).

Finally, it remains an open question whether other creative domains, such as film-making, invention, and entrepreneurship, also have uniquely successful conservative adventurers.

# Statements and Declarations

**Funding.** This work is supported by the ShanghaiTech start-up fund and the Science and Technology Commission of Shanghai Municipality (25692108200). QK is partially supported by the National Natural Science Foundation of China (72204206), City University of Hong Kong (Project No. 9610552, 7005968), and Hong Kong Institute for Data Science.

**Conflict of interest.** There is no financial/personal interest or belief that could affect the objectivity of the submitted research results. No conflict of interests exists.

**Data and Code Availability.** The processed data and code for analysis and plotting, as well as the pointers to the original data, are available at https://conservative-adventurers.vercel.app/.

**CRediT authorship contribution statement.** **Xingsheng Yang**: Methodology, Investigation, Data Curation, Writing - Original Draft, Writing - Review & Editing.

**Zhaoru Ke**: Methodology, Investigation, Data Curation, Writing - Original Draft, Writing - Review & Editing. **Xiangyi Xiao**: Investigation, Formal Analysis, Writing - Review & Editing. **Qing Ke**: Methodology, Supervision, Funding Acquisition. **Fengnan Gao**: Conceptualization, Investigation, Resources, Methodology, Supervision, Project Administration, Funding Acquisition, Writing - Original Draft, Writing - Review & Editing. **Haipeng Zhang**: Conceptualization, Investigation, Resources, Methodology, Supervision, Project Administration, Funding Acquisition, Writing - Original Draft, Writing - Review & Editing.

# Supporting Information for

## Conservative adventurers have more future academic impact

### This PDF file includes:

Supporting text
Figs. A1 to A24
Tables A1 to A40
SI References

# Supporting Information Text

## A1. Data Description

**A1.1. The APS Dataset.** We perform our analyzes primarily on the APS dataset[*], which is provided by the American Physical Society (APS) and includes information of all papers published in the Physical Review series of journals between 1893 and 2020, totalling more than 670,000 publications. For each paper, the dataset contains information such as the paper's digital object identifier (DOI), the date of publication, the name of the journal of publication, and the name and affiliation(s) of every author. In addition, the citations between publications in the Physical Review journals and the PACS codes of each paper are provided.

The Physics and Astronomy Classification Scheme[†] (PACS) is the classification criterion of papers used by the APS journals, which we apply in the definitions of the exploration propensity (EP) and the exploration distance (ED) in our study. The PACS codes in each paper are submitted by the authors and reviewed by the editors. The scheme first came in use in the early 1970s and was discontinued in around 2015. More than 90% of the publications were assigned PACS codes between 1976 and 2015, and one publication can have up to five PACS codes. A PACS code takes the form of six digits with a hierarchical structure. The highest level in a PACS code is marked by the first digit, ranging from `0` (general) to `9` (geophysics, astronomy and astrophysics). The second level is marked by the first two digits, such as `04` (general relativity and gravitation) and `32` (atomic properties and interactions with photons). The third level is recorded by the first four digits, e.g., `91.25` (geomagnetism and paleomagnetism; geoelectricity). Overall, there are 10 categories in the top level, 73 in the second, and 948 in the third. The fourth and fifth levels are designed to be represented by the first five digits and the entire six digits, respectively (e.g., `91.25.F` for "rock and mineral magnetism" and `91.25.fd` for "environmental magnetism"). All the 5,971 unique PACS codes cover the third level, 80% of them the fourth, and 10% the fifth. These unique 6-digit PACS codes can be seen as a mixture of the finest possible levels—20% of it being third levels, 70% fourth levels, and 10% fifth levels. We call each 6-digit PACS code a research *topic* and use the first two digits to represent a broader research *area*.

***A1.1.1. Author Name Disambiguation.*** Since the APS does not maintain a list of unique author identifiers, a name disambiguation is required to associate scientists with their publications. The name disambiguation procedure of scientific publications usually starts with assigning a unique identifier to each author of any publication, then proceeds to merge the identifiers that appear to correspond to the same person. We infer the authorship from the author names and metadata available in each publication, which is similar with the name disambiguation methods in prior works (1, 2). First, we compile a list of 2,333,999 authors by treating each author from any publication as if each were unique; then, we merge the records of the supposedly identical authors into one entry in the following steps.

(1) We first combine the author records with the same last names and compatible first names into one group. By "compatible", we mean either the first names are exactly matched when the complete first names are available, or the first names have same initials, when deciding for these only have initials.

(2) In each group, we measure the degree of similarity of authors from the following perspectives: whether they worked in similar and/or related institutions, whether they cited each other, had the same co-authors, and published in journals of similar research areas. Records with similarities above a certain threshold are merged.

With the above merging process, we obtain 395,678 authors. To evaluate the accuracy of our disambiguation procedure, we randomly sample some author pairs from all the author pairs which have the same last names and compatible first names. In detail, we sample 200 pairs of author identifiers that our procedure classifies as identical and 200 pairs of author identifiers that our procedure considers different. Out of these 400 pairs, we evaluate the outcomes of our procedure by manually looking up names on ResearchGate[‡] and Google Scholar[§] to recover the ground truth. As such, we obtain estimates of the false positive rate—the fraction of pairs that are merged wrongly—to be about 5% and the false negative rate—the fraction of pairs that are supposed to merge while they are not—to be around 10%. Compared to other related works (1, 2), our error rates are low. Besides, we note that the errors produced by disambiguation are not evenly distributed among authors. Previous studies have pointed out that name disambiguation is more difficult for authors with Asian names (1). In fact, in the evaluation just above, 90% of the errors are related to authors with Asian names. We have repeated our main analysis in the subset excluding the authors with Asian names, and have found that our experimental results remain essentially the same (see Section A7.1.2).

[*] https://publish.aps.org/datasets
[†] https://journals.aps.org/PACS
[‡] https://www.researchgate.net/
[§] https://scholar.google.com/

***A1.1.2. Data Selection.*** After disambiguating the author names, we obtained the publication records of all authors. We only consider the authors and their publications that meet the following criteria: (1) the publications were published during 1976-2015, when it was required to submit the PACS codes along with each publication; (2) we exclude the authors with any publication that has an incomplete record in the dataset, in the sense that not all fields (publish date, PACS codes) related to the paper were included; (3) the authors should be associated with at least ten papers that were published between 1976 and 2015. Criteria (1) and (2) ensure the accuracy of the author's EP and ED calculations and allow us to trace the author's initial performance; criterion (3) ensures that the author had a substantial scientific career rather than, say, only participated in some projects as short gigs. Note that enforcing (2) removes about 6% of all the authors, a small proportion, and 281,916 papers published by 25,237 scientists are left in the dataset after this step of screening.

***A1.1.3. Log-Citations.*** The most prominent problems with citation-based performance measures are that the number of citations for papers grows naturally as time elapses, and the heavy-tailed distribution of citations leads to erratic performance measurements in (3). As such, for any paper, we use the logarithm of the cumulative citations it received within five years after its publication to measure its scientific impact. In the following sections, we call it "5-year log-citations", denoted by $\log c_5$. Taking logarithms on the citation numbers makes them much less problematic as the effect of sometimes extraordinarily large citation numbers in averaging is mitigated, and since the logarithm transformation is strictly monotone increasing on $(0, \infty)$, the ordering of the papers' importance (as measured by 5-year citations) is preserved.

***A1.1.4. Gender Assignment.*** In our study, we infer a scientist's gender from the name. Specifically, we use a commercial service genderize.io[¶] widely used in scientific studies (4), which has built a database of names mapping first names to binary gender labels by integrating information from publicly available census datasets. For each first name, genderize.io gives a gender prediction and provides both an estimated probability of its labelling being correct and the count of samples in its database. We only accept its gender labelling of the first names with at least 10 samples and probabilities of at least 60% to be correct in its database, such that relatively precise gender assignments are carried out. After this process, we obtain 17,140 male and 1,904 female scientists in the APS dataset.

***A1.1.5. Limitations of the APS dataset.*** Although the APS dataset has been widely investigated in scientific studies, there are several limitations that should be kept in mind.

First, scientists may have published their manuscripts in any journals, but we only study their publications in the APS journals. Similarly, papers may receive citations from journals not in the APS journal list, but we only count the citations between pairs of the APS journals.

Second, the PACS codes were submitted by scientists and might later be reviewed and modified by editors, so they might not have been objectively written down in accordance with the classification scheme. Besides, the PACS codes and their exact meanings were subject to revisions by the APS from time to time. Documented updates were frequent on an academic timescale: PACS underwent at least 23 version revisions between 1975 and 2010, averaging a major update roughly every 1.52 years,[‖] and authors assign codes at submission under the then-current taxonomy. Nevertheless, the PACS had been an enduring and the recommended classification standard for the APS journals for a long time.

Third, some papers had incomplete records—some miss the PACS codes or the scientists' name. For the record, we note that over 95% of all the publication records of the scientists with at least 10 papers during 1976–2015 have the PACS codes on the records. More than 99% of the publications have complete records of all the other information (paper DOI, publish date, number of citations, authors' names and affiliations, and journal of publications). Our analyzes are performed only on those papers which have both records of PACS codes and authors' names.

**A1.2. The PubMed Dataset.** We also examine the PubMed dataset, which covers biomedicine publications. This dataset is constructed from the Microsoft Academic Graph (MAG) (5), by requiring the URLs of the papers to be from PubMed. It originally contains 29,776,639 papers.

For the PubMed dataset, there are two distinct coding systems that can be used to identify the areas and topics of papers. The first is a classification scheme provided by MAG. It is inferred based on paper content and other publicly available information (5) and applies to all papers in MAG. The second is the hierarchically-organized Medical Subject Headings (MeSH)[**] specific to the PubMed dataset and provided by the National Library of Medicine (NLM). In this study, we not only use the MAG coding systems to infer the areas and topics of papers within PubMed, but also utilize the MeSH system to build a separate dataset for robustness check purposes.

[¶] https://genderize.io/

[‖] https://www.isko.org/cyclo/physh.htm

[**] https://www.nlm.nih.gov/mesh/meshhome.html

Specifically, MAG's classification scheme consists of 6 levels (Figure A1). For papers in PubMed, there are 45 categories in the first level, 39,666 in the second level, 65,949 in the third, 91,552 in the fourth, and 17,643 in the fifth. We use the first-level fields as *areas* for computation for the EP and the second-level fields as *topics* for the ED. As a sizeable portion of the publications (about 17.1%) has no records of the second-level fields, we fill these missing values with the fields of other levels in the following steps.

1. If the publication has records of its third-level fields, we find the corresponding second-level fields of each of its third-level fields, and consider the corresponding second-level fields as the "true" second-level fields of the publication.

2. If a publication does not have records of its third-level fields, we use the fourth-level fields to obtain the corresponding third-level fields and then the third to the second.

3. Those with only higher-level fields in their records than the third-level ones can be dealt with similarly in a recursive fashion.

We also utilize the MeSH for inferring PubMed papers' areas and topics. This involves mapping a MeSH code to the MeSH tree[††] and obtaining the corresponding area and topic, from its ancestor and descendant, respectively. Given a MeSH code, we use its ancestor at the first level (the MeSH tree starts at the zeroth level) as the area it represents, and the descendant(s) at the tenth level as the topic(s). If all leaf nodes of a code are at levels less than ten, we just use these leaf nodes as the code's topics. Note that the NLM modifies the structure of the MeSH tree annually,[‡‡] and we decide to use the most recent (2016) MeSH tree. By doing this, 99.94% of the papers have at least one MeSH code. In total, we have 14 areas and 39,660 topics using MeSH, and they are of comparable order of magnitude as the results from using MAG's classification scheme.

The MAG provides citations between publications, similar to the APS dataset. Recalling $\log c_5$, we count only a paper's citations in the five-year period after its publication, and take its logarithm to measure the paper's impact. Note that here we are able to count a paper's citation from all papers that cite it, instead of from only those published in the APS journals in the APS dataset (see Section A1.1.5).

All the publication entries in the PubMed dataset have field IDs and publication dates, and more than 99% of them contain author IDs. Similarly to the APS dataset, we consider only papers published before the end of 2016 to make sure that the 5-year log-citations (after their publications) are calculable, and only authors with at least 10 publications are studied. (In the later case of using 10-year log-citations, we only consider papers published before the end of 2011 for similar reasons.)

Finally, using the MAG and MeSH coding systems respectively, we arrive at the $\text{PubMed}_1$ dataset comprising 1,218,355 scientists and 13,072,174 papers, and the $\text{PubMed}_2$ dataset with 1,260,125 scientists and 13,649,286 papers. The divergence in the number of papers between the two datasets is attributable to differences in code coverage rates for the papers.

**A1.3. The ACS Dataset.** We also construct a dataset of chemistry papers published in the 86 American Chemical Society (ACS) journals, from 1879 to 2021, by filtering the MAG dataset. Originally, it contains 1,325,257 papers. It is then processed in the same way as the PubMed dataset and the papers' areas and topics are decided by the MAG classification scheme (the MeSH system is only available for PubMed.). In total, the processed dataset has 21 areas and 16,257 topics, covering 62,316 scientists and 747,061 papers.

## A2. Exploration Metrics

**A2.1. Exploration Propensity.** The exploration propensity (EP) measures a scientist's likelihood of switching to an unfamiliar area. It is calculated as the proportion of her exploratory papers that explore unfamiliar research areas among all her papers. An exploratory paper is a paper that covers at least one area that is different from the areas of the papers in the *look-back* period, which can either consist of her past $J$ papers before this paper or the papers in the $K$ years before the publication year. When $J$ is set to be 1, the EP is the probability that a scientist explores different areas between any two consecutive publications. When $J$ is $\infty$, we consider all papers before it when deciding whether a paper is exploratory or not, even if the previously explored area might have seen so much progress that it is quite different from the area a few years ago. One example would be "machine learning", which nowadays mainly consists of deep learning and is drastically different from what machine learning used to be several years ago. By default, we use $J = 5$ and in the robustness check, we vary $J$ from 1 to 15 and $K$ from 1 to 15 as well. Besides, we also set them to $\infty$ to consider all the past papers for the robustness checks.

Formally, assuming that a scientist has published $L$ papers, we need to decide how many of them are exploratory. We denote her $i$-th paper as $p_i$ and when $p_i$ is not the first paper ($i \in \{2, \ldots, L\}$), we compare its set of areas $A_i$ with $A_{i,m}$, which is the set of past areas from her $m$ papers published in the *look-back* period. If $A_i \not\subset A_{i,m}$, $p_i$ is an exploratory paper since it contains areas unseen in the *look-back* period. The first paper, $p_1$, does not have previous papers, and thus we do not determine whether it is exploratory. Therefore, $(L-1)$ out of $L$ papers can be determined, and we calculate the EP as the fraction of exploratory papers among these $(L-1)$ papers. Specifically, for the APS dataset, we use the first two digits of the PACS codes as the area identifiers, of which there are 73 unique ones, while for the PubMed dataset, we use the first-level tags of 45 different areas.

[††] https://meshb-prev.nlm.nih.gov/treeView

[‡‡] https://www.nlm.nih.gov/databases/download/mesh.html

If we use the entire six digits of PACS codes (research "topics", as described in Section A1.1) when determining exploratory papers, most papers will be exploratory, and a large portion of scientists will appear to explore all the time with high EPs. The other extreme is to use the first digit that represents one of the 10 top level classifications as a PACS code's area, and the coarse-grained classifications would result in very few switches and many scientists with EPs being zero. We therefore use the first two digits, and we try four digits in Section A7.3.1.

**A2.2. Exploration Distance.** To complement the EP, we calculate the exploration distance (ED) of a scientist by an average distance of her papers to their corresponding previous papers. A paper's distance to another is determined by the average distance between the two papers' topics. And to compute the distance between two topics, the straightforward approach based on counting topic co-occurrences in the same papers (6) is not applicable in our scenario, since 98.31% of the six-digit PACS code pairs in the APS dataset do not co-occur, despite that they may be indirectly related. If the two-digit PACS area codes are used instead of the six-digit topic codes for the co-occurrences, only the crude area-level distances can be captured, and finer details in topics are omitted. Therefore, it would be more appropriate to use six-digit PACS codes to include subtle (dis)similarity at the topic-level. As such, we adopt the construction of a co-occurrence based code similarity graph using the entire dataset and measure the pairwise code distances according to how the pair's neighbors overlap (7–9). After such a graph is constructed, the distance between any pair of papers can be determined, based on which we calculate the EDs of the scientists. We elaborate on these operations in the following subsections.

***A2.2.1. Topic Co-occurrence Graph.*** Following (9), we consider all papers in the focal dataset and construct a weighted graph, in which each node represents a unique topic and an edge connects two topics that have appeared in the same papers. The edge weight between topic $i$ and $j$ is $w_{ij} = \sum_{p \in H} (n_p - 1)^{-1}$, where $H$ denotes the set of all papers containing $i$ and $j$, and $n_p$ is the number of topics in paper $p$. $(n_p - 1)$ is the number of co-occurring topics that each topic has in paper $p$ and each co-occurrence has $1/(n_p - 1)$ as the weight from this paper. Therefore, $w_{ij}$ depicts the co-occurrences of $i$ and $j$, with each co-occurrence weighted by the number of all topic co-occurrences of $i$ or $j$ on the corresponding paper. The strength of each topic, $s_i = \sum_j w_{ij}$, is equal to the number of papers containing that topic, excluding these with single topics. With this graph, the similarity between topic $i$ and $j$ can be calculated using the *weighted overlap* indicator that considers their overlapping neighbors.

***A2.2.2. Topic Distance.*** In an unweighted graph, the similarity between two nodes can be measured by the number of their overlapped neighbors divided by the number of nodes in the union set of their neighbors (10). (9) extends this to a weighted graph, and defines the *weighted overlap* similarity between $i$ and $j$ as:

$$O_{ij} = \frac{W_{ij}}{s_i + s_j - 2w_{ij} - W_{ij}} \qquad \text{[A1]}$$

where $w_{ij}$ is the weight of the edge between nodes $i$ and $j$, $s_i$ is the sum of weights from all $i$'s edges, and $W_{ij}$, defined by $\sum_{k \in \Lambda_i \cap \Lambda_j} \frac{w_{ik} + w_{kj}}{2}$, represents the weight from overlapped neighbors of $i$ and $j$, and $\Lambda_i$ is $i$'s neighboring nodes. The subtraction of $2w_{ij}$ in the denominator is to exclude the effect of direct links between $i$ and $j$, if there is any. By definition, $O_{ij} = 0$ if $i$ and $j$ do not have any overlapping neighbors, while $O_{ij} = 1$ if $i$ and $j$ have the same set of neighbors. We then define the topic distance between $i$ and $j$ as: $\text{TD}_{ij} = 1 - O_{ij}$, and $\text{TD}_{ij} \in [0, 1]$ by definition.

We choose to compute neighborhood overlap for ED instead of measuring how nodes are directly connected (i.e., the weighted co-occurrence for the graph edges), mainly because that very few (1.69%) pairs of nodes are directly connected. For these pairs of connected codes, the similarities measured from the neighborhood overlap correlate positively with the weighted co-occurrence (Fig. A2), indicating that the neighborhood overlap measurement is a reasonable replacement.

Some APS codes may disappear in certain years (9) and if the code graph is constructed from papers in a short period of time, it may miss some codes such that the distances for these missing codes from other years cannot be computed. To avoid this, we use papers of all years to build the graph, and all the codes that have ever appeared are in this graph. Here we show that the code distances are relatively stable, regardless of whether we construct the graph with all papers or just the papers from a certain period of time, by adopting the analysis from (11). First, we construct node graphs using papers from different time periods. For each period, we compute the distances for all possible node pairs and form a distance matrix only for the nodes that appear across all periods. We then calculate pairwise correlations for all the matrices, including the one built with all papers as used in our method. The correlations in Table A1 are fairly high, indicating the stability of node distances over time. The first row of Table A1 shows that our distance matrix that uses all papers aligns well with distance matrices built from papers of various time periods, and the correlations are mostly above 0.75. The only exception is 1981-1985 (0.544), during which the published papers only account for 1.21% of all the papers.

***A2.2.3. Paper Distance and Exploration Distance.*** Before diving into the ED, we need to compute paper distances on papers in consideration. Similar to Section A2.1 where we decide whether a paper $p_i$ is exploratory by comparing its set of areas ($A_i$) with that of the $m$ papers before it ($A_{i,m}$), the paper distance is the average of pairwise topic distances between $p_i$'s topics and these of the $m$ papers:

$$\mathrm{PD}_{p_i} = \frac{1}{|T_i| \times |T_{i,m}|} \sum_{t_j \in T_i} \sum_{t_k \in T_{i,m}} \mathrm{TD}_{t_j t_k}, \qquad \text{[A2]}$$

where $t_j$ is a topic in $p_i$'s topic set $T_i$, and $t_k$ is a topic in the $m$ past papers' topic set $T_{i,m}$. The ED of a scientist is computed as the average paper distances of all her published papers in a certain period, which often corresponds to the part before the split point.

**A2.3. Exploration Metrics in Relation to Other Factors.** We examine the relationships between the two exploration metrics and the factors that are often associated with systematic differences in human behaviors connected to academic career performances. Some factors, such as years into careers and different periods of time as starting points of careers, have been discussed in the main text.

***A2.3.1. Gender.*** Using the gender inferring tool in Section A1.1.4, we obtain 17,140 male and 1,904 female scientists in the APS dataset and plot the distributions of EP and ED for the two genders in Figure A3. Applications of the Kruskal–Wallis (KW) tests suggest that there are no statistically distinguishable differences between the EP distributions of both genders ($P = 0.3886$), while they seem to differ (statistically) in their ED distributions ($P = 0.0045$). The average ED of men and women are 0.570 and 0.561, respectively. In light of this, we will include gender as an independent variable in the robustness check Section A7 to see the role it plays in scientists' future performance.

***A2.3.2. Team Size.*** Team sizes affect the extent to which scientists explore (12). For each scientist, we calculate her average team size across all her papers, where the team size of a paper refers to its number of authors. We then plot the EP and ED distributions over different team sizes in Figure A4 and find that the two metrics peak when the team size is around 5 and 6.

***A2.3.3. Ivy League Experience.*** We identify scientists who have worked for or studied at Ivy League universities according to their affiliated institutions in their publications. Figure A5 shows the distributions of EPs and EDs for Ivy versus non-Ivy scientists, and the KW tests as expected suggest that these two groups of scientists have different exploration patterns—the ones from the Ivy group have larger EPs ($P = 1.40 \times 10^{-12}$) and EDs ($P = 3.60 \times 10^{-13}$).

## A3. Preliminary Analysis of the EP and ED

Existing literature have reported the correlation between scientists' academic performance and their tendencies to switch to new topics, among which (13) found a weak negative correlation between the scientists' 5-year citations per paper and their degrees of concentration on certain topics measured (reversely) by the switching probabilities. In what follows, we study the EP and ED and explore the relationships between them and their relations with confounding factors.

We start the section by plotting the histograms of career-long EPs and EDs of all scientists in Fig. A6.

**A3.1. Correlational Analysis.** As pointed out in Fig. 2 in the main text, we observed a negative correlation between the exploration propensity and the scientists' academic performance (Pearson correlation coefficient: $-0.1072$)—as a scientist's overall exploration propensity increased, her academic performance became worse. As our study unveils in the main text and in the sections below, having higher EP should be in fact associated with better academic performance in the future, and we remark that this is an exemplifying example of the failure of correlational analysis that statisticians have always warned us about.

Furthermore, there is a more substantial negative correlation between the scientists' exploration distances and their overall performance (Pearson correlation coefficient: $-0.2852$) (Fig. A7).

**A3.2. Relations Between Past Performance and Future Explorations.** When conceiving new projects, a scientist may determine the direction of the next paper(s) based on her experience in areas and topics that she has worked on. Two hypotheses arise. By some psychological accounts, engaging in mastery experience builds a person's self-efficacy (14), i.e., a scientist becomes more confident in tackling new challenges after achieving successes in the past, so she is more willing to explore new topics; or alternatively, it may prompt the scientist to commit to the areas of her existing expertise and enter a scientific comfort zone, thus she would focus on topics that she is already experienced in and more comfortable with. Either seems plausible.

We call the later one of a scientist's any two consecutive papers a *new publication*, and this is the opportunity where she could choose to switch topics and decide how far away the next topic would be from previous ones. We observe the connection between the scientist's performance measured by the 5-year log-citations per paper prior to her each new publication, and the outcomes of the publications measured by (the average of) whether an area switch takes place and (the average of) how far away the next paper's topics are from the last one. We find that scientists with better past performance are more likely to continue focusing on areas of familiarity, and vice versa, see Fig. A8.

## A4. Experiments

Given the negative correlation between a scientist's academic performance and her exploration propensity throughout their academic career (Section A3.1) and our observation of the relation between the prior performance and the future explorations (Section A3.2), we would hypothesize that the prior exploratory metrics can be used to explain and predict future successes. To test the validity of this idea, we design two types of tasks: choosing each (i) scientist's $i$-th ($i = 2, \dots, 15$) career year or (ii) $i$-th ($i = 2, \dots, 15$) article as split point, and using the exploration metrics prior to the point to predict their academic performance after the split point. For each type of task, we first apply regression analyzes to quantitatively verify the idea and then study the relations further via propensity score matching and weighting.

**A4.1. Regression.** Keeping in mind the systematic differences among the scientists, we introduce the following regression analyzes to study the explanatory and predictive powers of the aforementioned exploratory metrics on future performance while controlling for other features.

We construct the regression equation to study the relation between the future academic performance and the variable $\mathrm{EP}_{\mathrm{past}}$—the EP before the split point—while controlling for the other independent variables:

$$\begin{aligned}\mathrm{LogCit}_{\mathrm{future}} = \beta_0 + \beta_1 \mathrm{LogCit}_{\mathrm{past}} + \beta_2 P_{\mathrm{past}} \\ + \beta_3 \mathrm{year}_{\mathrm{first}} + \beta_4 \mathrm{area}_{\mathrm{first}} + \beta_5 \mathrm{EP}_{\mathrm{past}} + \mathrm{Noise},\end{aligned} \qquad [\mathrm{A3}]$$

where $\mathrm{LogCit}_{\mathrm{future}}$ represents scientists' average academic performance after the split point, and $\mathrm{LogCit}_{\mathrm{past}}$, $P_{\mathrm{past}}$, $\mathrm{year}_{\mathrm{first}}$, and $\mathrm{area}_{\mathrm{first}}$ record the average performance and the number of papers before the split point, and the publishing year and areas of the first paper, respectively. Note that $\mathrm{year}_{\mathrm{first}}$ and $\mathrm{area}_{\mathrm{first}}$ are both categorical variables and handled as such. The above regression output is reported in the column of Model Eq. (A3) in Table A2, where the past EP has a statistically significant positive impact on the future performance. The result is consistent with the (later) propensity score weighting/matching results in Section A4.2.

We then add the scientists' average ED before the split point into the mix of independent variables on top of Eq. (A3):

$$\begin{aligned}\mathrm{LogCit}_{\mathrm{future}} = \beta_0 + \beta_1 \mathrm{LogCit}_{\mathrm{past}} + \beta_2 P_{\mathrm{past}} + \beta_3 \mathrm{year}_{\mathrm{first}} \\ + \beta_4 \mathrm{area}_{\mathrm{first}} + \beta_5 \mathrm{EP}_{\mathrm{past}} + \beta_6 \mathrm{ED}_{\mathrm{past}} + \mathrm{Noise}.\end{aligned} \qquad [\mathrm{A4}]$$

The corresponding column in Table A2 indicates that after controlling for the performance and EP in the past (before the split point), the effect of the ED on the future performance is statistically significantly negative. This is later confirmed again by the propensity score weighting results: scientists who explore new topics at shorter distances enjoy certain advantages in the future.

We also propose the rudimentary regression model

$$\mathrm{LogCit}_{\mathrm{future}} = \beta_0 + \beta_1 \mathrm{EP}_{\mathrm{past}} + \beta_2 \mathrm{ED}_{\mathrm{past}} + \mathrm{Noise}, \qquad [\mathrm{A5}]$$

to investigate the basic relations between the future performance and $\mathrm{EP}_{\mathrm{past}}$ and $\mathrm{ED}_{\mathrm{past}}$ jointly. The output in Table A2 is consistent with what we have known, albeit in a crude fashion.

As illuminated in the above, scientists with higher EPs and lower EDs have better academic performances as measured by 5-year log-citations per paper. Similar to the main text, we go a step further and partitioned the scientists into four groups with four different kinds of patterns in exploratory metrics and put the group membership in the following regression study. Following the same framework as above and after calculating the EP and ED for every scientist, we label each scientist either high/low EP/ED, depending on whether her EP/ED metric lie in the top half or the bottom half of all. Then high-EP-low-ED scientists go to group A, low-EP-low-ED ones B, high-EP-high-ED ones C and the rest (low-EP-high-ED ones) D. Then we propose the next regression formula by

$$\begin{aligned}\mathrm{LogCit}_{\mathrm{future}} = \beta_0 + \beta_1 \mathrm{LogCit}_{\mathrm{past}} + \beta_2 P_{\mathrm{past}} \\ + \beta_3 \mathrm{year}_{\mathrm{first}} + \beta_4 \mathrm{area}_{\mathrm{first}} + \beta_5 \mathrm{group} + \mathrm{Noise},\end{aligned} \qquad [\mathrm{A6}]$$

where the variable group is the categorical variable indicating to which of the four groups a scientist belongs with group D as
the baseline. The regression result is shown in Table A2. We find that group A have the best future performance, followed by
C and B, and finally D. This again supports our analyzes in the main text via propensity score weighting about differences
among the four groups.

To verify the robustness of our results, we run regressions of Model Eq. (A3) and Model Eq. (A4) with different split points,
from 2 to 15 in number of papers and from 2 to 15 in career years, although some care is necessary to handle the EP that only
have values of 0 and 1 when the split point is set to be two papers. The regression results of Model Eq. (A3) are shown in
Table A3, where all the coefficients of the EP are (statistically) significantly positive. The more refined regression analyzes
with Model Eq. (A4) are shown in Table A4, and again all the coefficients of the EP are (statistically) significantly positive and
all the coefficients of the ED are (statistically) significantly negative. The consistency of results under exhaustive choices of
split points suggests the robustness of our results.

For the potentially unobserved confounding factors, we use E-values of the sensitivity analysis (15) to measure their potential
impact on the above regression results. Specifically, we use the "EValue" package[§§] in R to obtain the E-values of $EP_{past}$ and
$ED_{past}$, respectively. See the package documentation for detailed references and Table A5 for exact numbers. It suffices to
conclude that in the experiments with split points set in terms of career years, both $EP_{past}$ and $ED_{past}$ have large E-values,
which provides further evidence of the soundness of our discoveries on top of Section A6.

**A4.2. Propensity Score Matching and Weighting.** To further solidify the reported relations between the proposed exploration
metrics and future performance, we conduct more analyzes via propensity score matching (PSM) (16) and propensity score
weighting (PSW) (17). We design tasks similar to our regression analyzes, where the data are pre-processed similarly with the
length of career years and the number of papers serving as split points. However, when applying the PSM, we have to study
the EP and ED separately due to the limitation of the methodology itself.

***A4.2.1. The EP.*** Specifically, for each split point, the analysis on the EP via PSM is conducted in the following steps:

1. First, we divide the scientists into a treatment group with their EPs lying in the top half of all EPs and a control group
with their EPs in the bottom half, where the EPs are calculated (as usual) with each scientist's data before the split
point.

2. Then PSM is used to match scientists in the treatment and control groups, where the covariates (to be matched) are the
individual's performance, the number of papers before the split point, the PACS code and the year of the first paper.
After playing around with the `caliper` hyper-parameter, we obtain the matched groups, both being subsets of the original
treatment and control groups, respectively. The choice of `caliper` relies on human interventions such that the obtained
groups after matching are statistically indistinguishable from each other in terms of their past performance and other
covariates, and thus are *balanced*.

3. Finally, we compare the differences of the future performance of scientists between the matched treatment and control
groups.

In particular, we use the implementation of the `MatchIt`[¶¶] package from `R` to carry out the matching process in the above
second step. We mostly follow the default settings recommended by the package for tuning parameters, except for `caliper`.

The results of the above analyzes are detailed in the EP columns in Tables A6 and A8. The future performance of the
scientists in the treatment group are statistically significantly ($P < 0.01$) higher than those in the control group, regardless of
the choice of the split point. For example, we take the choice of the default split point of ten career years, and have 14,159
individuals in both the treatment and control groups in total, and after the matching 4,983 pairs of scientists remain. We
have ensured that the covariates such as the past performance are statistically indistinguishable between the groups after the
matching. On average, the future performance of the scientists in the treatment (high-EP) group is higher than those in the
control (low-EP) group by 4.71%.

***A4.2.2. The ED.*** We do similar analyzes for the ED. The methodology is virtually the same, except in two parts. The first is that
the variable of interest is no longer the EP, and the second is that we added the EP to the covariates to be matched via PSM.
The latter is to ensure that the obtained subsets of both groups have similar (and thus comparable) distributions in the EP.

As expected, the future performance of scientists in the high-ED treatment group is (statistically) significantly lower than
those in the low-ED control group, regardless of which split point is chosen. See the ED columns in Tables A6 and A8 for more details on the results. We point out that under the default split point of 10 career years, the future performance of the scientists in the treatment group is lower than those in the control group by 6.40%, on average.

[§§] https://cran.r-project.org/web/packages/EValue/index.html
[¶¶] https://cran.r-project.org/web/packages/MatchIt/

***A4.2.3. The EP and ED combined and PSW.*** The above results suggest that those with high EP (low ED) will have an edge in future academic performance, over their otherwise similar counterparts with low EP (high ED). But do the scientists with both high ES and low ED perform better than others? And if so, by how much?

Recall that in the main text and Section A4.1, we have divided the scientists into four groups A, B, C and D, each with different exploration habits quantified by both the EP and ED. We first conduct an elementary analysis via PSM by finding a one-to-one matching between group A and D, similar to what is done in Sections A4.2.1 and A4.2.2. The matching results show that the treatment group A performs statistically significantly better than the control group D, as expected. Again, we refer to the corresponding columns in Tables A6 and A8.

However, the PSM is in itself no longer sufficient to deal with the multiple-group scenario, and as a result, we introduce the propensity score weighting (18).

For simplicity of exposition, we start by observing two groups of data, one 'treatment' group and the other 'control'. In practice, whether a scientist is in the treatment group is intertwined with her other features, resulting in systematic difference between the two groups. The PSW scheme assigns scientists weights with the generalized boosted model (GBM), which depends on the estimated probability of each scientist being in either the treatment or the control group given her measured features. The GBM algorithm is an iterative scheme to combine the predictive power of a large collection of simple regression trees with proper weighting. At each iteration, the algorithm seeks to add to the model a small refinement in the form of a simple regression tree to improve the fitness of the current model to the data measured by the Bernoulli log-likelihood, where the simple regression tree is a recursive algorithm to find a partitioning of the covariate space to estimate the relation between the covariate space and the treatment assignment. At each recursion step, the simple regression tree algorithm finds a further level of partitioning of the current model such that the prediction error is minimized. The GBM keeps iterating to find a maximum-likelihood fit, if a proper stopping criterion is specified. Then we calculate the average treatment effect (ATE) of being in different groups, by performing a weighted linear regression on just the dummy variable indicating the group membership, with one of the groups as baseline. Alternatively, we can also compute the average treatment effect on the treated (ATT) by virtually the same procedure as above, except for minor differences in obtaining the weights via GBM. The ATE and ATT are both quantities in causal inference to measure such causal effects with subtle differences. Roughly speaking, the ATE measures the population (i.e. all scientists) difference in outcomes between the treatment and control, while the ATT measures the difference in outcomes if the scientists in the treatment group were in the control group.

Specifically, in our case, the covariates for calculating the propensity scores are the same as the independent variables in Section A4.2.1. The obtained ATEs for varying definitions of split points are reported in Tables A7 and A9 and Figs. A9 and A10, and they consistently suggest that group A outperforms any other groups and particularly by a significant margin group D, while groups B and C have similarly better performance than group D. Similarly, we compute the ATTs as a robustness check, and the results are presented in Figs. A11 and A12. Note that the results on the ATTs are virtually the same as those on the ATE.

We also compare group A (as the treatment) against all other three groups combined (as the control) with the same methodology, and the ATEs of being group A under varying split points are shown in Fig. A13.

**A4.3. Fixed Effect Analysis.** In this section, we control for individual fixed effects by constructing panel data, to explore from another perspective the connection between past EP and ED and future performance.

Specifically, following the Methods section in the main text, we divide the 21 calendar years (1995 – 2015) into 7 three-year periods (or 30 calendar years (1986 – 2015) into five 5-year periods) and perform the calculations of all relevant variables. Then, we solve the following regression equation using the Arellano–Bond estimator (19):

$$\begin{aligned}\text{LogCit}_{i,t} = \beta_0 + \beta_1 \text{LogCit}_{i,t-1} + \beta_2 \text{EP}_{i,t-1} + \beta_3 \text{ED}_{i,t-1} + \beta_4 \text{P}_{i,t-1} \\ + \beta_5 \text{Careeryear}_{i,t-1} + \mu_i + \eta_t + \text{Noise}.\end{aligned} \qquad \text{[A7]}$$

Note that the above Eq. (A7) is the same as the fixed-effect regression equation in the Methods section in the main text. Here, we treat Careeryear as an endogenous variable, i.e., a variable correlating with the observational noise in each time period.

The results in the Table A10 show that after controlling for individual fixed effects and a range of other covariates, the coefficients on EP and ED remain statistically significant, which is consistent with the findings in Table A2. Note that we also conduct the Hansen test and the AR tests (19), respectively, to justify our model in Eq. (A7). The Hansen test tells us that our model does not suffer from the over-identifying problem, which indicates our choice of the independent variables and endogenous variables is sound; the AR tests show the first-order autoregressive structure is valid, which permits us to employ system GMM in solving Eq. (A7).

**A4.4. Performance of “Drastic Changers”.** As mentioned in the main text, most scientists intentionally or unintentionally maintain consistent research agendas, while a very small portion of scientists exhibit drastic behavioral changes, such as transitioning from low-EP-high-ED to the completely opposite high-EP-low-ED (conservative adventurers) and vice versa. We may guess that these ‘drastic changers’ alter their research strategies on purpose. Here we examine how their performance correlates with these drastic changes.

To be more specific, for the default split point of 10 career years, we focus on group A and group D and observe the performance changes of these scientists who adopt the opposite strategies after the split point, i.e., scientist who were in the high-EP-low-ED (low-EP-high-ED) group before the split point and joined the low-EP-high-ED (high-EP-low-ED) group after the split point. We compare their performance after the split point to 1) their own performance before the split point and 2) the after-split performance of their counterparts, who were in the same group as the scientists before the split point and remain in the original group after the split point.

In the comparison of within-personal performance, scientists in group D (group A) who employ the opposite strategy after the split point, i.e., turning into group A (group D), comprises only 8.6% (6.9%) of the group. This dramatic change results in an increase of 4.77% (a decrease of 18.97%) on average in the individuals’ performance after the split point, as compared to their performance before the split.

We also conduct a comparative analysis of the future performance between scientists who change their research strategy and their counterparts who do not. Specifically, we focus on scientists in group D (group A) prior to the split point. Scientists who remain in the same group after the split point are assigned to the control group, while those who adopt the opposite strategies are allocated to the treatment group. We employ the PSW technique to measure the difference in after-split performance between the control and treatment groups while controlling for the usual covariates. The results indicate that the average treatment effect between the control group and the treatment group is 0.293 (-0.304) with a $P$ value of $4.55 \times 10^{-8}$ ($1.33 \times 10^{-6}$), which translated to a percentage is 34.14% (-26.23%).

**A4.5. Results on the PubMed Dataset.** To figure out how the past EPs and EDs of scientists are connected with future scientific impact in biomedicine, we conduct similar analyzes using regression and propensity score weighting on the PubMed dataset. Recall the difference between $\text{PubMed}_1$ and $\text{PubMed}_2$. In this section, we perform the same regression and PSW analyzes on both PubMed datasets, and the results are consistent. We therefore do not distinguish the two in the descriptions below.

First, we run the regression Eq. (A4) on the dataset with varying split points, from 2 to 15 in number of papers and in career years, respectively. As is shown in Tables A11 and A13, the regression coefficients of the EP and ED with all split points are (statistically) significant and have the right signs, which is consistent with our findings in the APS dataset (Table A4). The result suggests that conservative adventurers have higher future impact in biomedicine as well. When measuring scientists’ future impact, we also employ another method—the maximum 5-year log-citations after the split point as an alternative, and the results are robust (Table A12).

Furthermore, we utilize propensity score weighting to verify the advantage of being in group A, where the groups A, B, C and D are defined similarly. The experiments are similar to the above regression process, and the calculated ATEs for varying split points consistently show that group A performs better than any other group (see Figs. A14, A15, A17 and A18).

In the studies on the APS dataset in the main text, we see that group A has an expanding advantage as the thresholds on constructing high/low EP/ED groups (A, B, C and D) become extreme. As such, we test different thresholds in the PubMed dataset, and the results in Figs. A16 and A19 verify that the conclusions are similar to those before.

**A4.6. Results on the ACS Dataset.** In this section, we investigate the relationship between the past EPs and EDs of scientists and their subsequent scientific impact in the field of chemistry. To this end, we also employ regression and propensity score weighting techniques to analyze the ACS dataset.

First, we conduct a regression Eq. (A4) on the ACS dataset, utilizing a range of split points varying from 2 to 15 in the number of papers and career years. The results, as presented in Table A14, indicate that the regression coefficients of EP and ED are (statistically) significant and have the correct signs. The results are in agreement with those obtained from the APS dataset (Table A4) and PubMed dataset (Table A11). Our findings indicate that conservative adventurers have a higher future impact in chemistry. The results are robust when we employ “maximum 5-year log-citations” as an alternative method for measuring scientists’ future impact (Table A15).

In addition, we employ propensity score weighting to investigate the advantage of being in group A. The experimental procedures are similar to those of the aforementioned regression analysis, and the ATEs for different split points consistently demonstrate that group A outperforms all other groups (Figs. A20 and A21).

Furthermore, we also demonstrate that group A exhibits a widening advantage as the thresholds used to construct high- and low-EP/ED groups become increasingly extreme in the ACS dataset (Fig. A22).

## A5. Case Study

We select a representative case for each of the four groups in the main text. The representatives start their careers in similar topics and have comparable numbers of publications. Specifically, the areas of their first publications in the APS journals are the same, which all correspond to the PACS codes with their first two digits being "74"; the differences between their numbers of publications in the first ten years of their careers are no more than one. For clear illustrations, we select those specific cases so that individuals with high (low) EDs work on topics that have longer (shorter) average distance between their PACS codes in their own respective groups (cf. Section A2.2.2) and those with high (low) EPs have more (fewer) explored two-digit areas, comparing with other individuals within the same groups. Besides, the number of explored areas (marked by the first two digits of PACS codes) and explored topics (marked by the first six digits of PACS codes) should be comparable both between the two high-EP (low-EP) individuals with different EDs, and the average distances between PACS codes of their papers should be comparable both between the two high-ED (low-ED) individuals with different EPs.

## A6. Analysing Other Possible Latent Factors

As we mentioned in the main text and the Methods section, several conceivable mechanisms may provide alternative explanations to why conservative adventurers have greater future impact. Here, we study some possible hypotheses, from the viewpoints of (A) hot research areas, (B) paper novelty, (C) advantageous collaborations and (D) changing institutions, and demonstrate how we may exclude these hypotheses. The related results in the entire section are summarized in Fig. A23 and Tables A16 and A17. As usual, the *past* in this section corresponds to all the information of any scientist before the split point, while the *future* to that after the split point, the split point in all the experiments in this section is 10 career years or 10 papers. The independent variables (covariates) are inherited from Section A4.1.

**A6.1. Hypothesis A: Research Area.** As conservative adventurers explored new areas more frequently, they could be chasing hot areas or just have greater exposure to hot areas, thus they garner higher future impact. The question is in fact two-fold:

1. Are conservative adventurers more likely to publish on hot areas in the future?
2. Will working on hot areas in the future explain why conservative adventurers have higher future impact?

To answer the first question, we examine the effect of having different past exploration patterns, measured by the EP and ED, on the popularity of the areas of their future publications. Recall that a research *area* correspond to the first two-digit of PACS codes. For a specific year, we gauge the popularity of an area—$\mathrm{Popularity}_{\mathrm{area}}$, by the proportion of papers that are associated with the area among all the papers published in the same year. Since a paper may have several (unique) areas as identified by the first two digits in its PACS code(s), we define the area popularity of a paper—$\mathrm{Popularity}_{\mathrm{paper}}$, either by (1) the average area popularity of all the associated PACS codes, or (2) the maximal area popularity of those codes. With either definition, we run a regression analysis where the dependent variable is the average area popularity of a scientist's future papers—$\mathrm{Popularity}_{\mathrm{future}}$, against the usual independent variables as in Section A4.1. The regression equation is as follows:

$$\begin{aligned} \mathrm{Popularity}_{\mathrm{future}} = {} & \beta_0 + \beta_1 \mathrm{LogCit}_{\mathrm{past}} + \beta_2 P_{\mathrm{past}} \\ & + \beta_3 \mathrm{year}_{\mathrm{first}} + \beta_4 \mathrm{area}_{\mathrm{first}} + \beta_5 \mathrm{EP}_{\mathrm{past}} + \beta_6 \mathrm{ED}_{\mathrm{past}} + \mathrm{Noise}, \end{aligned} \quad \text{[A8]}$$

where the future area popularity is the average area popularity of a scientist's future papers. We find that conservative adventurers are more likely to publish high area-popularity papers in the future ($P < 0.001$, see Table A16 and Fig. A23), since the above regression reports that both the EP and ED have statistically significant coefficients and the coefficient of the EP (ED) is positive (negative). The results hold regardless of which definition of the $\mathrm{Popularity}_{\mathrm{paper}}$ is used.

Finally, we carry out a regression analysis of the future scientific impact when controlling for the $\mathrm{Popularity}_{\mathrm{future}}$, and find that the EP and ED remain statistically significant ($P < 0.001$, see Table A17), though area popularity may explain some of the difference. The regression equation is as follows:

$$\begin{aligned} \mathrm{LogCit}_{\mathrm{future}} = {} & \beta_0 + \beta_1 \mathrm{LogCit}_{\mathrm{past}} + \beta_2 P_{\mathrm{past}} \\ & + \beta_3 \mathrm{year}_{\mathrm{first}} + \beta_4 \mathrm{area}_{\mathrm{first}} + \beta_5 \mathrm{EP}_{\mathrm{past}} + \beta_6 \mathrm{ED}_{\mathrm{past}} \\ & + \mathrm{Popularity}_{\mathrm{future}} + \mathrm{Noise}. \end{aligned} \quad \text{[A9]}$$

**A6.2. Hypothesis B: Paper Novelty.** (20) suggested that the highest-impact publications have a combination of high novelty and high conventionality. Conservative adventurers may be more creative as they explored more frequently, so they may publish papers with higher novelty. At the same time, as they tended to explore closer areas, their publications may be somewhat more conventional in a sense. As such, publishing this type of paper may explain why conservative adventurers have greater future impact. To test the hypothesis, we evaluate the following two questions:

1. Do conservative adventurers tend to publish papers with both high novelty and high conventionality in the future?

2. Will publishing paper with both high novelty and high conventionality in the future explain why conservative adventurers have more future scientific impact?

We estimate the effect of the past exploration behavior (exploration propensity and exploration distance) on the probability of publishing papers with both high novelty and high conventionality by a regression analysis, similar to those in the above Section A6.1. We calculate the novelty and conventionality of each paper and label the papers of high novelty and high conventionality in the same way as (20). Then we run regressions where the dependent variable is the probability of a scientist publishing papers of both high novelty and conventionality. Similarly to the discovery in Section A6.1, we find that conservative adventurers are more likely to publish papers of both high novelty and conventionality in the future ($P < 0.001$, see Table A16 and Fig. A23).

Finally, we perform a regression analysis on the future impact while controlling for the probability of publishing papers with both high novelty and conventionality in the future. The results are again similar to those in Section A6.1 and suggest that the coefficients of EP and ED remain statistically significant ($P < 0.001$, see Table A17) with the correct signs, though publishing papers of both high novelty and conventionality may explain some of the difference in the future impact.

**A6.3. Hypothesis C: Advantageous Collaborations.** Are conservative adventurers prone to seek more advantageous collaborations? Some studies showed that teams often produced works of higher impact (21) than individuals, and having more distinguished co-authors may be partially responsible for greater future scientific impact. Besides, we measure scientists' EP and ED based on the codes of papers, and these codes may be "brought in" by their co-authors. Through measuring EP and ED, are we quantifying the scientists' behavior or their co-authors' behavior? We need to further quantify the co-authors' contribution to the paper topics. Here, we study whether 1) conservative adventurers having distinguished co-authors who contribute more to the papers, 2) co-authors bringing in their PACS codes or 3) conservative adventurers working with larger teams is the reason for higher future impact.

To begin with, we assess the author's contribution by whether she takes an important role in leading the research or managing correspondence. Similar to (22), we consider a scientist to be a *lead author* if she is the first or last author of the publication. Then we study whether the probability of having lead-author publications is different between conservative adventurers and other scientists. We find that having higher ED in the past predicts having higher proportions of lead-author publications in the future ($P < 0.001$, see Table A16 and Fig. A23). Finally, we measure the future impact of a scientist by only considering papers of which she was the lead author and thus has contributed the most to the publication. Our conclusions on the EP and ED are the same under this setting ($P < 0.001$, see Table A17).

For the second question, we measure the co-author's "contribution to topics" as a function of the co-author's likely importation of codes from her prior work to a focal paper (23). Following this idea, we first compute the number of unique codes that the focal author brings to the focal paper from her past 5 papers, denoted as $I_{focal}$. We then compute this code importation metric for all other authors combined except the focal author, denoted as $I_{other}$. Other co-authors' contribution is computed as $\frac{I_{other}}{I_{focal}+I_{other}}$. Note that here we try both 2 digits and 6 digits of PACS codes, to represent "area" contribution and "topic" contribution respectively, in the context of our main text.

We next perform the usual regression routine similar to Eq. (A8) and Eq. (A9), only on the newly constructed factor—co-authors' area/topic contribution in the future, and find that scientists with low EP and low ED are more likely to have co-authors with higher contribution in the future. When we set the future impact as dependent variable, and control for the co-authors' area/topic contribution in the future on top of the usual independent variables, the coefficients of EP and ED remain statistically significant ($P < 0.001$, see Table A17).

We also study whether the future team sizes have much to do with the EP and ED. We define the team size of each paper by the number of its authors. Then we run regressions where the dependent variable is the average team size of the scientist's future publications. We find that a lower ED is a statistically significant predictor of a larger average future team size ($P < 0.001$, see Table A16 and Fig. A23). Then we conduct one more regression analysis on the future scientific impact controlling further for the average future team size, and the obtained coefficients of the EP and ED remain statistically significant ($P < 0.001$, see Table A17) and have the correct signs.

**A6.4. Hypothesis D: Changing Institutions.** Conservative adventurers may work in different institutions in the future, where they may have higher impact because of being in institutions of higher reputation (24), and this may lead to greater future scientific impacts. Besides, they often have better opportunities in institutions of great reputation, e.g. Ivy League schools.

We measure a scientist's tendency of switching institutions, by dividing the number of institutions by the number of publications. By adding the tendency of changing institutions and whether working in the Ivy League in the future to the mix of dependent variables separately, we find that being conservative adventurers is a statistically significant predictor of working with more institutions in the future ($P < 0.001$, see Table A16 and Fig. A23), and having higher ED predicts well whether a scientist would work in the Ivy League in the future ($P < 0.001$, see Table A16 and Fig. A23). A further regression analysis, where these two factors are controlled for on top of the usual independent variables, finds that the coefficients of EP and ED remain statistically significant ($P < 0.001$, see Table A17).

**A6.5. Combining Hypotheses A–D.** To see if combining all the above hypotheses would explain away our main findings, we take all the corresponding variables in question mentioned above and perform a regression analysis with the following equation:

$$\begin{aligned} \text{LogCit}_{\text{future}} = \beta_0 &+ \beta_1 \text{LogCit}_{\text{past}} + \beta_2 P_{\text{past}} \\ &+ \beta_3 \text{year}_{\text{first}} + \beta_4 \text{area}_{\text{first}} + \beta_5 \text{EP}_{\text{past}} \\ &+ \beta_6 \mathbf{X}_{i,\text{attribute}}, \end{aligned} \quad \text{[A10]}$$

where $\mathbf{X}_{i,\text{attribute}}$ is a vector collecting all the attributes we studied in Section A6. We see in Table A17 that the coefficients of the EP and ED are always statistically significant, i.e., the established positive relation between having higher future research impacts and being conservative adventurers, cannot be explained away by the above hypotheses, separately or jointly.

**A6.6. Mediation Analysis.** We examine the correlation between the above confounding factors and past EP/ED and find that several confounding factors we study in this section are correlated with the exploration metrics (see Table A18 and Table A19). To accurately assess the direct and mediating effects between the EP/ED and the future impact, we employ mediation analysis for hypothesis A–D, which enables us to assess both the direct path coefficients linking the treatment variables (the EP or ED) to outcome variables (the future impact) and the indirect path coefficients mediated by intermediate variables (the confounding factors). By doing so, we can examine the presence of the mediating effect by the confounding factors, as well as quantify the respective magnitudes of the direct effect of exploratory metrics and the mediating effect of confounding factors on future impact.

In detail, we include $\text{LogCit}_{\text{past}}$, $P_{\text{past}}$, $\text{year}_{\text{first}}$, and $\text{area}_{\text{first}}$ as control variables, following the approach in the vanilla regression model (Eq. (A4)). We separately set one of the EP and ED as the treatment variable (while controlling for the other), and different confounding factors as the mediating variable, aiming to determine the magnitude of the direct and mediating effects. To calculate these effects, we utilize the "Mediation" package[***] in Python. The results of the mediation analysis are presented in Table A20 and Table A21. Notably, the average casual mediation effects (ACME) consistently exhibit small magnitudes, indicating a minor role of mediation in the overall effect. In contrast, the average direct effect (ADE) consistently shows statistical significance ($P < 0.01$) and accounts for a substantial portion of the total effect, affirming the existence of a direct relationship between EP/ED and future impact. Hence, the correlation between the exploratory metrics and future impact cannot be explained away by these mediating factors.

## A7. Robustness Checks

### A7.1. Different Subsets of Scientists.

***A7.1.1. The Minimum Number of Publications Requirement.*** In the main text, we select scientists with at least 10 publications to make sure they are not short-term researchers, and this selection inevitably (wrongfully) screens out some scientists. Here, we repeat our analysis under different selection rules (of authors with at least 5 publications and at least 20 publications) with other settings being default (10 career years or 10 papers being the split point), and find that the results remain statistically significant in the regressions (see Table A22).

***A7.1.2. Removing Scientists with Asian Last Names.*** As mentioned in Section A1.1.1, the disambiguation process may bring identification errors to scientists, especially to those with Asian names. Therefore, we follow the approach suggested by other studies (1) and exclude scientists with the top 200 most common Asian last names, in a list collected from Wikipedia. After such a removal, we obtain a smaller dataset with 21,963 scientists and perform the same regression analysis (see Equation (A4)). We find that the coefficients of the EP and ED are still statistically significant (see Table A23), and this suggests that our conclusions are robust against possible inaccuracies in name disambiguation.

[***] https://www.statsmodels.org/stable/generated/statsmodels.stats.mediation.Mediation.html

***A7.1.3. Different Research Period.*** Scientific publishing institutions, incentives, and dissemination technologies have changed across the twentieth and twenty-first centuries. A natural concern when using long bibliographic corpora is whether conclusions about topic-switching patterns and later impact are comparable across subperiods.

To address temporal heterogeneity directly, we repeat our core regressions on subsamples defined by the period in which a scientist published his or her first paper, separately for APS, ACS, and PubMed. Within each stratum, we estimate the same specification as in the main analysis and report coefficients on exploration frequency (EP) and exploration risk (ED). Split rules for the early versus late career window follow the main text (ten career years and ten publications as alternative split points). Table A24 reports the resulting coefficients and sample sizes.

We split the APS observation window (1976–2015) into 1976–2000 and 2000–2015. In Table A24, both periods exhibit a positive and significant EP coefficient and a negative and significant ED coefficient under each split rule, with magnitudes in the same ballpark across strata. Thus, the association consistent with the high-EP–low-ED (conservative adventurer) profile holds in each modern subsample, not only in the pooled sample.

For PubMed and ACS, we stratify into twentieth- versus twenty-first-century start cohorts. Both datasets again show positive EP and negative ED coefficients in both century strata, in line with the main results.

Across strata, the core conservative-adventurer pattern holds: positive EP and negative ED appear in twentieth- and twenty-first-century subsamples for all three datasets, so the main finding is not an artifact of pooling across a single calendar span. Effect magnitudes vary somewhat across periods; notably, the ED coefficient becomes more negative in twenty-first-century subsamples for PubMed (from $-0.28$ to $-0.36$) and especially for ACS (from $-0.19$ to $-0.48$), plausibly reflecting stronger specialization and competition in recent science. In the ACS twenty-first-century subsample with the ten-year career split, EP is not statistically significant ($\hat{\beta} = 0.06$, $p = 0.108$) because the separate, additive categorical fixed effects for the calendar year and the research area of the scientist's first paper consume substantial degrees of freedom in this subsample (leaving only about 7.3 observations per fixed-effect parameter); under the ten-paper split, which better captures completed early-career trajectories, EP is marginally significant ($\hat{\beta} = 0.05$, $p < 0.1$) with a larger sample ($N = 13{,}354$). ED remains strongly negative and significant across all ACS specifications. Overall, signs and significance align with the main results, and the conservative-adventurer pattern is visible in modern subsamples where the institutional context resembles contemporary science.

***A7.1.4. Demographic Robustness Check (Name Characteristics).*** To ensure our findings are not driven by potential biases in author name disambiguation algorithms—which historically face challenges with certain name origins—we categorized scientists into Asian and non-Asian groups following the approach detailed in Section "Removing Scientists with Asian Last Names." We then performed subgroup analyses on the APS dataset.

As presented in Table A25, the high-EP-low-ED pattern is significantly correlated with higher future academic impact in both the Asian and non-Asian subsamples, with qualitatively consistent effects across different career split standards. These findings demonstrate that our core mechanism is demographically robust and not an artifact of name-disambiguation discrepancies.

**A7.2. Controlling for the Genders of Scientists.** In Section A2.3.1, we find that men tend to have higher EDs. Here, we check whether the difference in future impact between scientists with different EPs and EDs is influenced by the confounding factor gender.

We add the gender of each scientist as a control variable to the regression Equation (A4). Through the gender assignment process mentioned in Section A1.1.4, we obtain a smaller set of scientists, each of whom is assigned a (predicted) gender. We run regressions on the dataset of 19,044 scientists—a subset of the original dataset, and find that even if we control for genders, the regression coefficients of the EP and ED exhibit phenomena consistent with our main findings (see Table A26).

**A7.3. Alternative Definitions of the EP and ED.** In this section, we change the definitions of EP and ED to test the robustness of the findings on the exploration metrics. First, we utilize different PACS code digits (2/4/6) for the calculation of EP and ED. Then, we replace the major components of computing the ED with other possibilities by constructing alternative topic graphs, trying a classic graph node distance measurement, and computing paper distance differently.

***A7.3.1. Different PACS code digits.*** We have always used *areas*—the first two digits of PACS codes in the APS dataset to calculate the EP, and *topics*—the entire six digits for the ED, which we call the 2-6 combination of PACS code digits for calculating the EP and ED. For the ED, we choose the intuitively more informative six digits since we want to capture finer-grained distances, and we prefer the ED to be discriminative. As a comparison, we plot the distributions of the ED calculated with the first 2, 4, and 6 digits in Figure A24, and they seem agreeable to our presumption that the distributions of the ED resulted from using 2 and 4 digits are more skewed and concentrated, i.e., many more scientists have similar EDs, and the EDs—as a metric—would be therefore carry less information.

Besides the 2-6 combination used in the paper, we try the 2-2, 2-4, 4-4 and 4-6 combinations. In Table A27, we see that the EPs always (statistically) significantly contribute to the future performance when calculated with the first 2 or 4 PACS digits. The table also shows that choosing the first 2 or 4 digits to calculate the ED is less than ideal, if we compare it with choosing the entire 6 digits. As such, we choose the way the ED is computed in our study to make the metric more informative.

***A7.3.2. Different Topic Graphs.*** In the main text, we use the co-occurrences of PACS codes in the same papers to construct the topic graph for measuring topic distances. Here, we try other popular alternatives, including the citation graph (25, 26) and the co-citing graph (13, 27).

- Citation graph: Unlike the co-occurrence based topic graph in the main text, the citation graph is directed. The edge from topic $i$ to $j$ indicates that at least one paper $p_i$ of topic $i$ cites a paper $p_j$ of topic $j$. Specifically, each of such citations contributes $w'_{i\to j}$ to the total edge weight from topic $i$ to $j$ by

$$w'_{i\to j} = \frac{1}{n_{p_i} \times r_{p_i} \times n_{p_j}}, \qquad \text{[A11]}$$

  where $n_{p_i}$ and $r_{p_i}$ denote the number of topics and the number of references in $p_i$, respectively. Then, the total weight from topic $i$ to topic $j$ equals to the sum of $w'_{i\to j}$ over all such citations by $w_{i\to j} = \sum w'_{i\to j}$.

  To calculate the distance between topic $i$ and $j$, we adopt the *weighted overlap* metric in the main text in the obtained *directed* citation graph, by considering the incoming and outgoing links of the node pair separately and taking the average of the two. This approach is similar to that in (28), with the difference being that we take the average to scale the similarity to $[0, 1]$, making it consistent with our other similarity metrics. Specifically, the *weighted overlap* of topic $i$ and $j$ based on their outgoing links is denoted as $O^{\text{out}}_{ij}$ and that based on incoming links as $O^{\text{in}}_{ij}$, and both are calculated by

$$O^{\text{out}}_{ij} = \frac{W^{\text{out}}_{ij}}{s^{\text{out}}_i + s^{\text{out}}_j - w_{i\to j} - w_{j\to i} - W^{\text{out}}_{ij}}, \qquad \text{[A12]}$$

$$O^{\text{in}}_{ij} = \frac{W^{\text{in}}_{ij}}{s^{in}_i + s^{\text{in}}_j - w_{i\to j} - w_{j\to i} - W^{\text{in}}_{ij}}, \qquad \text{[A13]}$$

  where $s^{\text{out}}_i$ denote the sum of weights from all $i$'s outgoing edges and $W^{\text{out}}_{ij}$ is the weight of $i$ and $j$'s overlapped outgoing neighbors. The *weighted overlap* between topic $i$ to topic $j$ in a directed graph is then calculated by

$$O'_{ij} = \frac{O^{out}_{ij} + O^{in}_{ij}}{2}. \qquad \text{[A14]}$$

- Co-citing graph: An edge between topic $i$ and $j$ indicates that papers of topic $i$ and papers of topic $j$ cite the same paper(s). Each time a paper of topic $i$ and a paper of topic $j$ cite the same paper(s), it contributes $w'_{ij}$ to total edge weight between topic $i$ and topic $j$, which is calculated by

$$w'_{ij} = \frac{1}{n_{p_i} \times n_{p_j}}, \qquad \text{[A15]}$$

  where $n_{p_i}$ ($n_{p_j}$) is the number of topics in $p_i$ ($p_j$). The total weight between topic $i$ and topic $j$ is $w_{ij} = \sum w'_{ij}$.

The results in Table A28 tell us that these different constructions of topic graphs do not affect our conclusions.

***A7.3.3. Different Node Distance Metrics.*** We try two alternative node distance metrics, one basically an unweighted version of the metric used in the main text and the other is based on graph embedding through neural networks.

In the main text, we have computed the *weighted overlap* of two nodes' neighbors as the two nodes' distance on the graph, as described in Section A2.2.2. It can be seen as a weighted graph version of the classic unweighted Jaccard similarity (29), and the similarity between node $i$ and $j$ can be defined by

$$\text{Jac}_{ij} = \frac{|\Lambda_i \cap \Lambda_j|}{|\Lambda_i \cup \Lambda_j|}, \qquad \text{[A16]}$$

where $\Lambda_i$ and $\Lambda_j$ are their sets of neighboring nodes, respectively. Here we adopt the unweighted version and take $(1 - \text{Jac}_{ij})$ to convert it to a distance metric. Using this alternative distance metric, the regression results in Table A29 suggest the same conclusions.

The other node distance metric is based on continuous lower-dimensional feature representations for the nodes learnt by node2vec (30), which takes into account higher-order connections between nodes. The distance between two nodes can then be calculated as the cosine distance for the corresponding feature vectors. As detailed in (30), we select various hyperparameters for node2vec, such as dimension $d = 128$, walks per node $r = 10$, walk length $l = 80$, context size $k = 20$, return parameter $p = 4$, and in-out parameter $q = 0.25$, based on link prediction experiments. We then run node2vec on our topic co-occurrence graph to acquire node representations. Topic similarity is computed using the following equation:

$$\text{Node2VecSim}_{ij} := \text{MinMax}(\cos(f(i), f(j))), \tag{A17}$$

where MinMax and cos refer to the min-max scaling and cosine similarity operations, respectively, and $f(i)$ corresponds to the representation vector of node $i$ acquired above. We then obtain the distance metric by subtracting $\text{Node2VecSim}_{ij}$ from 1. We observe that when utilizing the EP calculated through node2vec for regression, we have robust results consistent with our main findings; see Table A29.

***A7.3.4. Different Paper Distance.*** When computing the ED of a scientist, we have calculated a paper $p_i$'s distance from the $m$ papers before it. The method used in the main text (see Section A2.2.3) first constructs two sets, among which $T_i$ contains the topics of $p_i$ and $T_{i,m}$ consists of all topics from the $m$ papers before $p_i$. The task then becomes calculating the distance between these two sets. For a topic $t_j$ in $T_i$, the method calculates its distance to all possible topics in $T_{i,m}$ and takes the average to obtain $t_j$'s distance to the set $T_{i,m}$. This is repeated over all topics in $T_i$ and their distances to $T_{i,m}$ are averaged to obtain the paper distance between $p_i$ and the $m$ papers before $p_i$. As an alternative, we use the classic Hausdorff distance (31) to calculate the distance between two sets, which takes the minimum distance from $t_j$ to all topics in set $T_{i,m}$ as $t_j$'s distance to $T_{i,m}$, and finds the maximum among all the node-to-set distances to be $p_i$'s paper distance from the $m$ papers before it. It is calculated by

$$\text{PD}_{p_i} = \max_{t_j \in T_i} \min_{t_k \in T_{i,m}} \text{TD}_{t_j t_k}. \tag{A18}$$

Table A30 shows consistent regression results using the Hausdorff distance, compared with these in the main text.

**A7.4. Citation Measurements.** In the main text and the above, we measure a paper's impact by taking the logarithm of the number of citations it receives within five years after it is published (5-year log-citations, $\log c_5$), and here we try 10-year log-citations as an alternative. Furthermore, considering the fact that papers published in different periods of time (1) and in different research areas (32) may have different trends of popularity, we take into account possible factors that may affect the outcomes of interest. Finally, adopting a different angle of measuring academic achievements as suggested in (1), we could pay attention to a scientist's highest impact work in the future, which is the one with the highest $\log c_5$ or $\log c_{10}$.

***A7.4.1. 10-Year Citations.*** We try the 10-year log-citations ($\log c_{10}$) to measure the impact of papers over a longer period, as in (1). Note that when using 10-year citations as citation measurement, only papers published before the end of 2010 are used in the regression. Table A31 shows that our conclusions remain the same.

***A7.4.2. Normalized Citations.*** We normalize the citations with two existing methods to measure a paper's impact relatively to its peers, to offset the influence of papers being published in different years and having different areas. Specifically, the first method (32) divides a paper's number of citations by the estimated number of citations that a paper published in the same year with the same research areas would receive. Since the focal paper $p$ may have several research areas (that may have duplicates), we compute $e_i$ as the 5-year log-citations averaged over all papers published in the same year that are associated with area $i$, and compute the arithmetic mean of all the $e_i$ of $p$ for the estimated log-citations $e$ that an average peer paper of $p$ would receive

$$\begin{aligned} e_i &= \frac{1}{m} \sum_{j}^{m} c_{i,j}, \\ e &= \frac{1}{n} \sum_{i}^{n} e_i, \end{aligned} \tag{A19}$$

where $c_{i,j}$ is the 5-year log-citations received by paper $j$ in area $i$, $m$ is the number of papers in $i$ in the same year and of the same area as the paper $p$, and $n$ is the number of areas paper $p$ covers. Paper $p$'s normalized 5-year log-citations $c_{\text{norm}}$ is then calculated by $c_{\text{norm}} = c_{\text{raw}}/e$, where $c_{\text{raw}}$ is the (raw) log-citations of $p$.

To accommodate the fact that papers are often associated with more than one area, the second method (33) proposes that $c_{i,j}$ should be weighted by $f_{i,j}$, which is the fraction that area $i$ takes up among all areas of paper $j$. The normalizing factor $e$ admits the harmonic mean of $e_i$:

777

$$
\begin{aligned}
e_i &= \frac{\sum_j^m c_{i,j} f_{i,j}}{\sum_j^m f_{i,j}}, \\
e &= \left( \frac{\sum_i^n e_i^{-1}}{n} \right)^{-1}, \\
c_{\text{norm}} &= c_{\text{raw}}/e.
\end{aligned} \quad [A20]
$$

Table A31 shows that under these two methods of normalizing citations, the main findings remain true.

***A7.4.3. Percentiles among Peers.*** One way to take care of the influences on the paper citations from both research areas and
time periods is comparing a paper's citations with other papers in the same area and published in the same year, and using
its relative percentile as the impact measurement (1). Here we calculate the percentiles under $\log c_5$ and if a paper's 5-year
log-citations is at the $k$-th percentile, it indicates that $k\%$ of the papers from the same area and year have fewer $\log c_5$ citations
than the paper in question. Since papers are usually assigned with more than one area, we calculate the maximal and mean
percentile in all its areas. analyzes show that under these two percentile measurements, our conclusions still hold (Table A31).

***A7.4.4. Maximum Future Citations.*** In this section, we use the maximum 5-year log-citations or the maximum 10-year log-citations
among all the scientist's publications after the split point to measure her future impact. This measurement (1) is meant to
capture a scientist's highest impact—another measurement of academic success—in the future. Table A31 shows that we
observe similar findings that are consistent with those reported above, if we adopt this alternate way of measuring a scientist's
future accomplishments.

**A7.5. Timeframe Alignment Analysis.** In our main analysis, the observation windows differ across datasets due to the historical
availability of topic classification infrastructures: the APS dataset begins in 1976 (coinciding with the systematic implementation
of PACS codes), whereas the PubMed and ACS datasets utilize MAG field-of-study codes that allow retroactive assignment back
to 1900. A natural methodological concern is whether the differing temporal spans—specifically the inclusion of early-to-mid
20th-century data in PubMed and ACS—introduce temporal bias or drive the observed variations in the effects of topic-switching
behaviors.

To directly address this concern and rule out the possibility that our findings are artifacts of misaligned timeframes, we
conducted a timeframe alignment robustness check. We restricted the PubMed and ACS datasets strictly to scientists whose
careers began in or after 1976, perfectly aligning their observation windows with the APS dataset. Within these temporally
truncated samples, we re-estimated our core regression specifications to assess the impact of EP and ED on future academic
performance.

Table A32 reports the estimated coefficients and sample sizes for this aligned analysis under both the ten-career-year and
ten-paper split rules. The results demonstrate that under this identical timeframe constraint, the core findings remain highly
consistent with the main text. For both the PubMed and ACS datasets, the coefficient for EP remains positive and statistically
significant, while the coefficient for ED remains negative and statistically significant.

These consistent estimates confirm that the conservative-adventurer advantage (the high-EP–low-ED pattern) is robust to
the specific choice of observation windows. The fundamental trade-off between topic-switching frequency and distance holds
similarly across datasets when observed within the same modern historical span, confirming that our overarching conclusions
are not driven by the longer historical baseline in the original biomedical and chemistry cohorts.

**A7.6. Different Look-back Windows.** When calculating the EPs and EDs, we backtrack the focal paper's past $J$ papers or past
$K$ years, which is called the *look-back* period, to decide whether it is an exploratory paper or to compute its paper distance. In
the main text, we have used $J = 5$ papers, and here we repeat the experiments by varying the look-back period to be the past
$J$ papers and the last $K$ years from 1 to 15 as well as setting it to $\infty$. The results in Tables A33 and A34 are consistent with
these in the main text.

**A7.7. Propensity Score Matching.**

***A7.7.1. Varying Grouping Quantiles.*** In Sections A4.2.1 and A4.2.2, we select the scientists with either of their exploration metrics
(EP and ED) in the top 50% of all relevant values to be the *high* group, and those with their metrics in the bottom 50% the
*low* group. Here we change the definition of being *high* (*low*) in either metric to being in the top (bottom) 40%, 30%, 25%, 20%
and 15%, respectively, and repeat the experiments. The results are consistent with those in the main text (see Tables A35
and A36).

***A7.7.2. Shuffling after Matching—The Null Model.*** In order to verify that the differences between the matched groups are not due to
the random nature of quantities involved, we create synthetic datasets by shuffling scientists and their papers after the groups
are matched, so that the resulting future performance are randomized. This will be referred to as the null model. We calculate
the gap between the randomized future performance of the treatment group (with "high" metric) and that of the control group
(with "low" metric) to see whether the synthetic datasets give a gap more extreme (larger in absolute values in this case) than
the actually observed one (cf. $P$values in permutation tests). This can be repeated many times, and if more extreme gaps
occur a lot, our finding are not robust enough. Specifically, we uniformly at random reassign scientists to papers and make sure
that each scientist is assigned the same number of papers as she actually had (in a certain year) and each paper has the same
number of authors as it actually had. We repeat the experiment 1,000 times, and observe none of such "extreme random cases"
(see the paper-level rows in Table A37), which suggests the robustness of our findings.

Alternatively, instead of randomly shuffling the author-paper relations, it is possible to shuffle future performances of
scientists uniformly at random. Such results are reported in the author-level rows of Table A37, and at the worst 4 out of 1,000
times reported more extreme intergroup gaps than what is actually observed. Again, this is a strong indicator that our results
are robust against what any randomness allows for.

**A7.8. Propensity Score Weighting.**

***A7.8.1. Shuffling after Weighting—The Null Model.*** Similarly to what was done in A7.7.2, we may create synthetic datasets as the null
model by reshuffling uniformly at random the author-paper relations—the paper-level reshuffling, or the future performances of
scientists—the author-level reshuffling. With both types of reshuffling, we repeat the propensity score weighting methodology
of Section A4.2.3 on the synthetic datasets. The results are summarized in Table A38. Simply put, the high/low EP/ED
groups are indistinguishable in future performances under the null model, and this is again in accordance with the conviction
that our findings are more than reliable.

**A7.9. Perturbing Independent Variables.** To further ensure the robustness of our regression results against potential variations
in the independent variables, we introduce artificial Gaussian noise $\epsilon \sim N(0, \sigma^2)$ and incorporate it into each scientist's
independent variables of EP, ED, and $\text{LogCit}_{\text{past}}$ separately. Through a series of experiments, we systematically adjust the
value of $\sigma$ to determine the maximum level of perturbation that can be introduced while still maintaining consistent regression
results. The detailed regression outcomes are presented in Table A39, where we see the regression results are robust when
adding different levels of perturbation to the independent variables.

**A7.10. Survival Bias.** To see whether the "conservative adventuring" behavior may offer risky advice such as productivity drops
to scholars, we conduct regression analyses using future publication count as the dependent variable and past EP and ED as
independent variables, both under 10 papers and 10 career years set-up. The results, as shown in Table A40, indicate that the
regression coefficients are not statistically significant, suggesting no clear association between the conservative adventuring
behavior and a decrease in future productivity drops.

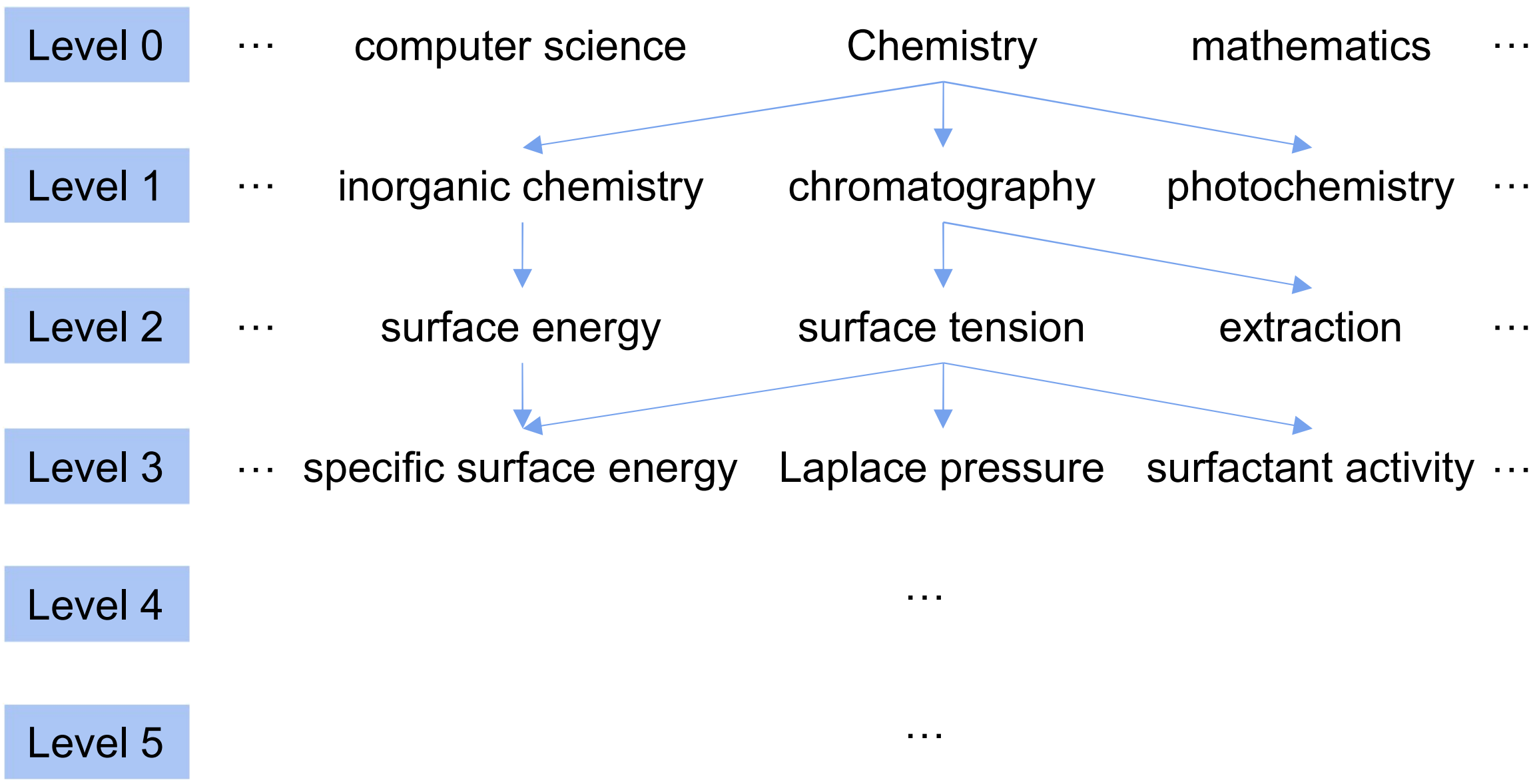

**Fig. A1.** An illustrative sample of the classification scheme of the MAG dataset.

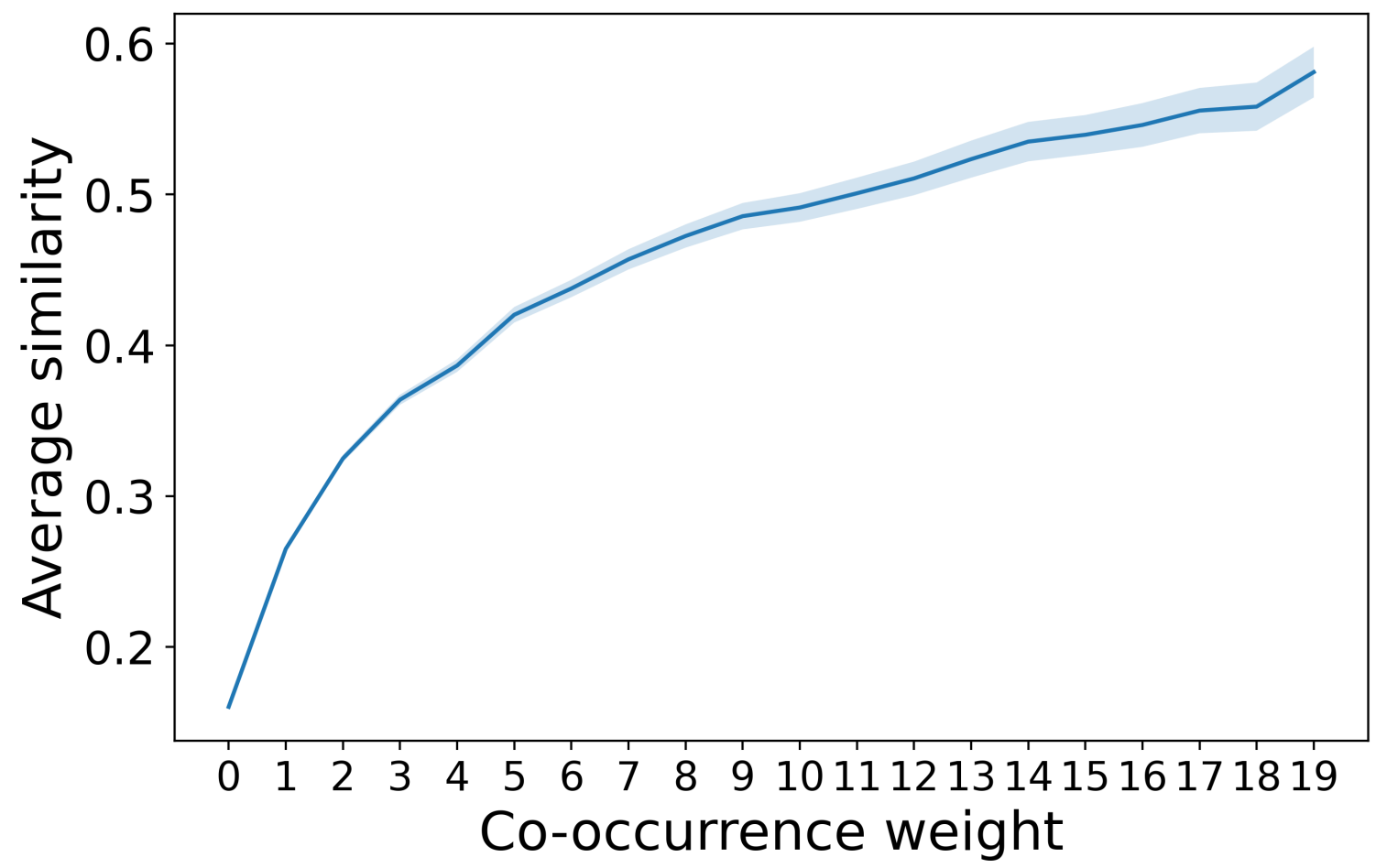

**Fig. A2. The relationship between the co-occurrence weight and the neighborhood overlapping similarity used in the paper.** We divide the connected code pairs (98.67% of all connected code pairs) with co-occurrence weight ranging from 0 and 20 into 20 groups by aligning their weights to the nearest integer and calculate the average similarity and corresponding 95% confidence bands for the code pairs within each group. We see a clear monotone increasing relation between the two metrics.

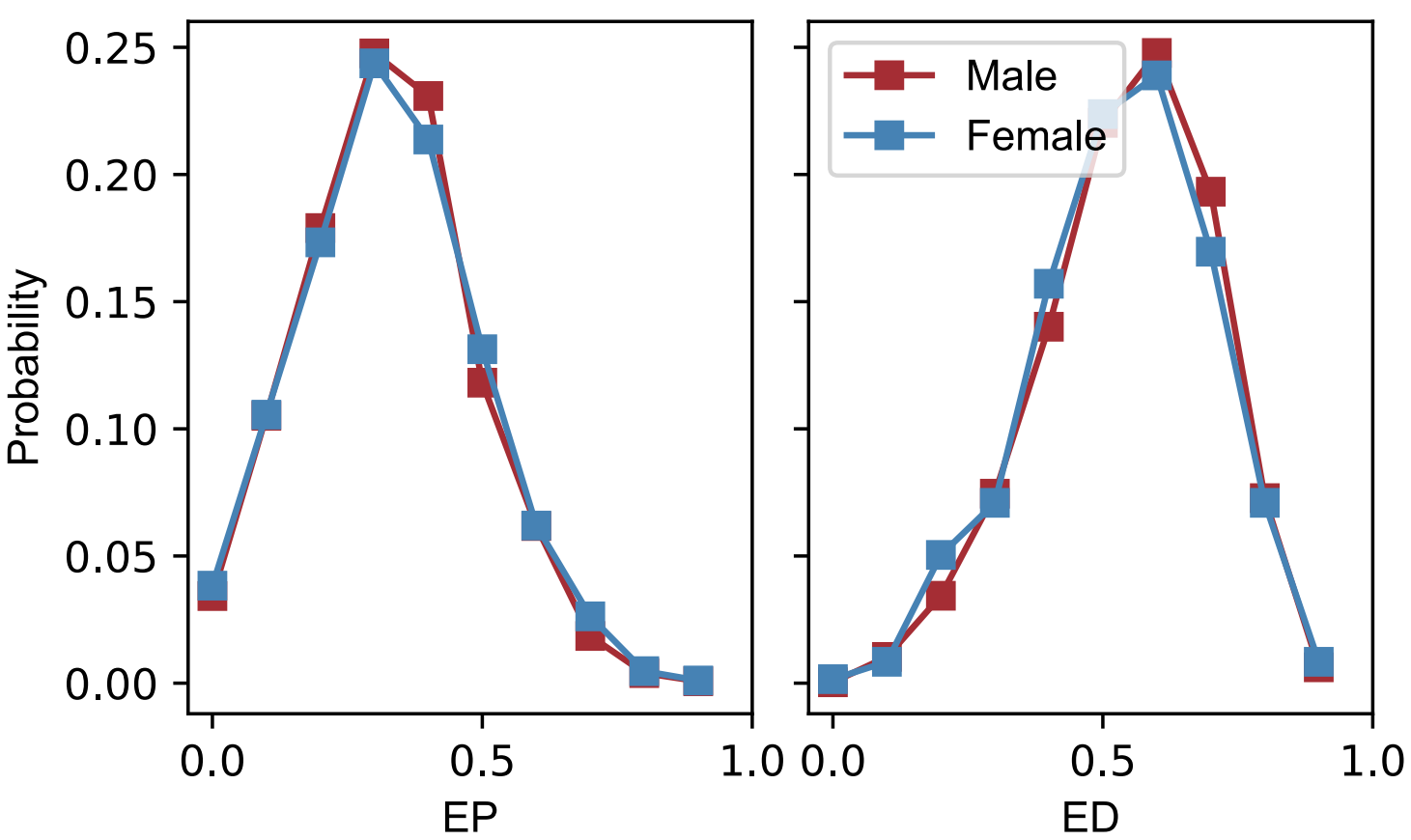

**Fig. A3. Distributions of EPs and EDs for male and female scientists.** We have calculated and aggregated the career-long EPs and EDs of every male and female scientist, respectively.

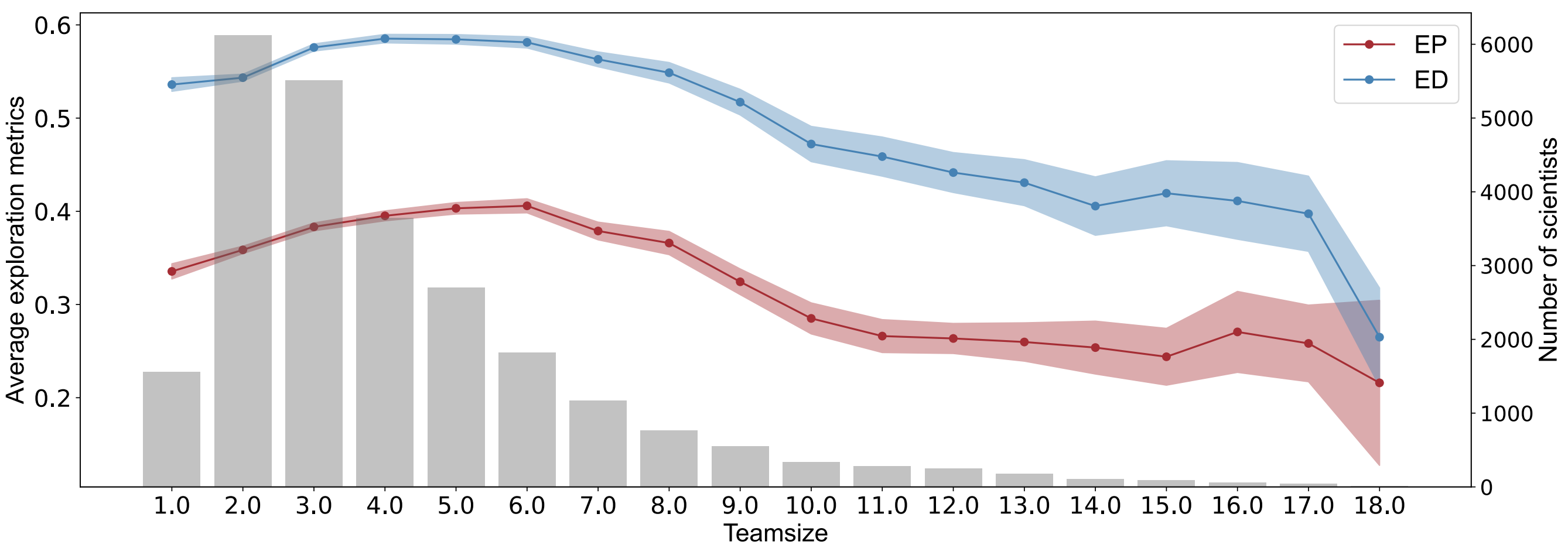

**Fig. A4. The average EPs and EDs for scientists with different team sizes and the distribution of average team sizes of scientists.**

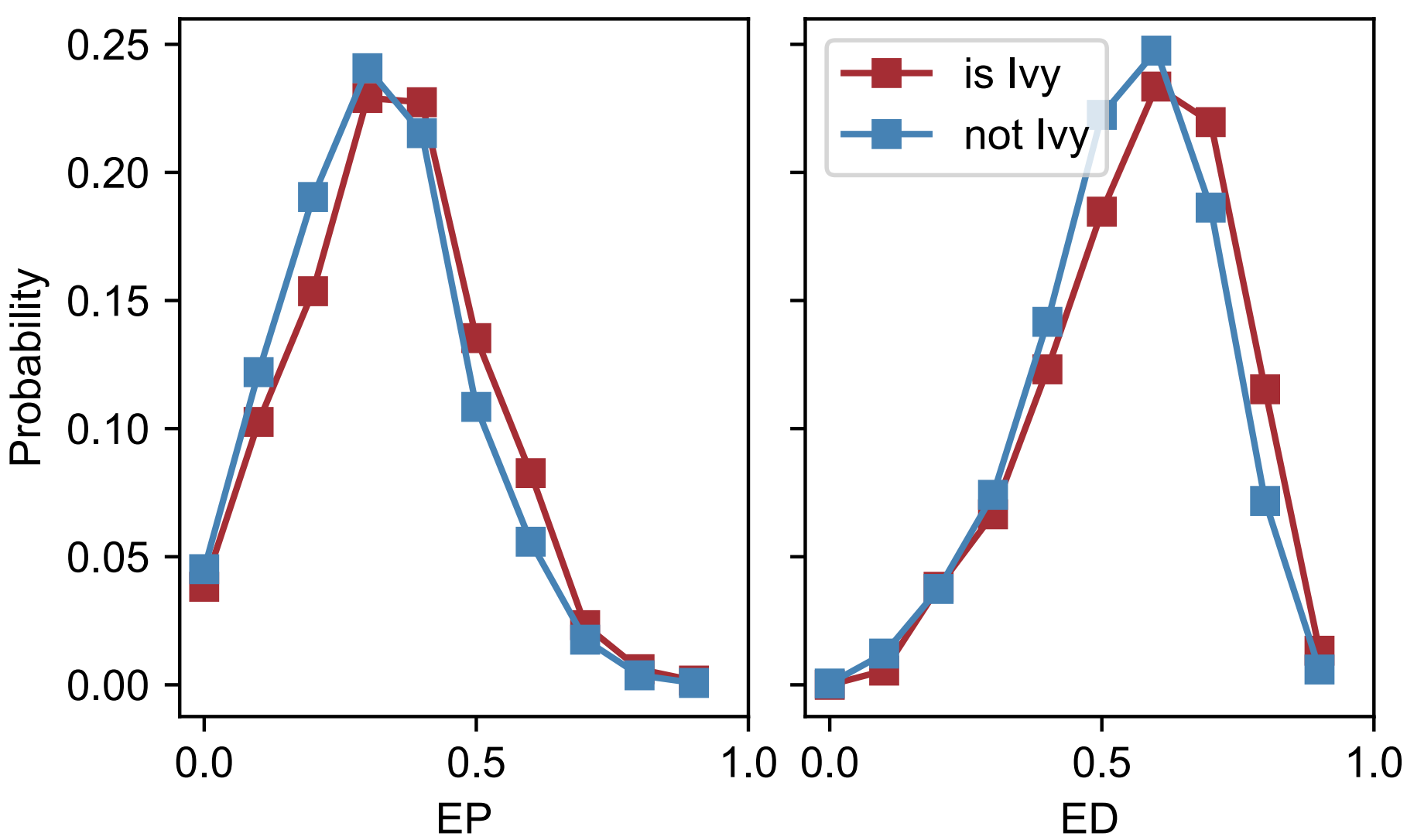

**Fig. A5. Distributions of EPs and EDs for Ivy versus non-Ivy scientists.** We calculate the career-long EPs and EDs for every scientist with and without Ivy League affiliations, respectively.

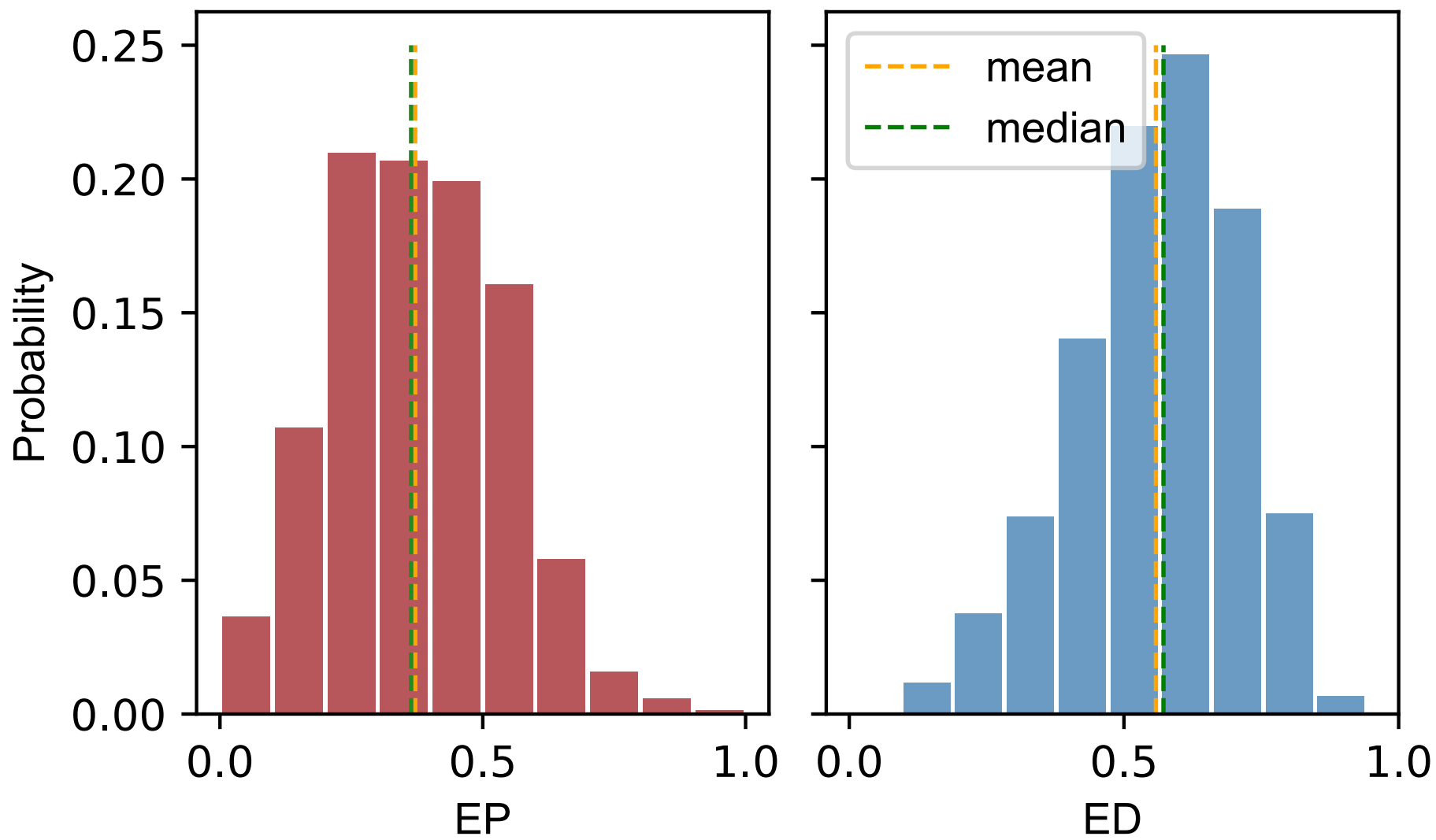

**Fig. A6. The distributions of EPs and EDs for scientists with at least 10 papers.** Both the EP and ED are calculated in each scientist's entire career.

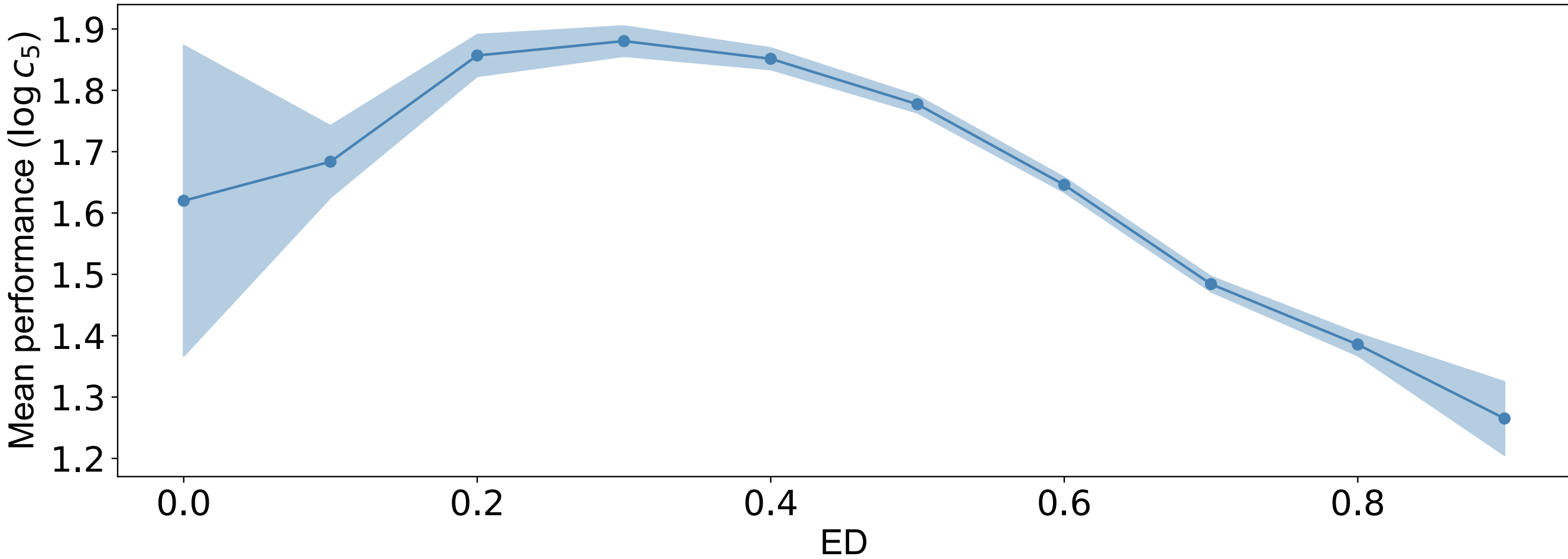

**Fig. A7. Association between scientists' ED and academic performance.** The figure is otherwise similar to Fig. 2a in the main text, except that the horizontal axis corresponds to the current object of interest—the (career-long) ED, instead of the EP.

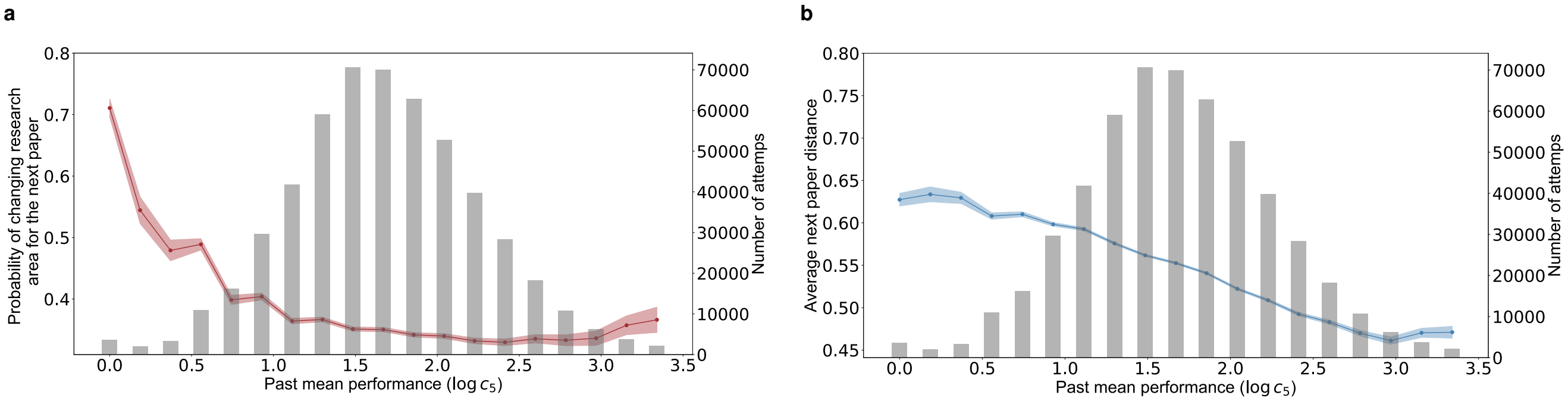

**Fig. A8. Relationships between the prior performance and the next explorations.** For each publication of any scientist (except the first one), we calculate the prior research performance by averaging the 5-year log-citations of all her publications prior to this publication, then we determine whether she explores in this publication, and calculate how far she explores. For all publications of similar prior research performance that are grouped together by having the same nearest integer multiple of $0.1$, we calculate the probability of exploration by counting how many times explorations take place and calculate the average distance of the next paper from the past (five) papers during all these publications. (a) Scientists' past performance versus the probability of changing research area in the next paper. Each red point marks the mean of the $\log c_5$ per paper of the scientists in that group, with 95% confidence bands. The light grey bars in the figure indicate the distribution of scientists' past performance at every (new) publication. Note that part of (a) has appeared in the main text. (b) Scientist's past performance versus the next paper distance. Scientists are less likely to be bold in their explorations as the prior performance improves.

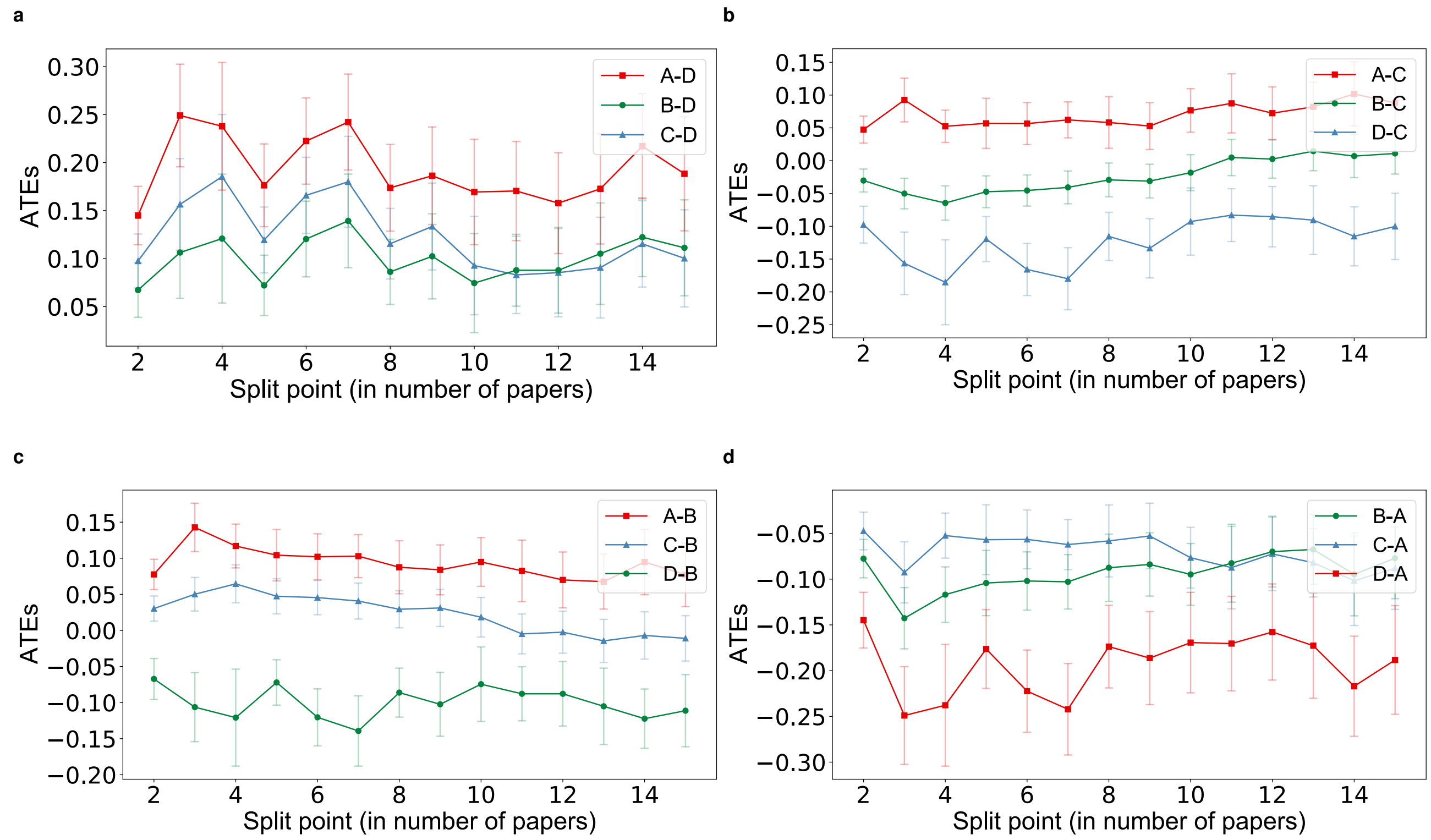

**Fig. A9. Average treatment effects with different baseline groups and varying numbers of papers as split point.** For completeness, each plot has a different group as the baseline group: (a) has group D as the baseline, (b) group C, (c) group B, and (d) group A.

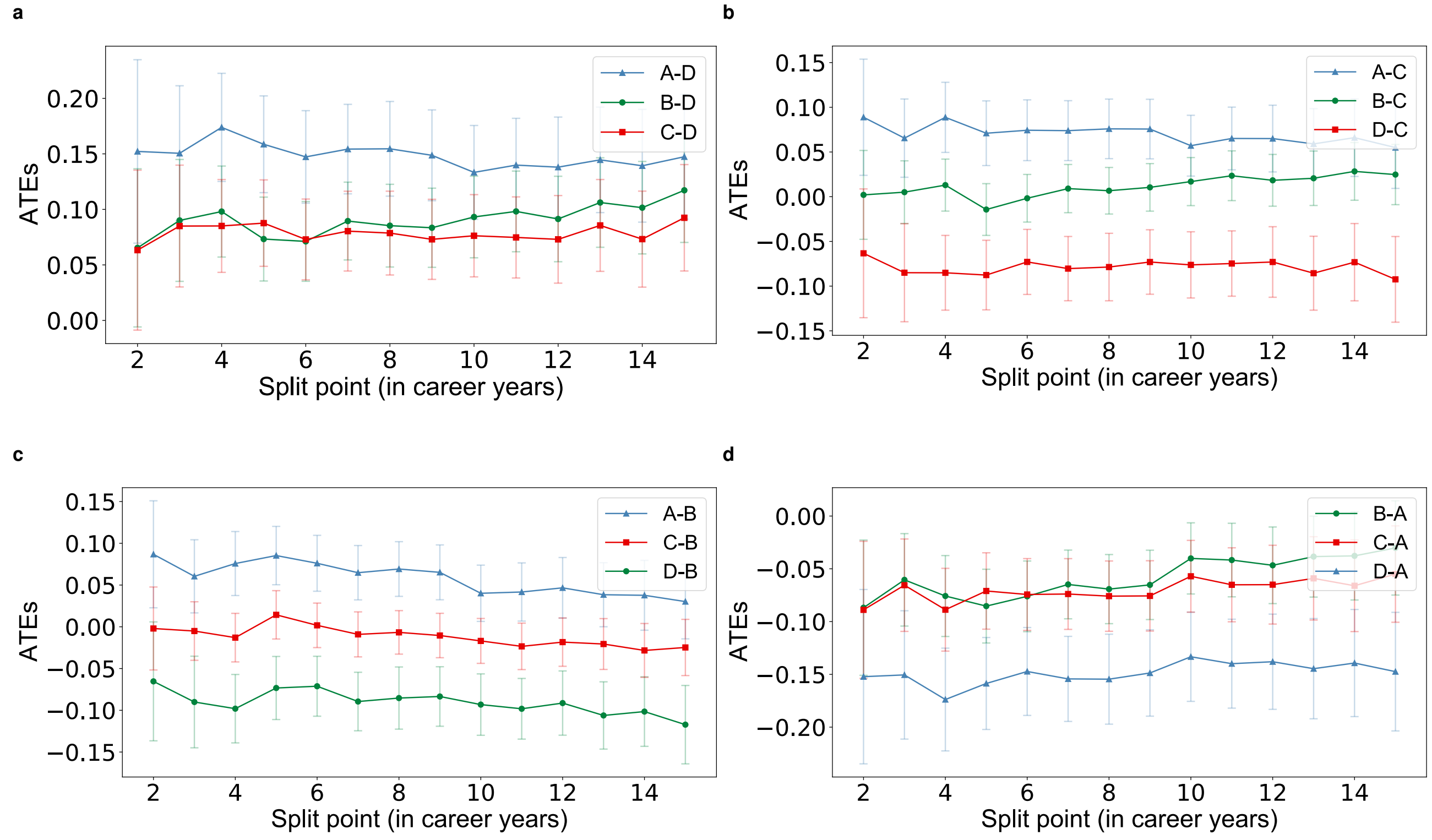

**Fig. A10. Average treatment effects with different baseline groups and varying career years as split point.** (a) has group D as the baseline, (b) group C, (c) group B, and (d) group A. Note that (b) has appeared in the main text.

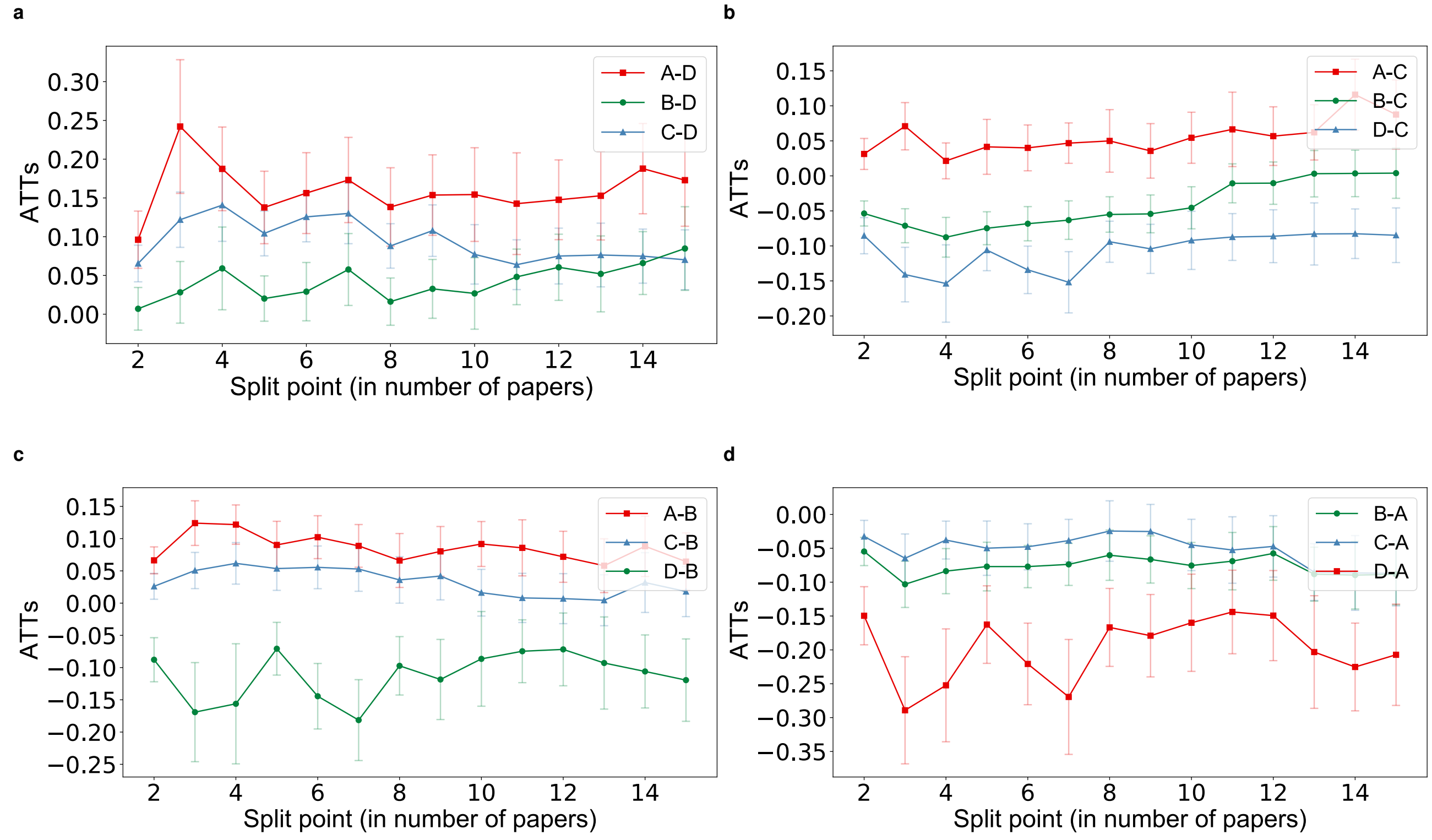

**Fig. A11. Average treatment effects for different treated groups and varying numbers of papers as split point.** Note that (a) has group D as the baseline, (b) group C, (c) group B, and (d) group A.

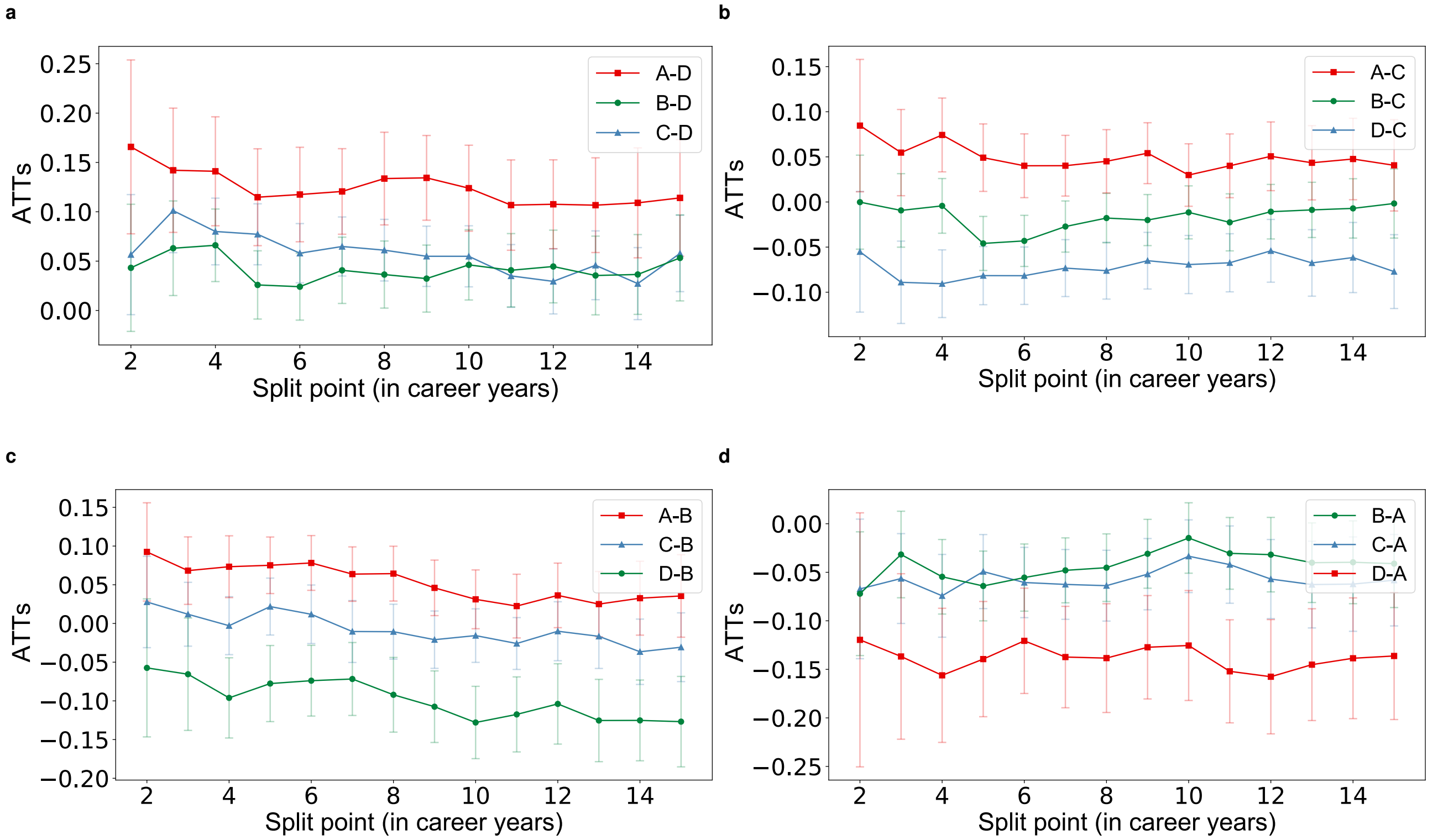

**Fig. A12. Average treatment effects for different treated groups and varying career years as split point.** Note that (a) has group D as the baseline, (b) group C, (c) group B, and (d) group A.

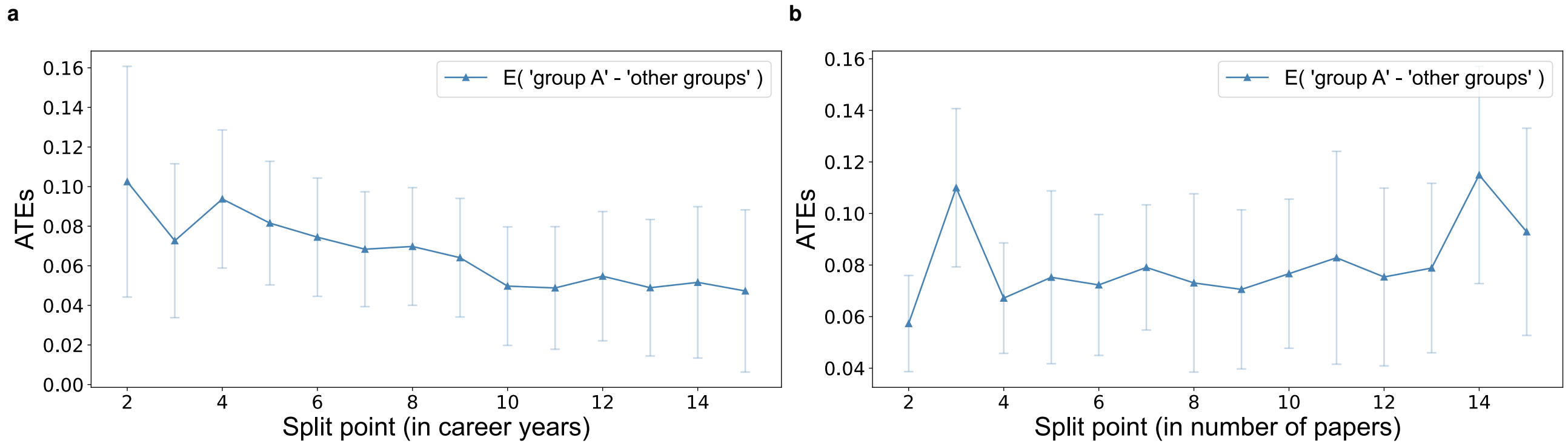

**Fig. A13. Average treatment effects for varying split points in a simplified setup**. Group A functions as the treatment group, and the other three groups make up the control. Note that (a) has appeared in the main text as part of Fig. 4c.

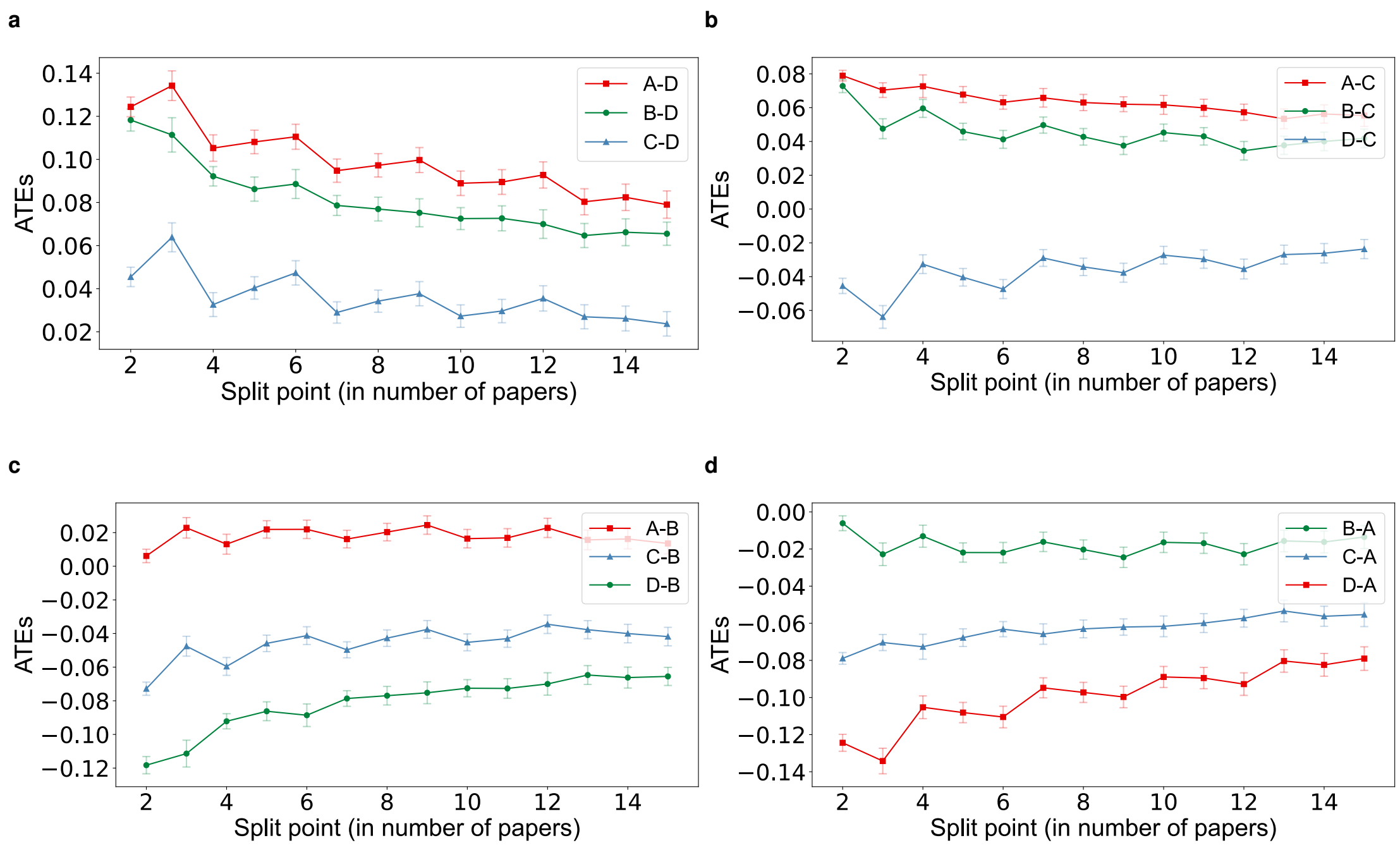

**Fig. A14. Average treatment effects with different baseline groups and varying numbers of papers as split point in the $\mathrm{PubMed}_1$ dataset.** Note that (a) has group D as the baseline, (b) group C, (c) group B, and (d) group A.

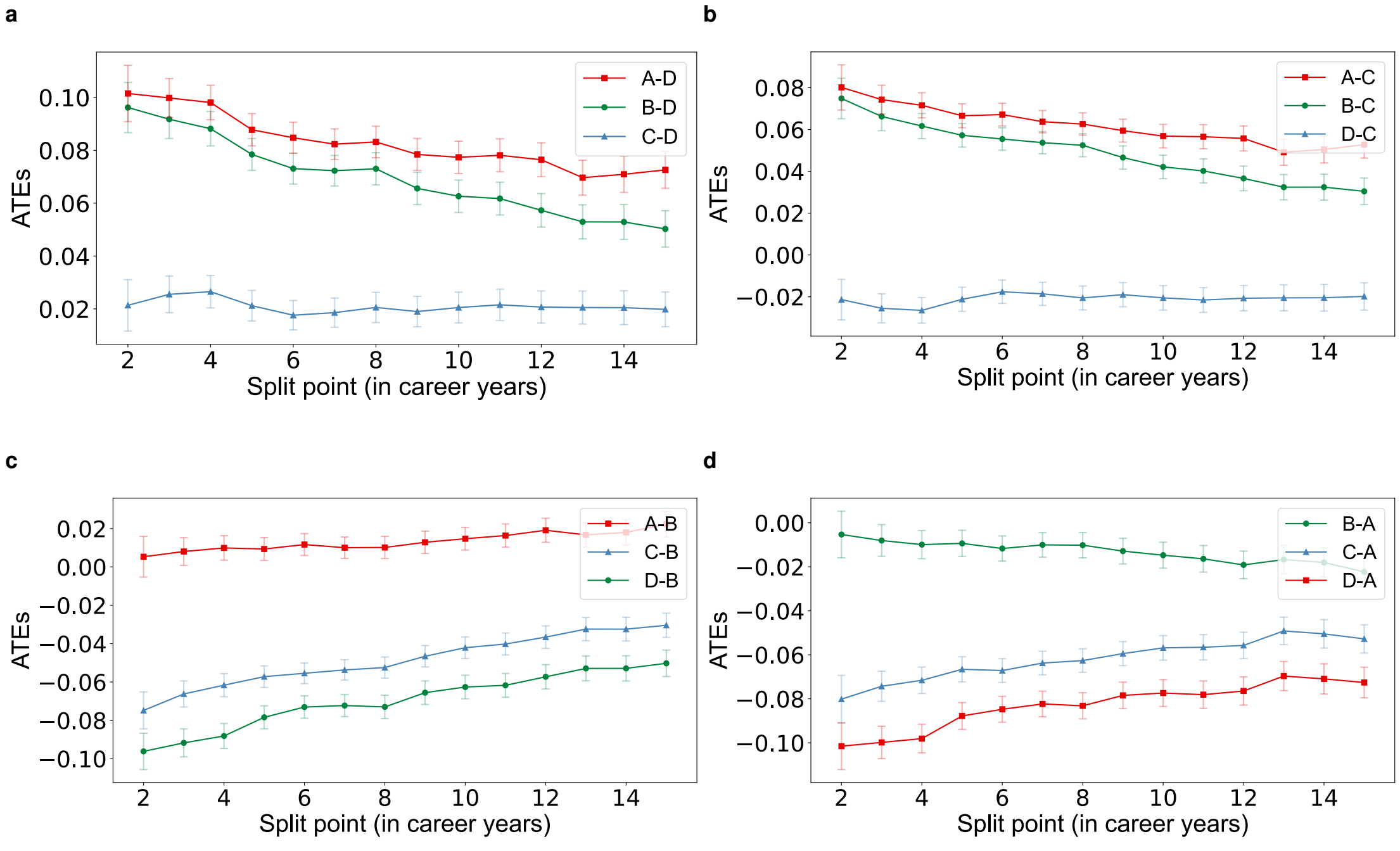

**Fig. A15. Average treatment effects with different baseline groups and varying career years as split point in the $\mathrm{PubMed}_1$ dataset.** Note that (a) has group D as the baseline, (b) group C, (c) group B, and (d) group A.

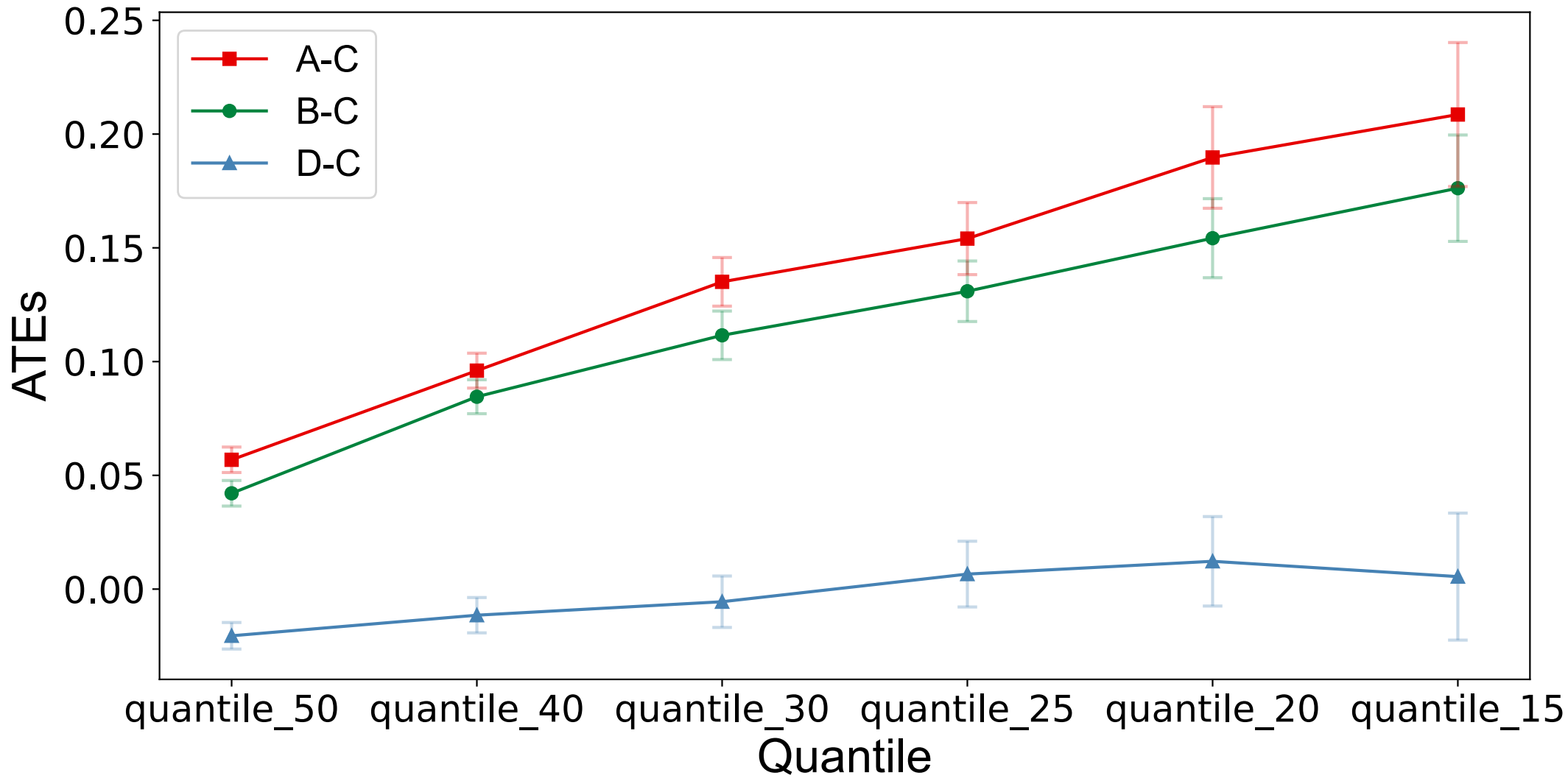

**Fig. A16. Average treatment effects in the $\mathrm{PubMed}_1$ dataset with varying quantiles**. The split point is 10 career year and group C is the baseline.

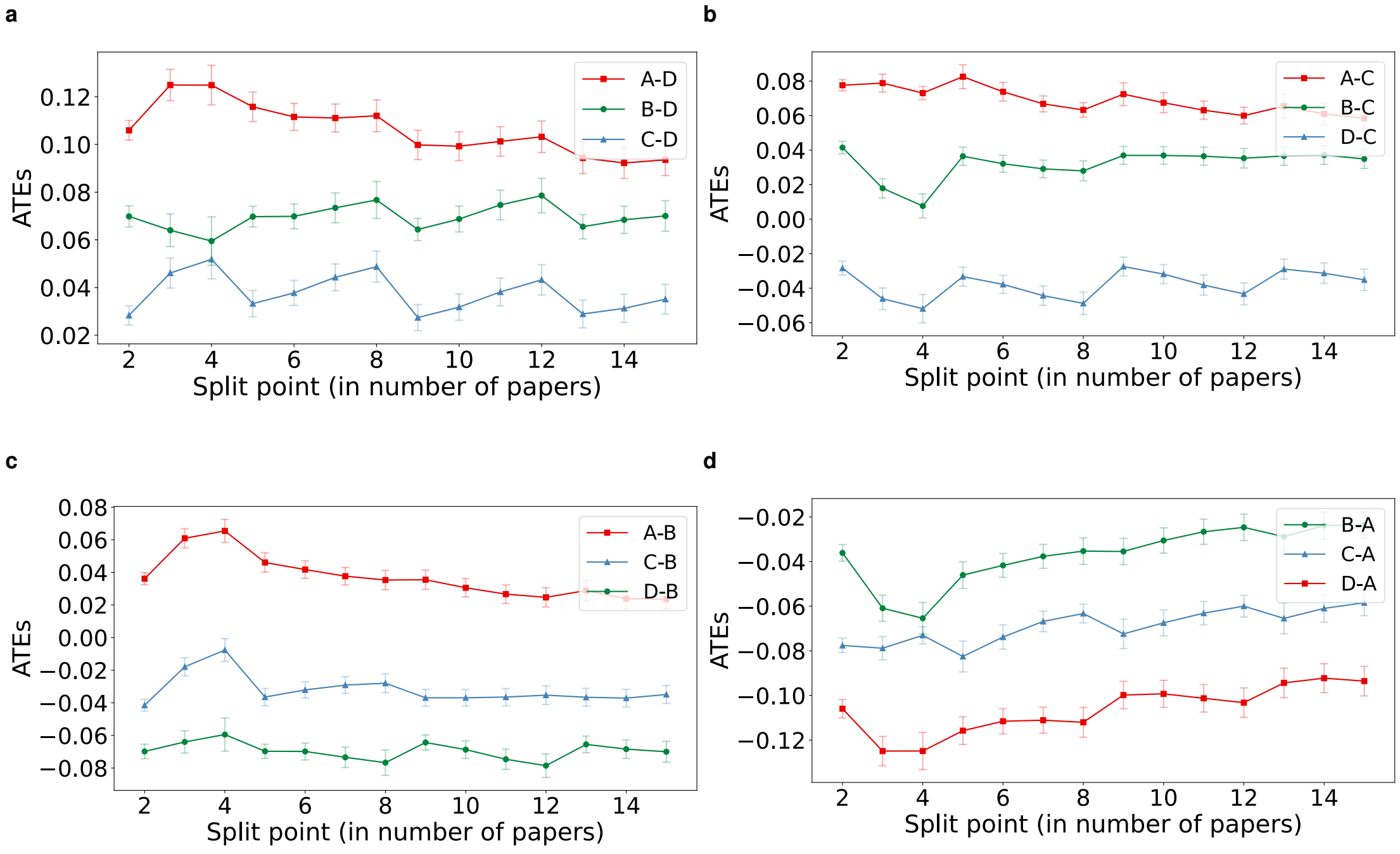

**Fig. A17. Average treatment effects with different baseline groups and varying numbers of papers as split point in the $\mathrm{PubMed_2}$ dataset.** Note that (a) has group D as the baseline, (b) group C, (c) group B, and (d) group A.

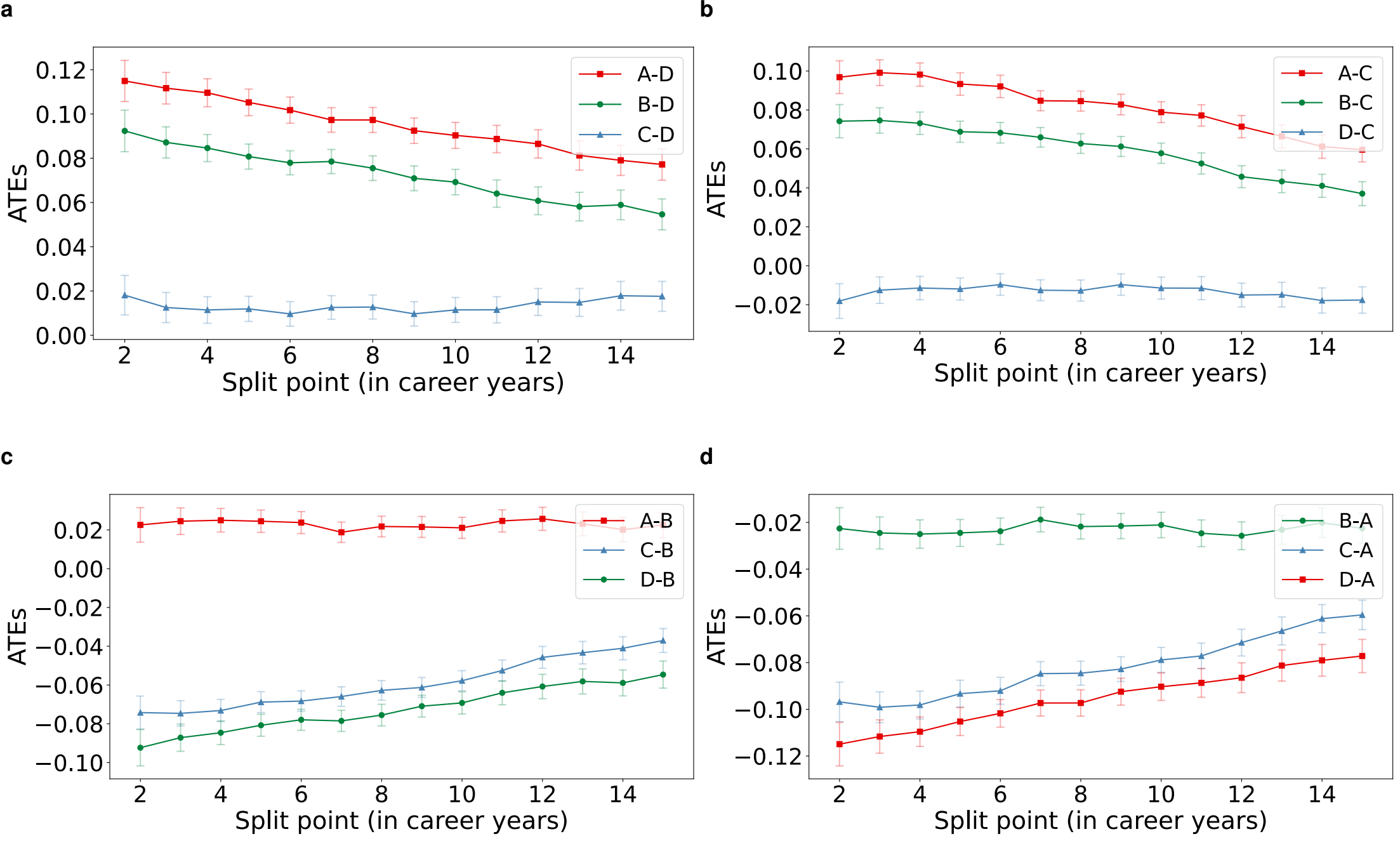

**Fig. A18. Average treatment effects with different baseline groups and varying career years as split point in the $\mathrm{PubMed}_2$ dataset.** Note that (a) has group D as the baseline, (b) group C, (c) group B, and (d) group A.

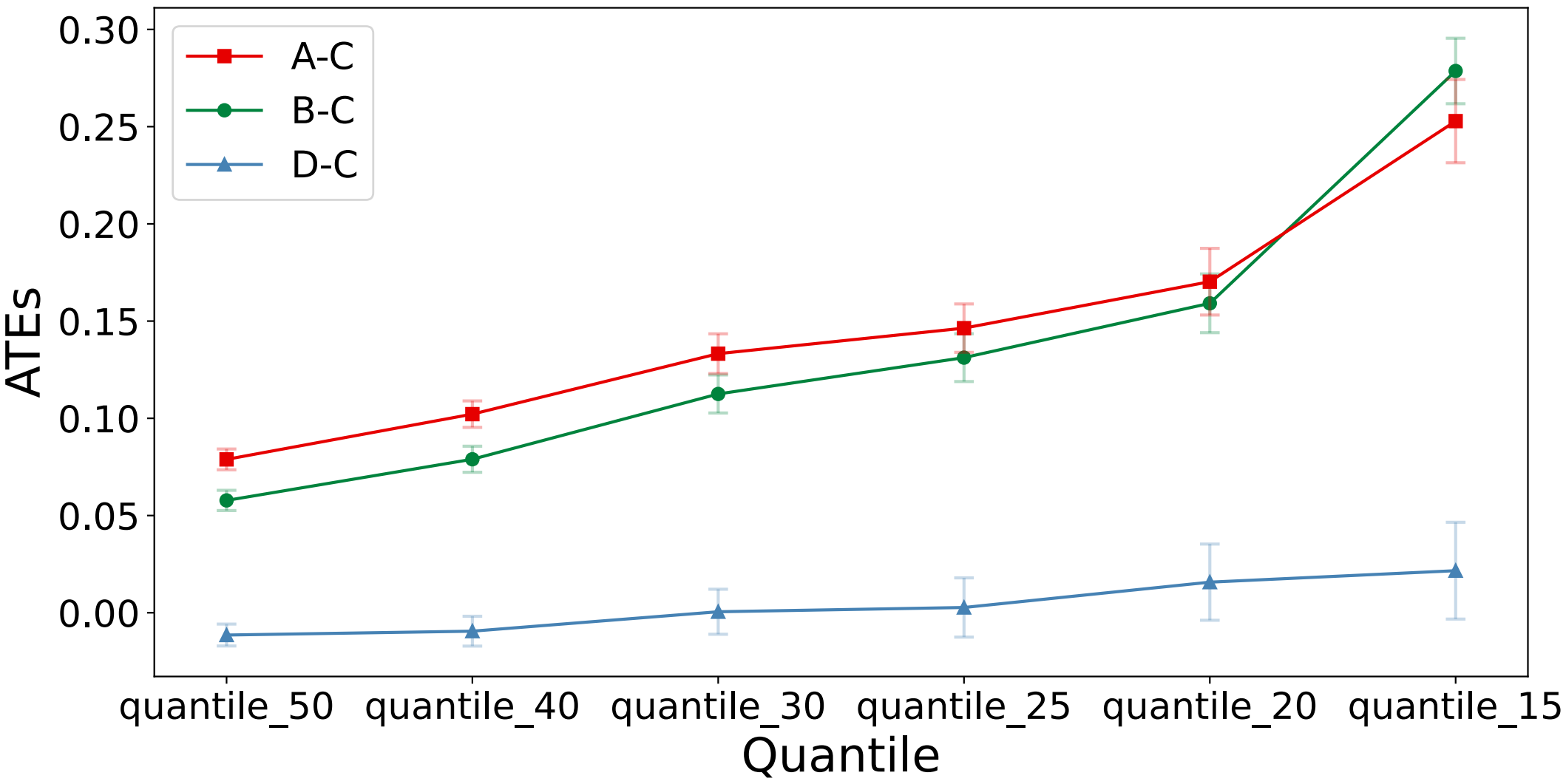

**Fig. A19. Average treatment effects in the $\mathrm{PubMed}_2$ dataset with varying quantiles**. The split point is 10 career year and group C is the baseline.

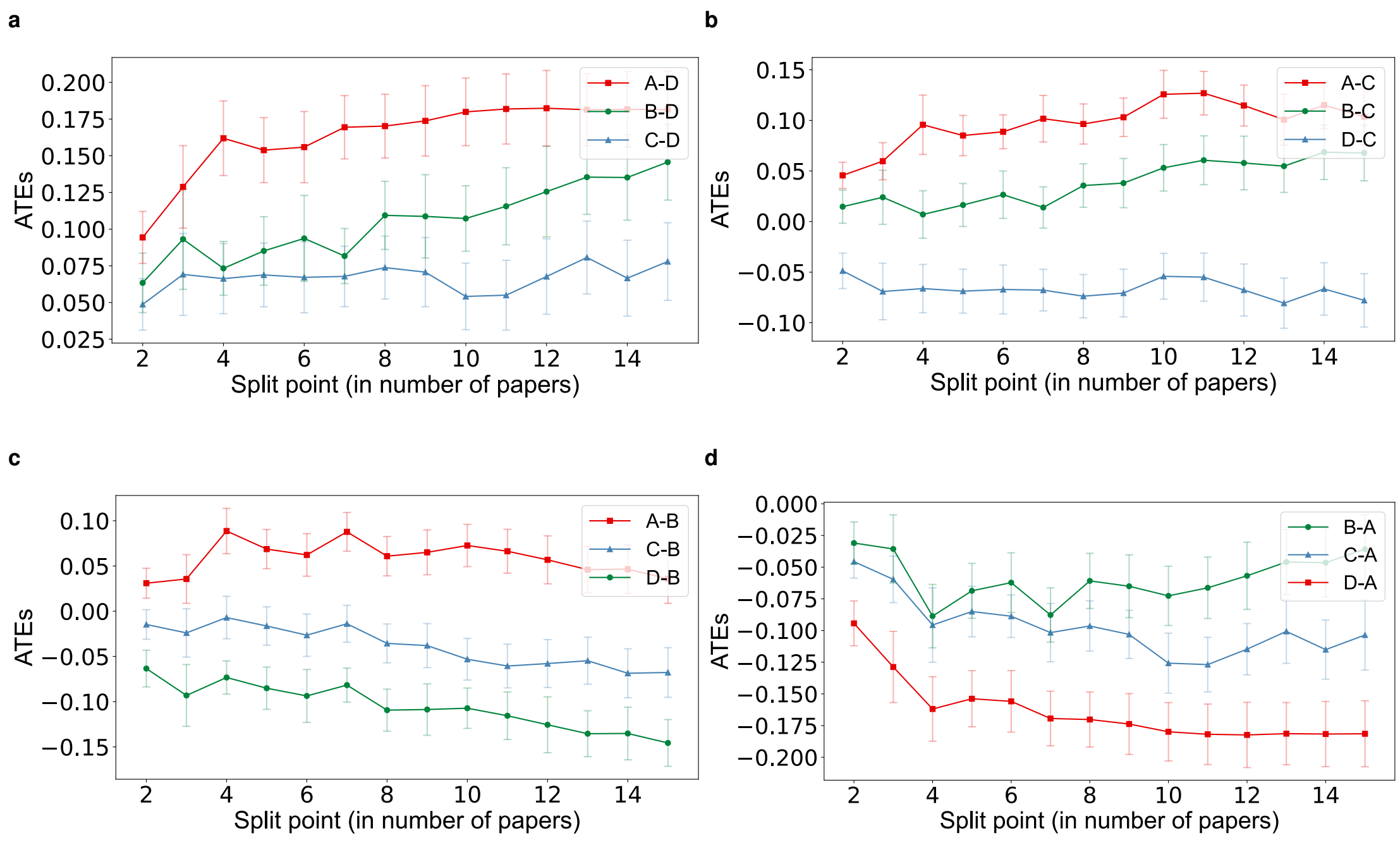

**Fig. A20. Average treatment effects with different baseline groups and varying numbers of papers as split point in the ACS dataset.** Note that (a) has group D as the baseline, (b) group C, (c) group B, and (d) group A.

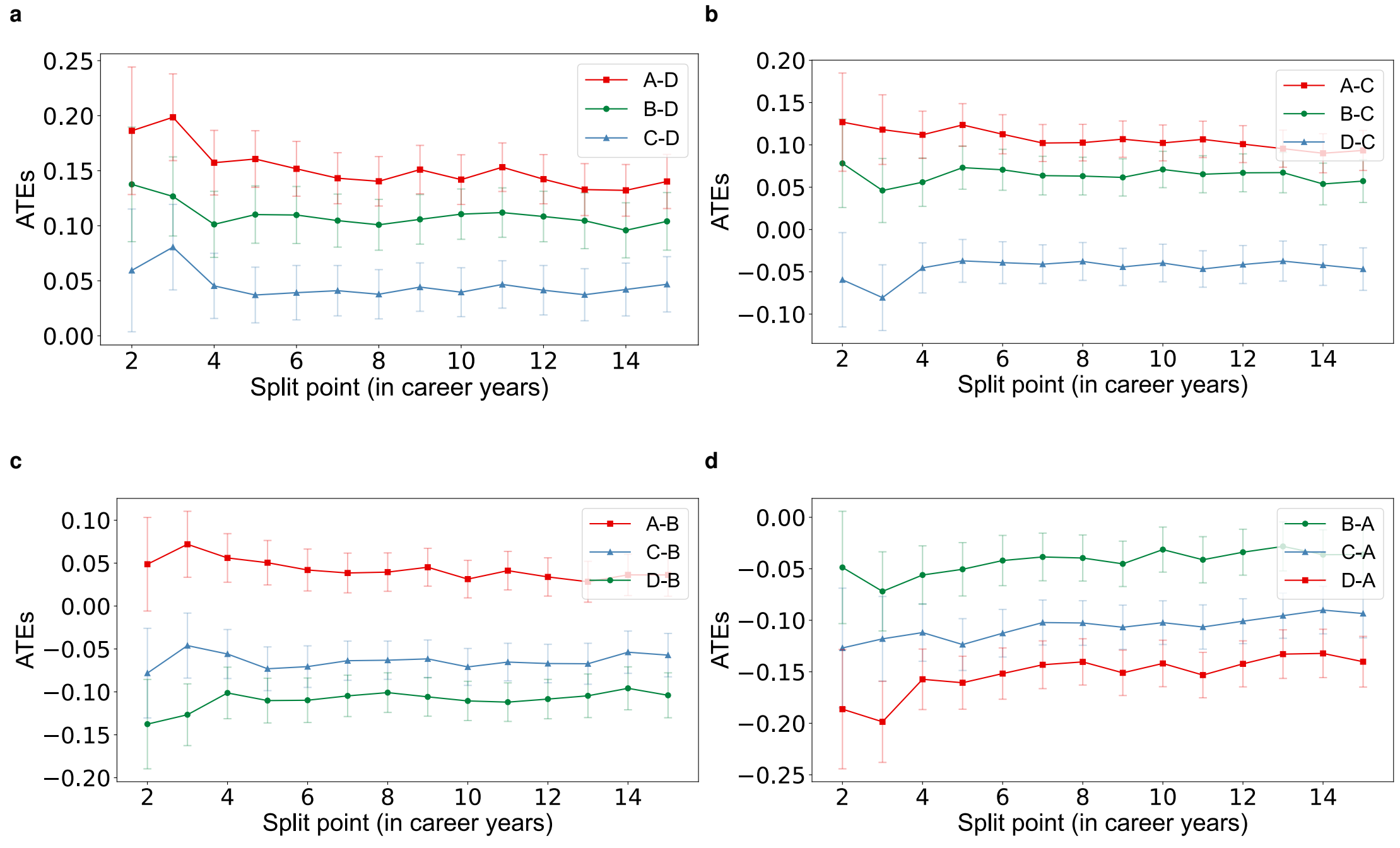

**Fig. A21. Average treatment effects with different baseline groups and varying career years as split point in the ACS dataset.** Note that (a) has group D as the baseline, (b) group C, (c) group B, and (d) group A.

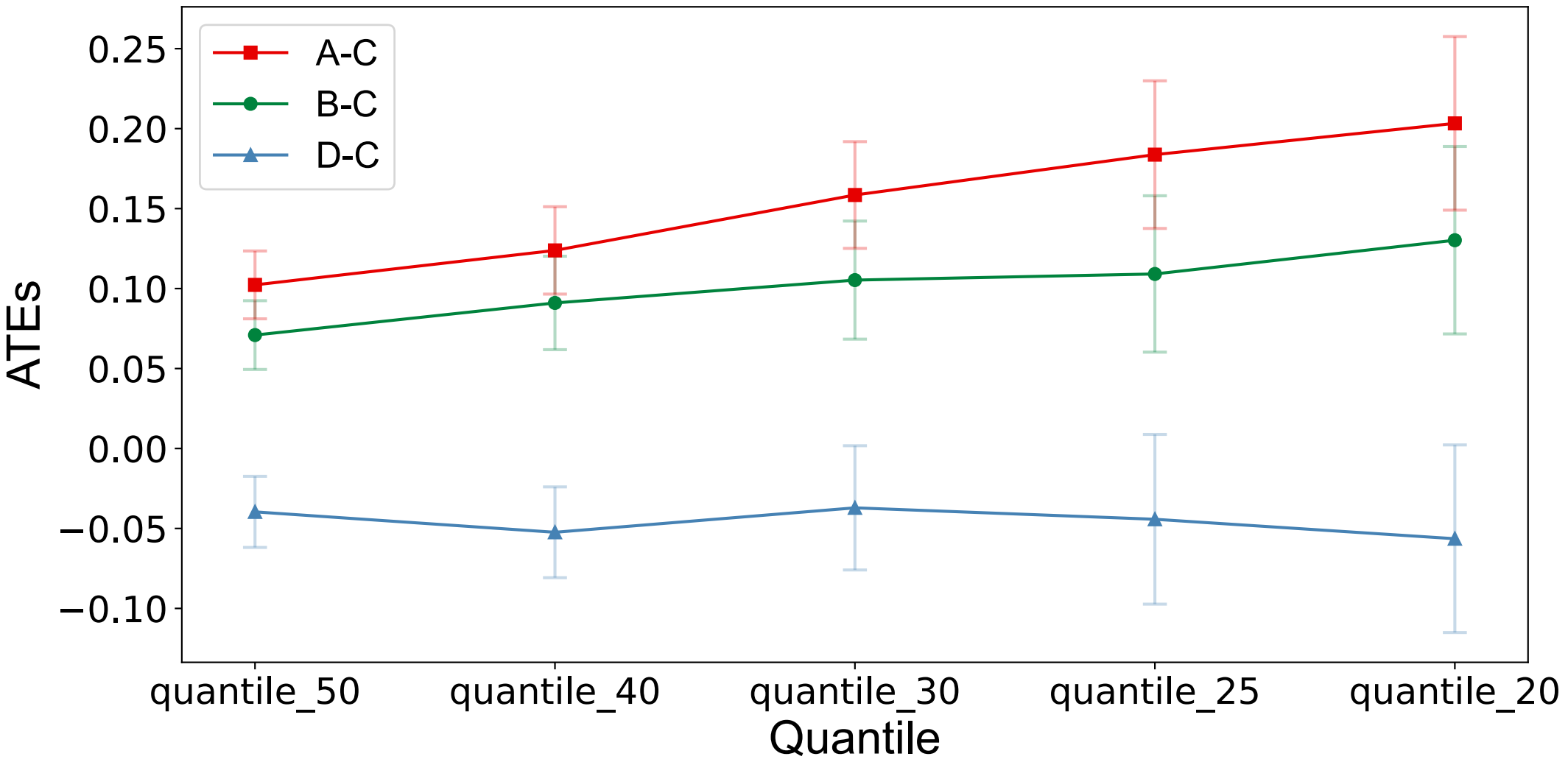

**Fig. A22. Average treatment effects in the ACS dataset with varying quantiles**. The split point is 10 career year and group C is the baseline.

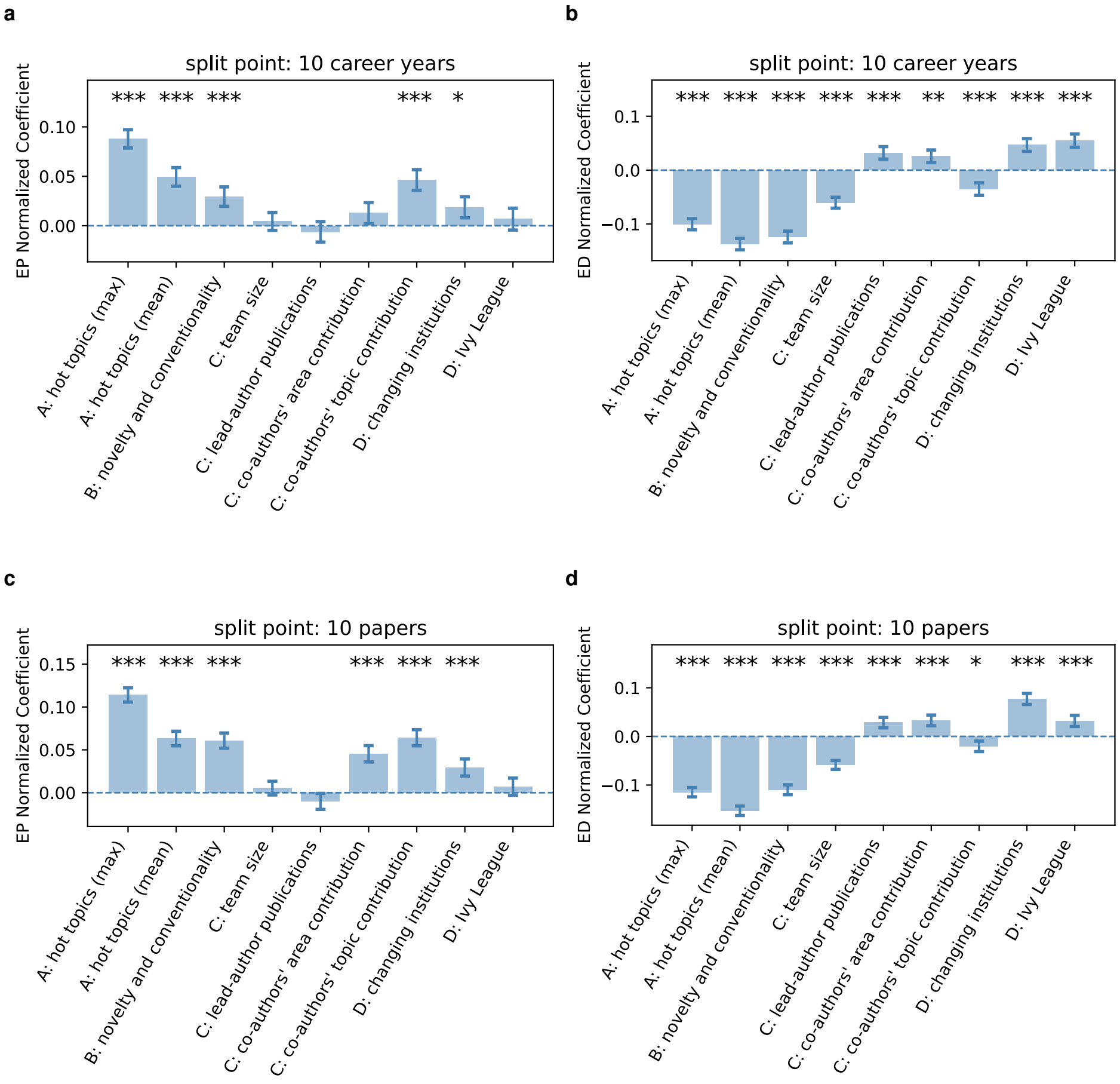

**Fig. A23. Possible mechanisms to explain why conservative adventurers have more impact in the future.** We present the regression analyzes to investigate if scientists with different EPs and EDs behave differently in these dimensions in the future: area popularity of future publications, probability of publishing both novel and conventional papers, team size, probability of publishing lead-author papers, co-author's area and topic contribution, tendency of changing institutions and whether she would work in an Ivy League university. (a-b) are results with 10 career years as split points, (c-d) are results with 10 papers as split points. The normalized coefficients are calculated by the equation mentioned in the caption of Table A16. ($^{*}P < 0.1$; $^{**}P < 0.05$; $^{***}P < 0.01$. Error bars are based on standard errors.)

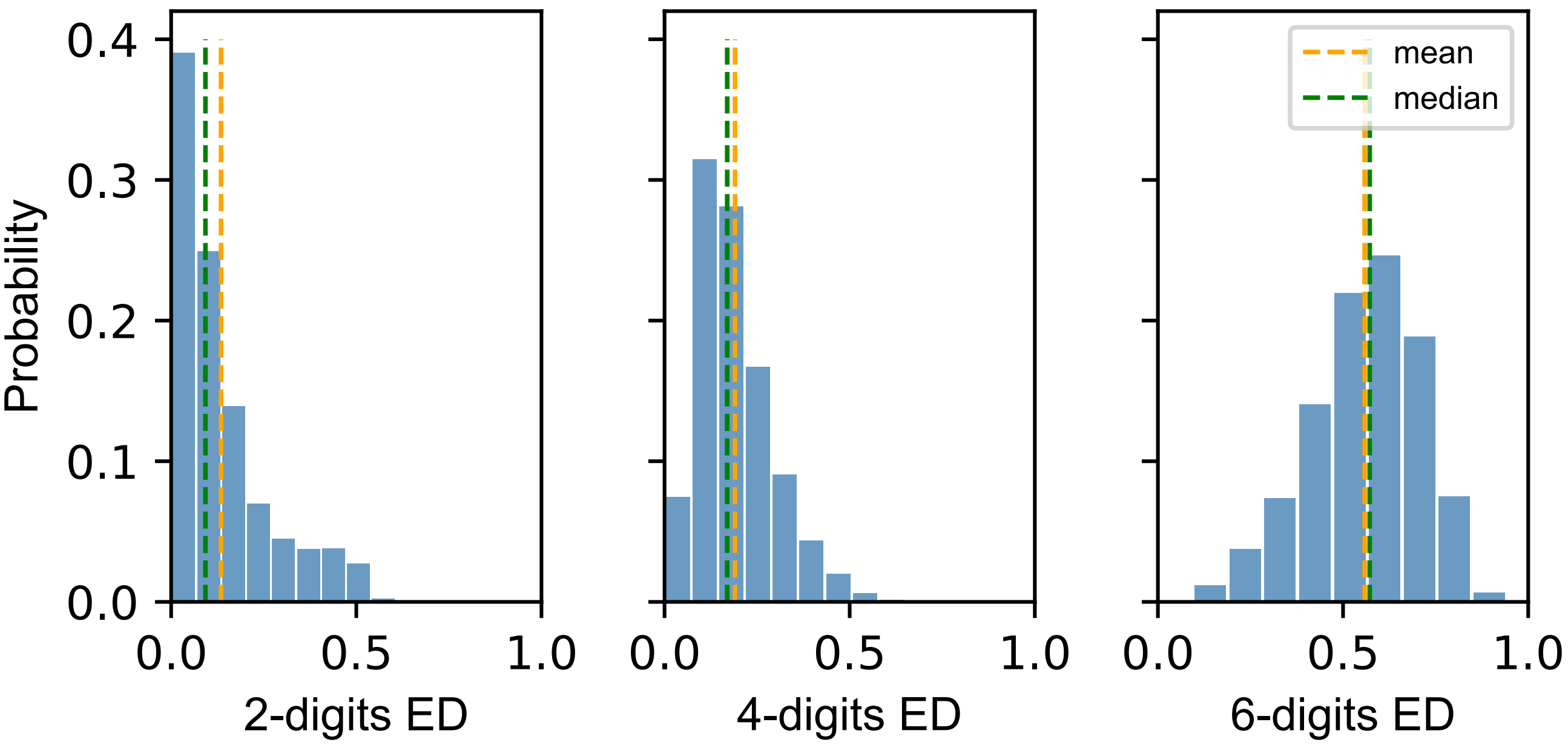

Fig. A24. Distributions of the ED calculated with 2, 4, and 6 digits of PACS codes, respectively.

**Table A1. Pearson correlation coefficient between code distance matrices calculated using papers from different time periods**

| Time periods | 1981-2015 | 1981-1985 | 1986-1990 | 1991-1995 | 1996-2000 | 2001-2005 | 2006-2010 | 2011-2015 |
|---|---|---|---|---|---|---|---|---|
| 1981-2015 | — | 0.544 | 0.751 | 0.793 | 0.849 | 0.883 | 0.892 | 0.886 |
| 1981-1985 | — | — | 0.706 | 0.661 | 0.646 | 0.652 | 0.65 | 0.649 |
| 1986-1990 | — | — | — | 0.808 | 0.778 | 0.758 | 0.748 | 0.734 |
| 1991-1995 | — | — | — | — | 0.805 | 0.76 | 0.75 | 0.728 |
| 1996-2000 | — | — | — | — | — | 0.823 | 0.806 | 0.783 |
| 2001-2005 | — | — | — | — | — | — | 0.869 | 0.832 |
| 2006-2010 | — | — | — | — | — | — | — | 0.851 |
| 2011-2015 | — | — | — | — | — | — | — | — |

We divide the entire dataset of papers into seven 5-year periods: 1981-1985, 1986-1990, 1991-1995, 1996-2000, 2001-2005, 2006-2010, and 2011-2015, respectively, and the proportions of papers in each period are 1.21%, 6.02%, 11.62%, 15.51%, 19.45%, 22.77%, and 23.34%, respectively. We then calculate the code distance matrices for each period and subsequently compute the correlation coefficients between the different code distance matrices.

**Table A2. Regression results with varying regression equations**

| | 10 career years as split point | | | | 10 papers as split point | | | |
|---|---|---|---|---|---|---|---|---|
| Regression Eq. | Eq. (A3) | Eq. (A4) | Eq. (A5) | Eq. (A6) | Eq. (A3) | Eq. (A4) | Eq. (A5) | Eq. (A6) |
| past EP | 0.17*** | 0.27*** | 0.29*** | | 0.19*** | 0.30*** | 0.43*** | |
| | (0.02) | (0.03) | (0.03) | | (0.02) | (0.03) | (0.03) | |
| past ED | | -0.25*** | -0.92*** | | | -0.25*** | -1.03*** | |
| | | (0.04) | (0.04) | | | (0.04) | (0.03) | |
| group A | | | | 0.09*** | | | | 0.10*** |
| | | | | (0.02) | | | | (0.01) |
| group B | | | | 0.06*** | | | | 0.03** |
| | | | | (0.01) | | | | (0.01) |
| group C | | | | 0.07*** | | | | 0.06*** |
| | | | | (0.01) | | | | (0.01) |
| past average $\log c_5$ | 0.47*** | 0.47*** | | 0.47*** | 0.53*** | 0.52*** | | 0.53*** |
| | (0.01) | (0.01) | | (0.01) | (0.01) | (0.01) | | (0.01) |
| past no. of papers | 0.00*** | 0.00*** | | 0.00*** | | | | |
| | (0.00) | (0.00) | | (0.00) | | | | |
| first year | YES | YES | | YES | YES | YES | | YES |
| first area | YES | YES | | YES | YES | YES | | YES |
| const | 0.59*** | 0.71*** | 1.97*** | 0.60*** | 0.63*** | 0.73*** | 2.04*** | 0.67*** |
| | (0.11) | (0.11) | (0.02) | (0.11) | (0.08) | (0.08) | (0.02) | (0.08) |
| Sample Size | 14,526 | 14,526 | 14,526 | 14,526 | 18,154 | 18,154 | 18,154 | 18,154 |
| $R^2$ | 0.28 | 0.29 | 0.04 | 0.28 | 0.34 | 0.34 | 0.05 | 0.34 |

Standard errors in parentheses. (*$P < 0.1$; **$P < 0.05$; ***$P < 0.01$.)
Regression shows how the past EP and ED correlate with future scientific impact when we control for the past average $\log c_5$, the past publication numbers, the first year of career and the area of the first paper.
"Yes" means the corresponding variable is controlled for in the regression study.
Note that, in order to reduce the noise of $\text{EP}_{\text{past}}$, $\text{ED}_{\text{past}}$ and $\text{LogCit}_{\text{future}}$, we select the scientists who had at least 5 publications before the split point and 3 publications after the split point if the split point is defined in terms of career years. If the split point is defined in terms of papers, we select the samples who had at least 3 publications after the split point.

**Table A3. Regression results of Model** Eq. (A3) **with varying split points**

| Split point | Career years as split point | | Number of papers as split point | |
|---|---|---|---|---|
| | EP coeff | Sample Size | EP coeff | Sample Size |
| 2 | 0.10*** | 3,944 | 0.06*** | 25,237 |
| 3 | 0.11*** | 8,197 | 0.09*** | 25,237 |
| 4 | 0.14*** | 11,823 | 0.12*** | 25,237 |
| 5 | 0.16*** | 14,197 | 0.15*** | 25,237 |
| 6 | 0.15*** | 15,373 | 0.17*** | 25,237 |
| 7 | 0.16*** | 15,879 | 0.18*** | 25,237 |
| 8 | 0.18*** | 15,844 | 0.18*** | 22,440 |
| 9 | 0.16*** | 15,312 | 0.20*** | 20,107 |
| 10 | 0.17*** | 14,526 | 0.19*** | 18,154 |
| 11 | 0.18*** | 13,628 | 0.17*** | 16,448 |
| 12 | 0.17*** | 12,693 | 0.17*** | 15,035 |
| 13 | 0.17*** | 11,624 | 0.18*** | 13,679 |
| 14 | 0.17*** | 10,616 | 0.17*** | 12,502 |
| 15 | 0.17*** | 9,518 | 0.17*** | 11,467 |

Regression results under different split points show how the past EP correlate with future scientific impact when we control for the past average $\log c_5$, the past publication numbers, the first year of career and the area of the first paper. The regression equation is always Eq. (A3), unless stated otherwise. We list here only the coefficients of the EP in each regression. ($^*P < 0.1$; $^{**}P < 0.05$; $^{***}P < 0.01$.)

**Table A4. Regression results of Model** Eq. (A4) **with varying split points**

| Split point | Career years as split point | | | Number of papers as split point | | |
|---|---|---|---|---|---|---|
| | EP coeff | ED coeff | Sample Size | EP coeff | ED coeff | Sample Size |
| 2 | 0.18*** | -0.21*** | 3,944 | 0.07*** | -0.05*** | 25,237 |
| 3 | 0.21*** | -0.27*** | 8,197 | 0.11*** | -0.08*** | 25,237 |
| 4 | 0.24*** | -0.26*** | 11,823 | 0.16*** | -0.12*** | 25,237 |
| 5 | 0.26*** | -0.27*** | 14,197 | 0.20*** | -0.16*** | 25,237 |
| 6 | 0.24*** | -0.22*** | 15,373 | 0.23*** | -0.19*** | 25,237 |
| 7 | 0.25*** | -0.23*** | 15,879 | 0.26*** | -0.21*** | 25,237 |
| 8 | 0.28*** | -0.24*** | 15,844 | 0.27*** | -0.22*** | 22,440 |
| 9 | 0.24*** | -0.22*** | 15,312 | 0.29*** | -0.22*** | 20,107 |
| 10 | 0.27*** | -0.25*** | 14,526 | 0.30*** | -0.25*** | 18,154 |
| 11 | 0.29*** | -0.25*** | 13,628 | 0.28*** | -0.24*** | 16,448 |
| 12 | 0.27*** | -0.25*** | 12,693 | 0.29*** | -0.24*** | 15,035 |
| 13 | 0.28*** | -0.26*** | 11,624 | 0.31*** | -0.27*** | 13,679 |
| 14 | 0.29*** | -0.26*** | 10,616 | 0.31*** | -0.27*** | 12,502 |
| 15 | 0.30*** | -0.28*** | 9,518 | 0.30*** | -0.26*** | 11,467 |

Regression results under different split points show how the past EP and ED correlate with future scientific impact when we control for the past average $\log c_5$, the past publication numbers, the first year of career and the area of the first paper. The regression equation is always Eq. (A4), unless stated otherwise. We list here only the coefficients of the EP and ED in each regression. ($^*P < 0.1$; $^{**}P < 0.05$; $^{***}P < 0.01$.)

**Table A5. E-values for the regression analyzes with varying split points**

| Split point | Career years as split point | | Number of papers as split point | |
|---|---|---|---|---|
| | EP | ED | EP | ED |
| 2 | 1.95 | 2.12 | 1.49 | 1.37 |
| 3 | 2.13 | 2.44 | 1.71 | 1.54 |
| 4 | 2.25 | 2.33 | 1.91 | 1.73 |
| 5 | 2.33 | 2.39 | 2.11 | 1.91 |
| 6 | 2.24 | 2.14 | 2.24 | 2.01 |
| 7 | 2.29 | 2.18 | 2.32 | 2.08 |
| 8 | 2.44 | 2.28 | 2.39 | 2.12 |
| 9 | 2.27 | 2.13 | 2.50 | 2.12 |
| 10 | 2.41 | 2.28 | 2.53 | 2.27 |
| 11 | 2.49 | 2.31 | 2.45 | 2.22 |
| 12 | 2.42 | 2.29 | 2.49 | 2.24 |
| 13 | 2.45 | 2.35 | 2.61 | 2.41 |
| 14 | 2.49 | 2.34 | 2.59 | 2.40 |
| 15 | 2.53 | 2.43 | 2.57 | 2.39 |

**Table A6. PSM results with varying career years as split point**

| Career year | EP | | | | ED | | | | Combining the EP and ED | | | |
|---|---|---|---|---|---|---|---|---|---|---|---|---|
| | Pairs | Treat | Control | P | Pairs | Treat | Control | P | Pairs | Treat | Control | P |
| 2 | 1,313 | 1.772 | 1.725 | ** | 809 | 1.751 | 1.762 | NS | 192 | 1.896 | 1.725 | *** |
| | | (0.017) | (0.016) | (*) | | (0.021) | (0.021) | (NS) | | (0.043) | (0.043) | (***) |
| 3 | 2,793 | 1.753 | 1.689 | *** | 1,772 | 1.692 | 1.734 | ** | 314 | 1.822 | 1.691 | *** |
| | | (0.011) | (0.011) | (***) | | (0.014) | (0.014) | (**) | | (0.031) | (0.033) | (***) |
| 4 | 4,108 | 1.725 | 1.668 | *** | 2,634 | 1.666 | 1.722 | *** | 761 | 1.746 | 1.618 | *** |
| | | (0.009) | (0.009) | (***) | | (0.011) | (0.012) | (***) | | (0.021) | (0.021) | (***) |
| 5 | 4,870 | 1.702 | 1.637 | *** | 3,163 | 1.631 | 1.681 | *** | 826 | 1.745 | 1.618 | *** |
| | | (0.009) | (0.009) | (***) | | (0.010) | (0.011) | (***) | | (0.020) | (0.020) | (***) |
| 6 | 5,281 | 1.665 | 1.611 | *** | 3,451 | 1.608 | 1.662 | *** | 926 | 1.705 | 1.598 | *** |
| | | (0.008) | (0.008) | (***) | | (0.010) | (0.010) | (***) | | (0.019) | (0.019) | (***) |
| 7 | 5,438 | 1.643 | 1.588 | *** | 3,630 | 1.587 | 1.635 | *** | 1,003 | 1.666 | 1.562 | *** |
| | | (0.008) | (0.008) | (***) | | (0.009) | (0.010) | (***) | | (0.018) | (0.018) | (***) |
| 8 | 5,420 | 1.629 | 1.568 | *** | 3,588 | 1.560 | 1.608 | *** | 949 | 1.641 | 1.530 | *** |
| | | (0.008) | (0.008) | (***) | | (0.010) | (0.010) | (***) | | (0.018) | (0.019) | (***) |
| 9 | 5,253 | 1.601 | 1.550 | *** | 3,413 | 1.534 | 1.586 | *** | 992 | 1.594 | 1.510 | *** |
| | | (0.008) | (0.008) | (***) | | (0.010) | (0.010) | (***) | | (0.018) | (0.018) | (***) |
| 10 | 4,983 | 1.586 | 1.540 | *** | 3,250 | 1.518 | 1.580 | *** | 994 | 1.570 | 1.506 | ** |
| | | (0.008) | (0.008) | (***) | | (0.010) | (0.010) | (***) | | (0.018) | (0.018) | (***) |
| 11 | 4,679 | 1.576 | 1.543 | *** | 3,005 | 1.519 | 1.562 | *** | 989 | 1.572 | 1.484 | *** |
| | | (0.008) | (0.009) | (***) | | (0.010) | (0.011) | (***) | | (0.018) | (0.017) | (***) |
| 12 | 4,308 | 1.569 | 1.528 | *** | 2,783 | 1.516 | 1.561 | *** | 848 | 1.584 | 1.460 | *** |
| | | (0.009) | (0.009) | (***) | | (0.011) | (0.011) | (***) | | (0.020) | (0.019) | (***) |
| 13 | 3,967 | 1.552 | 1.518 | *** | 2,572 | 1.505 | 1.553 | *** | 829 | 1.574 | 1.476 | *** |
| | | (0.009) | (0.009) | (**) | | (0.011) | (0.012) | (***) | | (0.020) | (0.019) | (***) |
| 14 | 3,608 | 1.561 | 1.522 | *** | 2,332 | 1.500 | 1.551 | *** | 755 | 1.533 | 1.450 | *** |
| | | (0.010) | (0.010) | (***) | | (0.012) | (0.012) | (***) | | (0.022) | (0.020) | (***) |
| 15 | 3,252 | 1.553 | 1.514 | *** | 2,041 | 1.494 | 1.555 | *** | 689 | 1.577 | 1.451 | *** |
| | | (0.010) | (0.010) | (***) | | (0.013) | (0.013) | (***) | | (0.024) | (0.021) | (***) |

For each choice of split point, we present the number of matched pairs, the mean and standard error (in brackets) of the future performance in the treatment and control groups, and the $P$values of paired t-tests and Kruskal–Wallis tests (in brackets). (NS, not significant; $*P < 0.1$; $**P < 0.05$; $***P < 0.01$.)

**Table A7. Average treatment effect in different groups with varying career years as split point**

| Career year | E(C-D) | E(B-D) | E(A-D) |
|---|---|---|---|
| 2 | 0.06332* | 0.06536* | 0.15219*** |
| | (-0.00869, 0.13533) | (-0.00585, 0.13657) | (0.0696, 0.23478) |
| 3 | 0.085*** | 0.09007*** | 0.15051*** |
| | (0.03016, 0.13984) | (0.03521, 0.14493) | (0.08971, 0.21131) |
| 4 | 0.08509*** | 0.09807*** | 0.17384*** |
| | (0.0433, 0.12689) | (0.05709, 0.13904) | (0.12515, 0.22252) |
| 5 | 0.08764*** | 0.07329*** | 0.15863*** |
| | (0.04873, 0.12655) | (0.03553, 0.11106) | (0.11507, 0.2022) |
| 6 | 0.07294*** | 0.07119*** | 0.14727*** |
| | (0.03656, 0.10932) | (0.0353, 0.10708) | (0.10562, 0.18891) |
| 7 | 0.08042*** | 0.08949*** | 0.1543*** |
| | (0.04449, 0.11636) | (0.05446, 0.12453) | (0.114, 0.1946) |
| 8 | 0.07867*** | 0.08535*** | 0.15456*** |
| | (0.04095, 0.11639) | (0.04811, 0.12259) | (0.11193, 0.19718) |
| 9 | 0.07304*** | 0.08347*** | 0.1487*** |
| | (0.03698, 0.1091) | (0.04783, 0.11911) | (0.1078, 0.18959) |
| 10 | 0.07626*** | 0.09318*** | 0.13332*** |
| | (0.03923, 0.11329) | (0.05642, 0.12994) | (0.09108, 0.17556) |
| 11 | 0.07475*** | 0.09819*** | 0.13987*** |
| | (0.03823, 0.11127) | (0.06186, 0.13451) | (0.09774, 0.182) |
| 12 | 0.07302*** | 0.09136*** | 0.13803*** |
| | (0.03358, 0.11247) | (0.05288, 0.12983) | (0.09292, 0.18313) |
| 13 | 0.08556*** | 0.10618*** | 0.14461*** |
| | (0.04421, 0.12691) | (0.06589, 0.14647) | (0.09712, 0.19209) |
| 14 | 0.07323*** | 0.10154*** | 0.13927*** |
| | (0.03001, 0.11645) | (0.0599, 0.14318) | (0.08851, 0.19003) |
| 15 | 0.09245*** | 0.1172*** | 0.14741*** |
| | (0.04454, 0.14036) | (0.07025, 0.16414) | (0.09123, 0.20359) |

PSW results with varying career years as split point. (*$P < 0.1$; **$P < 0.05$; ***$P < 0.01$.) The groups A, B, C and D are defined in Section A4.1.

**Table A8. PSM results with varying numbers of papers as split point**

| Number of papers | EP | | | | ED | | | | Combining the EP amd ED | | | |
|---|---|---|---|---|---|---|---|---|---|---|---|---|
| | Pairs | Treat | Control | P | Pairs | Treat | Control | P | Pairs | Treat | Control | P |
| 2 | 9,814 | 1.719 | 1.663 | *** | 6,023 | 1.668 | 1.689 | ** | 165 | 1.903 | 1.636 | *** |
| | | (0.005) | (0.005) | (***) | | (0.007) | (0.007) | (**) | | (0.041) | (0.039) | (***) |
| 3 | 4,744 | 1.722 | 1.625 | *** | 6,016 | 1.643 | 1.676 | *** | 1,376 | 1.794 | 1.714 | *** |
| | | (0.008) | (0.008) | (***) | | (0.007) | (0.007) | (***) | | (0.015) | (0.015) | (***) |
| 4 | 3,610 | 1.744 | 1.646 | *** | 5,953 | 1.634 | 1.659 | ** | 1,904 | 1.696 | 1.610 | *** |
| | | (0.009) | (0.009) | (***) | | (0.007) | (0.007) | (***) | | (0.013) | (0.012) | (***) |
| 5 | 2,849 | 1.798 | 1.707 | *** | 5,862 | 1.622 | 1.649 | *** | 1,253 | 1.777 | 1.707 | *** |
| | | (0.011) | (0.011) | (***) | | (0.007) | (0.007) | (***) | | (0.016) | (0.017) | (***) |
| 6 | 5,992 | 1.692 | 1.603 | *** | 5,740 | 1.601 | 1.639 | *** | 1,696 | 1.736 | 1.634 | *** |
| | | (0.008) | (0.008) | (***) | | (0.008) | (0.008) | (***) | | (0.014) | (0.014) | (***) |
| 7 | 5,277 | 1.714 | 1.613 | *** | 5,584 | 1.591 | 1.630 | *** | 1,847 | 1.668 | 1.573 | *** |
| | | (0.008) | (0.008) | (***) | | (0.008) | (0.008) | (***) | | (0.014) | (0.014) | (***) |
| 8 | 3,166 | 1.772 | 1.693 | *** | 4,927 | 1.598 | 1.633 | *** | 427 | 1.883 | 1.673 | *** |
| | | (0.011) | (0.011) | (***) | | (0.008) | (0.008) | (***) | | (0.029) | (0.029) | (***) |
| 9 | 5,262 | 1.687 | 1.609 | *** | 4,291 | 1.602 | 1.649 | *** | 1,369 | 1.727 | 1.610 | *** |
| | | (0.008) | (0.008) | (***) | | (0.009) | (0.009) | (***) | | (0.016) | (0.016) | (***) |
| 10 | 4,403 | 1.717 | 1.641 | *** | 3,831 | 1.615 | 1.658 | *** | 1,270 | 1.701 | 1.606 | *** |
| | | (0.009) | (0.009) | (***) | | (0.010) | (0.009) | (***) | | (0.016) | (0.016) | (***) |
| 11 | 2,526 | 1.790 | 1.736 | *** | 3,411 | 1.623 | 1.660 | *** | 506 | 1.818 | 1.700 | *** |
| | | (0.012) | (0.012) | (***) | | (0.010) | (0.010) | (***) | | (0.026) | (0.027) | (***) |
| 12 | 4,157 | 1.699 | 1.642 | *** | 3,091 | 1.639 | 1.682 | *** | 637 | 1.695 | 1.588 | *** |
| | | (0.009) | (0.009) | (***) | | (0.010) | (0.011) | (***) | | (0.023) | (0.022) | (***) |
| 13 | 3,599 | 1.742 | 1.683 | *** | 2,814 | 1.640 | 1.699 | *** | 947 | 1.715 | 1.613 | *** |
| | | (0.010) | (0.010) | (***) | | (0.011) | (0.011) | (***) | | (0.019) | (0.018) | (***) |
| 14 | 2,060 | 1.813 | 1.738 | *** | 2,556 | 1.652 | 1.722 | *** | 533 | 1.790 | 1.631 | *** |
| | | (0.013) | (0.012) | (***) | | (0.011) | (0.012) | (***) | | (0.024) | (0.024) | (***) |
| 15 | 3,019 | 1.743 | 1.688 | *** | 2,325 | 1.656 | 1.735 | *** | 521 | 1.760 | 1.610 | *** |
| | | (0.011) | (0.010) | (***) | | (0.012) | (0.012) | (***) | | (0.025) | (0.024) | (***) |

PSM results for EP, ED, combining EP and ED in different numbers of papers as split points. For each choice of split point, we present the number of matched pairs, the mean and standard error (in brackets) of the future performance in the treatment and control groups, and the $P$values of paired-sample t-tests and Kruskal–Wallis tests (in brackets). ($*P < 0.1$; $**P < 0.05$; $***P < 0.01$.)

**Table A9. Average treatment effect with varying numbers of papers as split point**

| Number of papers | E(C-D) | E(B-D) | E(A-D) |
|---|---|---|---|
| 2 | 0.0975*** | 0.06725*** | 0.14483*** |
| | (0.06941, 0.12559) | (0.03888, 0.09562) | (0.11441, 0.17524) |
| 3 | 0.15646*** | 0.10633*** | 0.24904*** |
| | (0.10877, 0.20414) | (0.05856, 0.1541) | (0.19562, 0.30247) |
| 4 | 0.18537*** | 0.12084*** | 0.2378*** |
| | (0.12063, 0.25011) | (0.05371, 0.18796) | (0.17128, 0.30431) |
| 5 | 0.11941*** | 0.07209*** | 0.17634*** |
| | (0.08514, 0.15368) | (0.04065, 0.10352) | (0.13324, 0.21943) |
| 6 | 0.16588*** | 0.1204*** | 0.22243*** |
| | (0.12624, 0.20552) | (0.08099, 0.15981) | (0.17748, 0.26738) |
| 7 | 0.17998*** | 0.13927*** | 0.24218*** |
| | (0.13266, 0.2273) | (0.09054, 0.188) | (0.19218, 0.29218) |
| 8 | 0.11549*** | 0.08618*** | 0.17371*** |
| | (0.07871, 0.15226) | (0.05226, 0.12011) | (0.12854, 0.21888) |
| 9 | 0.13348*** | 0.10238*** | 0.1863*** |
| | (0.08824, 0.17872) | (0.05805, 0.14671) | (0.13546, 0.23713) |
| 10 | 0.09276*** | 0.0745*** | 0.16935*** |
| | (0.04149, 0.14402) | (0.02287, 0.12612) | (0.11443, 0.22427) |
| 11 | 0.08301*** | 0.08783*** | 0.17041*** |
| | (0.04279, 0.12324) | (0.05039, 0.12527) | (0.11877, 0.22205) |
| 12 | 0.08535*** | 0.08783*** | 0.15778*** |
| | (0.03935, 0.13135) | (0.04314, 0.13253) | (0.10521, 0.21034) |
| 13 | 0.09062*** | 0.10515*** | 0.17269*** |
| | (0.03817, 0.14307) | (0.05226, 0.15804) | (0.11517, 0.23022) |
| 14 | 0.11532*** | 0.12231*** | 0.21708*** |
| | (0.07031, 0.16034) | (0.0812, 0.16341) | (0.16231, 0.27184) |
| 15 | 0.10024*** | 0.11123*** | 0.18833*** |
| | (0.04964, 0.15083) | (0.06124, 0.16122) | (0.12892, 0.24774) |

PSW results with varying numbers of papers as split points. (*$P < 0.1$; **$P < 0.05$; ***$P < 0.01$.) The groups A, B, C and D are defined in Section A4.1.

**Table A10. Regression results of the system GMM**

| Different period | 5-year period | 3-year period |
|---|---|---|
| L.LogCit | 0.198*** | 0.166*** |
| | (0.0184) | (0.0121) |
| L.EP | 1.091*** | 0.881*** |
| | (0.295) | (0.205) |
| L.ED | -2.207*** | -2.773*** |
| | (0.807) | (0.74) |
| L.P | 0.0528*** | 0.0702*** |
| | (0.00629) | (0.00722) |
| L.Careeryear | -0.0205*** | -0.0349*** |
| | (0.00353) | (0.00373) |
| Constant | 1.946*** | 2.623*** |
| | (0.362) | (0.343) |
| Time Fixed Effect | Yes | Yes |
| Author Fixed Effect | Yes | Yes |
| Observations | 36,297 | 53,484 |
| Number of authors | 18,766 | 21,031 |
| Hansen | 14.02 | 13.79 |
| | (0.372) | (0.245) |
| AR(1) | -21.29 | -33.47 |
| | (0) | (0) |
| AR(2) | -0.824 | 1.37 |
| | (0.41) | (0.171) |

Standard errors are reported below the estimated coefficients of the system GMM. (*$P < 0.1$; **$P < 0.05$; ***$P < 0.01$.) "L" in variable names stands for lag or lagged. Regression shows how the lagged EP and ED correlate with future scientific impact when we control for scientific impact, publication number, career year, time fixed effect, and author fixed effect in the lagged period.

"Yes" means the corresponding variable is controlled for in the regression study.

In the below part of the table, the Hansen statistic tests if the proposed model suffers from over-identification, and the AR(1) and AR(2) are tests for autocorrelation in differences. The corresponding p-values for these tests are shown in parentheses. In our case, the Hansen test shows no signs of over identifying issues, and the AR tests show that there is first-order residual autocorrelation but no second-order autocorrelation. Combining the tests shows that our setup (choices of lagged response, independent, dependent and endogenous variables) in the model satisfies the conditions for applying system GMM.

**Table A11. Regression results on the $\mathrm{PubMed}_1$ dataset**

| Split point | Career years as split points | | | Number of papers as split points | | |
|---|---|---|---|---|---|---|
| | EP coeff | ED coeff | Sample Size | EP coeff | ED coeff | Sample Size |
| 2 | 0.03*** | -0.23*** | 267,703 | 0.02*** | -0.17*** | 1,218,355 |
| 3 | 0.04*** | -0.22*** | 449,043 | 0.03*** | -0.19*** | 1,218,355 |
| 4 | 0.05*** | -0.21*** | 572,547 | 0.04*** | -0.19*** | 1,218,355 |
| 5 | 0.06*** | -0.19*** | 644,001 | 0.05*** | -0.18*** | 1,218,355 |
| 6 | 0.06*** | -0.20*** | 678,146 | 0.06*** | -0.17*** | 1,218,355 |
| 7 | 0.07*** | -0.18*** | 685,769 | 0.07*** | -0.16*** | 1,218,355 |
| 8 | 0.07*** | -0.18*** | 676,301 | 0.08*** | -0.16*** | 1,108,314 |
| 9 | 0.08*** | -0.17*** | 654,789 | 0.09*** | -0.16*** | 1,016,137 |
| 10 | 0.09*** | -0.17*** | 626,712 | 0.09*** | -0.17*** | 937,477 |
| 11 | 0.10*** | -0.16*** | 593,621 | 0.09*** | -0.17*** | 869,808 |
| 12 | 0.10*** | -0.16*** | 558,425 | 0.10*** | -0.17*** | 811,250 |
| 13 | 0.10*** | -0.15*** | 522,716 | 0.10*** | -0.18*** | 759,211 |
| 14 | 0.11*** | -0.15*** | 487,054 | 0.10*** | -0.18*** | 712,870 |
| 15 | 0.12*** | -0.15*** | 451,955 | 0.10*** | -0.19*** | 671,552 |

Regression results under different setting of split points show how the past EP and ED correlate with future scientific impact when controlling for the past average $\log c_5$, the past publication numbers, the first year of career and area of the first paper. Note that, in order to reduce the noise brought by the scientist's past EP and ED and future performance in the regression, we select the scientists who had at least 5 publications before the split point and 3 publications after the split point. ($*P < 0.1$; $**P < 0.05$; $***P < 0.01$.)

**Table A12. Regression results on the $\mathrm{PubMed}_1$ dataset with an alternative performance measurement**

| Split point | Career years as split points | | | Number of papers as split points | | |
|---|---|---|---|---|---|---|
| | EP coeff | ED coeff | Sample Size | EP coeff | ED coeff | Sample Size |
| 2 | 0.23*** | -0.17*** | 267,703 | 0.06*** | -0.13*** | 1,218,355 |
| 3 | 0.31*** | -0.16*** | 449,043 | 0.12*** | -0.15*** | 1,218,355 |
| 4 | 0.34*** | -0.17*** | 572,547 | 0.16*** | -0.15*** | 1,218,355 |
| 5 | 0.37*** | -0.12*** | 644,001 | 0.20*** | -0.14*** | 1,218,355 |
| 6 | 0.39*** | -0.12*** | 678,146 | 0.24*** | -0.12*** | 1,218,355 |
| 7 | 0.41*** | -0.10*** | 685,769 | 0.28*** | -0.11*** | 1,218,355 |
| 8 | 0.42*** | -0.09*** | 676,301 | 0.31*** | -0.11*** | 1,108,314 |
| 9 | 0.43*** | -0.08*** | 654,789 | 0.33*** | -0.12*** | 1,016,137 |
| 10 | 0.45*** | -0.07*** | 626,712 | 0.35*** | -0.14*** | 937,477 |
| 11 | 0.46*** | -0.06*** | 593,621 | 0.36*** | -0.15*** | 869,808 |
| 12 | 0.48*** | -0.05*** | 558,425 | 0.38*** | -0.15*** | 811,250 |
| 13 | 0.49*** | -0.03*** | 522,716 | 0.40*** | -0.17*** | 759,211 |
| 14 | 0.51*** | -0.03*** | 487,054 | 0.41*** | -0.18*** | 712,870 |
| 15 | 0.52*** | -0.02* | 451,955 | 0.42*** | -0.20*** | 671,552 |

The regression equation is always Eq. (A4).
Regression results under different settings of split points show how the past EP and ED correlate with future scientific impact measured by the maximum 5-year log-citations after split points, when we control for the past average $\log c_5$, the past publication numbers, the first year of career and area of the first paper. ($^*P < 0.1$; $^{**}P < 0.05$; $^{***}P < 0.01$.)

**Table A13. Regression results on the $\mathrm{PubMed_2}$ dataset**

| Split point | Career years as split points | | | Number of papers as split points | | |
|---|---|---|---|---|---|---|
| | EP coeff | ED coeff | Sample Size | EP coeff | ED coeff | Sample Size |
| 2 | 0.08*** | -0.37*** | 272,841 | 0.03*** | -0.22*** | 1,260,125 |
| 3 | 0.08*** | -0.37*** | 469,409 | 0.05*** | -0.26*** | 1,260,125 |
| 4 | 0.08*** | -0.36*** | 606,146 | 0.07*** | -0.27*** | 1,260,125 |
| 5 | 0.09*** | -0.35*** | 685,944 | 0.09*** | -0.28*** | 1,260,125 |
| 6 | 0.10*** | -0.34*** | 724,003 | 0.11*** | -0.27*** | 1,260,125 |
| 7 | 0.11*** | -0.33*** | 731,949 | 0.12*** | -0.26*** | 1,260,125 |
| 8 | 0.12*** | -0.32*** | 720,138 | 0.13*** | -0.27*** | 1,145,845 |
| 9 | 0.12*** | -0.30*** | 694,829 | 0.13*** | -0.27*** | 1,050,425 |
| 10 | 0.13*** | -0.29*** | 661,940 | 0.14*** | -0.28*** | 969,054 |
| 11 | 0.14*** | -0.28*** | 623,877 | 0.15*** | -0.27*** | 899,121 |
| 12 | 0.16*** | -0.27*** | 583,888 | 0.16*** | -0.27*** | 837,921 |
| 13 | 0.17*** | -0.25*** | 544,020 | 0.16*** | -0.27*** | 783,907 |
| 14 | 0.17*** | -0.24*** | 504,690 | 0.16*** | -0.27*** | 736,002 |
| 15 | 0.17*** | -0.23*** | 466,476 | 0.17*** | -0.26*** | 692,976 |

Regression results under different setting of split points show how the past EP and ED correlate with future scientific impact when controlling for the past average $\log c_5$, the past publication numbers, the first year of career and area of the first paper. Note that, in order to reduce the noise brought by the scientist's past EP and ED and future performance in the regression, we select the scientists who had at least 5 publications before the split point and 3 publications after the split point. ($^*P < 0.1$; $^{**}P < 0.05$; $^{***}P < 0.01$.)

Table A14. Regression results on the ACS dataset

| Split point | Career years as split points | | | Number of papers as split points | | |
|---|---|---|---|---|---|---|
| | EP coeff | ED coeff | Sample Size | EP coeff | ED coeff | Sample Size |
| 2 | 0.11*** | -0.24*** | 17,319 | 0.04*** | -0.12*** | 62,486 |
| 3 | 0.11*** | -0.26*** | 24,162 | 0.06*** | -0.15*** | 62,486 |
| 4 | 0.12*** | -0.27*** | 29,042 | 0.08*** | -0.16*** | 62,486 |
| 5 | 0.13*** | -0.27*** | 32,270 | 0.11*** | -0.17*** | 62,486 |
| 6 | 0.14*** | -0.25*** | 33,920 | 0.13*** | -0.17*** | 62,486 |
| 7 | 0.14*** | -0.24*** | 34,345 | 0.15*** | -0.19*** | 62,486 |
| 8 | 0.14*** | -0.22*** | 34,188 | 0.16*** | -0.21*** | 54,756 |
| 9 | 0.15*** | -0.22*** | 33,281 | 0.17*** | -0.23*** | 48,689 |
| 10 | 0.15*** | -0.21*** | 31,826 | 0.18*** | -0.25*** | 43,680 |
| 11 | 0.16*** | -0.21*** | 30,278 | 0.18*** | -0.28*** | 39,479 |
| 12 | 0.16*** | -0.19*** | 28,524 | 0.19*** | -0.28*** | 35,770 |
| 13 | 0.17*** | -0.18*** | 26,769 | 0.18*** | -0.30*** | 32,641 |
| 14 | 0.17*** | -0.18*** | 25,085 | 0.18*** | -0.32*** | 29,927 |
| 15 | 0.18*** | -0.17*** | 23,426 | 0.20*** | -0.34*** | 27,533 |

Regression results under different setting of split points show how the past EP and ED correlate with future scientific impact when controlling for the past average $\log c_5$, the past publication numbers, the first year of career and area of the first paper. Note that, in order to reduce the noise brought by the scientist's past EP and ED and future performance in the regression, we select the scientists who had at least 5 publications before the split point and 3 publications after the split point. (*$P < 0.1$; **$P < 0.05$; ***$P < 0.01$.)

**Table A15. Regression results on the ACS dataset with an alternative performance measurement**

| Split point | Career years as split points | | | Number of papers as split points | | |
|---|---|---|---|---|---|---|
| | EP coeff | ED coeff | Sample Size | EP coeff | ED coeff | Sample Size |
| 2 | 0.34*** | -0.48*** | 17,319 | 0.09*** | -0.31*** | 62,486 |
| 3 | 0.35*** | -0.51*** | 24,162 | 0.14*** | -0.38*** | 62,486 |
| 4 | 0.35*** | -0.56*** | 29,042 | 0.20*** | -0.41*** | 62,486 |
| 5 | 0.37*** | -0.55*** | 32,270 | 0.27*** | -0.43*** | 62,486 |
| 6 | 0.39*** | -0.56*** | 33,920 | 0.30*** | -0.44*** | 62,486 |
| 7 | 0.40*** | -0.54*** | 34,345 | 0.35*** | -0.47*** | 62,486 |
| 8 | 0.41*** | -0.50*** | 34,188 | 0.39*** | -0.51*** | 54,756 |
| 9 | 0.41*** | -0.49*** | 33,281 | 0.40*** | -0.57*** | 48,689 |
| 10 | 0.42*** | -0.46*** | 31,826 | 0.43*** | -0.62*** | 43,680 |
| 11 | 0.43*** | -0.44*** | 30,278 | 0.44*** | -0.65*** | 39,479 |
| 12 | 0.42*** | -0.42*** | 28,524 | 0.46*** | -0.66*** | 35,770 |
| 13 | 0.43*** | -0.42*** | 26,769 | 0.44*** | -0.70*** | 32,641 |
| 14 | 0.43*** | -0.43*** | 25,085 | 0.45*** | -0.74*** | 29,927 |
| 15 | 0.42*** | -0.43*** | 23,426 | 0.47*** | -0.77*** | 27,533 |

The regression equation is always Eq. (A4).

Regression results under different settings of split points show how the past EP and ED correlate with future scientific impact measured by the maximum 5-year log-citations after split points, when we control for the past average $\log c_5$, the past publication numbers, the first year of career and area of the first paper. (*$P < 0.1$; **$P < 0.05$; ***$P < 0.01$.)

**Table A16. Normalized regression coefficients of the EP and ED with different factors as the dependent variable**

| Different factors | 10 career years as split point | | 10 papers as split point | |
|---|---|---|---|---|
| | EP | ED | EP | ED |
| A: hot areas (max) | 0.09*** | -0.10*** | 0.11*** | -0.11*** |
| A: hot areas (mean) | 0.05*** | -0.14*** | 0.06*** | -0.15*** |
| B: novelty and conventionality | 0.03*** | -0.12*** | 0.06*** | -0.11*** |
| C: lead-author publications | -0.01 | 0.043*** | -0.01 | 0.03*** |
| C: co-authors' area contribution | 0.01 | 0.02** | 0.04*** | 0.03*** |
| C: co-authors' topic contribution | 0.04*** | -0.03*** | 0.06*** | -0.02* |
| C: team size | 0.00 | -0.06*** | 0.01 | -0.06*** |
| D: changing institutions | 0.02** | 0.05*** | 0.03*** | 0.08*** |
| D: Ivy League | 0.01 | 0.05*** | 0.01 | 0.03*** |

We run several regressions to determine if scientists with different EPs and EDs behave differently in these dimensions in the future: area popularity of future publications, probability of publishing both novel and conventional papers, team size, probability of publishing lead-author papers, co-author's area and topic contribution, tendency of changing institutions, whether working in the Ivy League and co-authors' area/topic contribution. In the table, we list the normalized coefficients of the EP and ED in each regression. (*$P < 0.1$; **$P < 0.05$; ***$P < 0.01$.)

We compute the normalized coefficients of the independent variable $i$ with the following equation (34)

$$\text{coef}_{i,\text{norm}} = \text{coef}_{i,\text{raw}} \times \frac{\text{SD}_i}{\text{SD}_{\text{dependent}}},$$

where $\text{coef}_{i,\text{raw}}$ is the raw regression coefficient of the variable $i$, $\text{SD}_i$ is the standard deviation of $i$, $\text{SD}_{\text{dependent}}$ is the standard deviation of the dependent variable, and $\text{coef}_{\text{norm}}$ is the normalized coefficient of $i$.

**Table A17. Coefficients of the EP and ED when controlling for different factors**

| Different factors | 10 career years as split point | | 10 papers as split point | |
|---|---|---|---|---|
| | EP | ED | EP | ED |
| vanilla regression | 0.27*** | -0.25*** | 0.30*** | -0.25*** |
| A: hot areas (max) | 0.24*** | -0.21*** | 0.27*** | -0.21*** |
| A: hot areas (mean) | 0.25*** | -0.19*** | 0.28*** | -0.18*** |
| B: novelty and conventionality | 0.26*** | -0.18*** | 0.28*** | -0.20*** |
| C: lead-author publications | 0.27*** | -0.28*** | 0.29*** | -0.26*** |
| C: co-authors's area contribution | 0.27*** | -0.26*** | 0.29*** | -0.26*** |
| C: co-authors's topic contribution | 0.25*** | -0.23*** | 0.28*** | -0.24*** |
| C: team size | 0.27*** | -0.22*** | 0.30*** | -0.24*** |
| D: changing institutions | 0.28*** | -0.22*** | 0.31*** | -0.21*** |
| D: Ivy League | 0.27*** | -0.25*** | 0.30*** | -0.25*** |
| combining all the above | 0.28*** | -0.11** | 0.28*** | -0.11** |

We run several regressions to test if the hypotheses explain the observation that conservative adventurers gain more impact in the future. Specifically, we firstly control for each of the factors separately on top of our vanilla regression (see Eq. (A4)), and then we control for all of them in one regression study. We list here only the coefficients of the EP and ED in each regression. ($*P < 0.1$; $**P < 0.05$; $***P < 0.01$.) Vanilla regression refers to the (original) regression analysis with Eq. (A4) without any of the variables below.

**Table A18. Pearson correlation coefficients between confounding factors and other independent variables**

| | $\mathrm{EP}_{\mathrm{past}}$ | $\mathrm{ED}_{\mathrm{past}}$ | $P_{\mathrm{past}}$ | $\mathrm{LogCit}_{\mathrm{past}}$ |
|---|---|---|---|---|
| A: hot areas (max) | 0.02 | -0.12 | 0.09 | 0.07 |
| A: hot areas (mean) | -0.03 | -0.16 | 0.09 | 0.10 |
| B: novelty and conventionality | -0.13 | -0.27 | 0.11 | 0.13 |
| C: co-authors' area contribution | 0.03 | 0.02 | 0.05 | 0.07 |
| C: co-authors' topic contribution | 0.00 | -0.06 | 0.10 | 0.09 |
| C: team size | -0.07 | -0.12 | 0.05 | -0.01 |
| D: changing institutions | 0.09 | 0.08 | -0.19 | -0.09 |
| D: Ivy League | 0.02 | 0.04 | 0.02 | 0.05 |

The correlation coefficients of the independent variables in each regression in Hypothesis A–E. The split point is 10 career years.

**Table A19. Pearson correlation coefficients between confounding factors and other independent variables with 10 papers as split point**

| | $EP_{past}$ | $ED_{past}$ | $LogCit_{past}$ |
|---|---|---|---|
| A: hot areas (max) | 0.04 | -0.10 | 0.06 |
| A: hot areas (mean) | -0.02 | -0.15 | 0.09 |
| B: novelty and conventionality | -0.11 | -0.25 | 0.12 |
| C: co-authors' area contribution | 0.07 | 0.03 | 0.05 |
| C: co-authors' topic contribution | 0.03 | -0.04 | 0.07 |
| C: team size | -0.07 | -0.13 | 0.00 |
| D: changing institutions | 0.07 | 0.08 | -0.07 |
| D: Ivy League | 0.01 | 0.00 | 0.09 |

The correlation coefficients of the independent variables in each regression in Hypothesis A–E. Here, the split point is 10 papers.

**Table A20. Mediation effects of different factors on the future impact**

| Mediators | EP as treatment variable | | | ED as treatment variable | | |
|---|---|---|---|---|---|---|
| | ACME | ADE | Total Effect | ACME | ADE | Total Effect |
| A: hot areas (max) | 0.03*** | 0.24*** | 0.27*** | -0.04*** | -0.21*** | -0.25*** |
| A: hot areas (mean) | 0.02*** | 0.25*** | 0.27*** | -0.06*** | -0.19*** | -0.25*** |
| B: novelty and conventionality | 0.01*** | 0.26*** | 0.27*** | -0.06*** | -0.18*** | -0.25*** |
| C: co-authors' area contribution | 0.01 | 0.27*** | 0.27*** | 0.01*** | -0.26*** | -0.25*** |
| C: co-authors' topic contribution | 0.02*** | 0.25*** | 0.27*** | -0.02*** | -0.23** | -0.25*** |
| C: team size | 0.00 | 0.27*** | 0.27*** | -0.02*** | -0.23*** | -0.25*** |
| D: changing institutions | -0.01 | 0.28*** | 0.27*** | -0.02*** | -0.22*** | -0.25*** |
| D: Ivy League | 0.00 | 0.27*** | 0.27*** | 0.01*** | -0.25*** | -0.25*** |

We perform mediation analysis of the past exploratory metrics on the future impact with several confounding factors as mediators and 10 career years as split point. The EP and ED are separately set as the treatment variable. ACME stands for average causal mediation effects, and ADE stands for average direct effects. (*$P < 0.1$; **$P < 0.05$; ***$P < 0.01$.)

**Table A21. Mediation effects of different factors on the future impact with 10 papers as split point**

| Mediators | EP as treatment variable | | | ED as treatment variable | | |
|---|---|---|---|---|---|---|
| | ACME | ADE | Total Effect | ACME | ADE | Total Effect |
| A: hot areas (max) | 0.03*** | 0.26*** | 0.30*** | -0.04*** | -0.20*** | -0.24*** |
| A: hot areas (mean) | 0.02*** | 0.28*** | 0.30*** | -0.06*** | -0.18*** | -0.25*** |
| B: novelty and conventionality | 0.02*** | 0.28*** | 0.30*** | -0.05*** | -0.20*** | -0.25*** |
| C: co-authors' area contribution | 0.01*** | 0.28*** | 0.30*** | 0.01*** | -0.25*** | -0.24*** |
| C: co-authors' topic contribution | 0.02*** | 0.27*** | 0.30*** | -0.01*** | -0.24*** | -0.24*** |
| C: team size | 0.00 | 0.30*** | 0.30*** | -0.01*** | -0.23*** | -0.24*** |
| D: changing institutions | -0.01*** | 0.31*** | 0.30*** | -0.03*** | -0.21*** | -0.24*** |
| D: Ivy League | 0.00 | 0.30*** | 0.30*** | 0.01** | -0.25*** | -0.25*** |

We perform mediation analysis of the past exploratory metrics on the future impact with several confounding factors as mediators and 10 papers as split point. The EP and ED are separately set as the treatment variable. ACME stands for average causal mediation effects, and ADE stands for average direct effects. (*$P < 0.1$; **$P < 0.05$; ***$P < 0.01$.)

**Table A22. Regression coefficients of the EP and ED on different subsets of scientists, with at least 5 or 20 publications**

| Different subsets | 10 career years as split point | | 10 papers as split point | |
|---|---|---|---|---|
| | EP | ED | EP | ED |
| at least 5 publications | 0.229*** | -0.179*** | 0.298*** | -0.245*** |
| at least 20 publications | 0.288*** | -0.265*** | 0.247*** | -0.168*** |

We study scientists with at least 5 or 20 publications with regression equation Eq. (A4), respectively, as opposed to the usual 10 papers considered in the main text and for the rest of the supplement. We list here only the coefficients of the EP and ED in each regression. (*$P < 0.1$; **$P < 0.05$; ***$P < 0.01$.)

**Table A23. Regression coefficients of the EP and ED on the subset excluding scientists with common Asian last names**

| Different subsets | 10 career years as split point | | 10 papers as split point | |
|---|---|---|---|---|
| | EP | ED | EP | ED |
| Without scientists having Asian last names | 0.251*** | -0.251*** | 0.290*** | -0.293*** |

We exclude scientists who have common Asian last names for regression Eq. (A4), with the same regression setup as in the main text. We list here only the coefficients of the EP and ED in each regression. (*$P < 0.1$; **$P < 0.05$; ***$P < 0.01$.)

**Table A24. Regression coefficients of the EP and ED of scientists who started their careers in different periods**

| Dataset | Period | 10 career years as split points | | | 10 papers as split points | | |
|---|---|---|---|---|---|---|---|
| | | EP coeff | ED coeff | Sample size | EP coeff | ED coeff | Sample size |
| APS | 1976-2000 | 0.26*** | -0.26*** | 11,332 | 0.31*** | -0.25*** | 12,202 |
| | 2000-2015 | 0.28*** | -0.18* | 3,194 | 0.27*** | -0.21** | 5,952 |
| PubMed | 20th Century | 0.15*** | -0.28*** | 501,423 | 0.18*** | -0.24*** | 606,627 |
| | 21st Century | 0.08*** | -0.36*** | 160,517 | 0.07*** | -0.36*** | 362,417 |
| ACS | 20th Century | 0.17*** | -0.19*** | 26,559 | 0.24*** | -0.19*** | 30,306 |
| | 21st Century | 0.06 | -0.48*** | 5,252 | 0.05* | -0.63*** | 13,354 |

We study scientists start her career at different period with regression equation Eq. (A4). We list only the coefficients of the EP and ED in each regression. (*$P < 0.1$; **$P < 0.05$; ***$P < 0.01$.)

**Table A25. Regression Coefficients of EP and ED for Scientists in Asian and Non-Asian Groups**

| Country | 10 career years as split points | | 10 papers as split points | |
|---|---|---|---|---|
| | **EP coeff** | **ED coeff** | **EP coeff** | **ED coeff** |
| Asian | 0.383*** | -0.246* | 0.282*** | -0.009 |
| Non-Asian | 0.251*** | -0.251*** | 0.290*** | -0.293*** |

We study scientists in Asian and Non-Asian Groups with regression equation Eq. (A4). We list only the coefficients of the EP and ED in each regression. (*$P < 0.1$; **$P < 0.05$; ***$P < 0.01$.)

**Table A26. Regression coefficients of the EP and ED when controlling for scientists' genders**

| Control factor | 10 career years as split point | | 10 papers as split point | |
|---|---|---|---|---|
| | EP | ED | EP | ED |
| gender | 0.240*** | -0.323*** | 0.250*** | -0.301*** |

We add the gender of each scientist as a control variable in the regression equation Eq. (A4), with the same regression setup as in the main text. We list here only the coefficients of the EP and ED in each regression. (*$P < 0.1$; **$P < 0.05$; ***$P < 0.01$.)

**Table A27. Regression coefficients of the EP and ED with varying definitions of EP and ED**

| Different definitions of EP and ED | 10 career years as split point | | 10 papers as split point | |
|---|---|---|---|---|
| | EP | ED | EP | ED |
| 2-digit EP and 2-digit ED | 0.173*** | -0.035 | 0.196*** | -0.063 |
| 2-digit EP and 4-digit ED | 0.196*** | -0.100* | 0.195*** | -0.025 |
| 4-digit EP and 4-digit ED | 0.153*** | -0.045 | 0.173*** | 0.011 |
| 4-digit EP and 6-digit ED | 0.220*** | -0.209*** | 0.261*** | -0.224*** |
| 2-digit EP and 6-digit ED | 0.270*** | -0.245*** | 0.298*** | -0.245*** |

We vary the numbers of PACS code digits for computing EP and ED. The regression equation is Eq. (A4), with the same setup as in the main text. We list here only the coefficients of the EP and ED in each regression. (*$P < 0.1$; **$P < 0.05$; ***$P < 0.01$.)
It shows that using 2 or 4 digits for the EP does not affect the significance of the results, while the ED needs to be calculated with 6 digits. See the discussion in Section A7.3.1.

**Table A28. Regression coefficients of EP and ED under alternative topic graphs for measuring topic distances**

| Topic Graph | 10 career years as split point | | 10 papers as split point | |
|---|---|---|---|---|
| | EP | ED | EP | ED |
| citation | 0.255*** | -0.267*** | 0.277*** | -0.258*** |
| co-citing | 0.180*** | -0.317** | 0.199*** | -0.313*** |

We test the citation graph and co-citing graph as described in Section A7.3.2. The regression equation is Eq. (A4), with the same setup as in the main text. We list here only the coefficients of the EP and ED in each regression. (*$P < 0.1$; **$P < 0.05$; ***$P < 0.01$.)

**Table A29. Regression coefficients of EP and ED under different node distance metrics**

| Distance Metric | 10 career years as split point | | 10 papers as split point | |
|---|---|---|---|---|
| | EP | ED | EP | ED |
| Jaccard | 0.233*** | -0.204*** | 0.222*** | -0.119*** |
| node2vec | 0.272*** | -0.502*** | 0.258*** | -0.412*** |

Instead of using the weighted overlap of two nodes' neighbors as their distance on the graph, we explore the distance metric using unweighted Jaccard and node2vec algorithms, as described in Section A7.3.3. The regression equation is Eq. (A4), with the same setup as in the main text. We list here only the coefficients of the EP and ED in each regression. (*$P < 0.1$; **$P < 0.05$; ***$P < 0.01$.)

**Table A30. Regression coefficients of EP and ED when using Hausdorff distance for paper distance**

| paper distance | 10 career years as split point | | 10 papers as split point | |
|---|---|---|---|---|
| | EP | ED | EP | ED |
| Hausdorff (max-min) | 0.323*** | -0.274*** | 0.336*** | -0.278*** |

Different from taking pairwise average in the main text, we try Hausdorff distance (max-min distance) when measuring paper distance, as described in Section A7.3.4. The regression equation is Eq. (A4), with the same setup as in the main text. We list here only the coefficients of the EP and ED in each regression. (*$P < 0.1$; **$P < 0.05$; ***$P < 0.01$.)

**Table A31. Regression coefficients of the EP and ED under varying definitions of academic impact**

| Varying definitions of future performance | 10 career years as split point | | 10 papers as split point | |
|---|---|---|---|---|
| | EP | ED | EP | ED |
| $\log c_{10}$ | 0.229*** | -0.184*** | 0.291*** | -0.271*** |
| normalization: method 1 (32) | 0.134*** | -0.057** | 0.135*** | -0.086*** |
| normalization: method 2 (33) | 0.134*** | -0.068*** | 0.135*** | -0.082*** |
| percentile: max | 0.034*** | -0.014** | 0.044*** | -0.025*** |
| percentile: mean | 0.029*** | -0.018*** | 0.036*** | -0.025*** |
| maximum $\log c_5$ | 0.495*** | -0.242*** | 0.476*** | -0.408*** |
| maximum $\log c_{10}$ | 0.239*** | -0.223*** | 0.460*** | -0.403*** |

As discussed in Section A7.4, we test different measurements of future academic impact, as alternatives to $\log c_5$ used in the main text. The regression equation is Eq. (A4), with the same setup as in the main text. We list here only the coefficients of the EP and ED in each regression. ($^*P < 0.1$; $^{**}P < 0.05$; $^{***}P < 0.01$.)

**Table A32. Regression coefficients of EP and ED with timeframes strictly aligned to post-1976.**

| Dataset | Period | 10 career years as split points | | | 10 papers as split points | | |
|---|---|---|---|---|---|---|---|
| | | EP coeff | ED coeff | Sample size | EP coeff | ED coeff | Sample size |
| PubMed | 1976-2021 | 0.13*** | -0.30*** | 596,904 | 0.13*** | -0.30*** | 883,293 |
| ACS | 1976-2021 | 0.16*** | -0.33*** | 26,328 | 0.17*** | -0.37*** | 36,378 |

We list only the coefficients of the EP and ED in each regression. (*$P < 0.1$; **$P < 0.05$; ***$P < 0.01$.)

**Table A33. Regression coefficients of the EP and ED under varying numbers of papers as the look-back period**

| look-back papers | 10 career years as split point | | 10 papers as split point | |
|---|---|---|---|---|
| | EP | ED | EP | ED |
| 1 | 0.218*** | -0.295*** | 0.250*** | -0.324*** |
| 2 | 0.255*** | -0.305*** | 0.272*** | -0.312*** |
| 3 | 0.266*** | -0.280*** | 0.290*** | -0.289*** |
| 4 | 0.275*** | -0.263*** | 0.293*** | -0.261*** |
| 5 | 0.270*** | -0.245*** | 0.298*** | -0.245*** |
| 6 | 0.272*** | -0.236*** | 0.302*** | -0.234*** |
| 7 | 0.265*** | -0.222*** | 0.300*** | -0.221*** |
| 8 | 0.262*** | -0.213*** | 0.305*** | -0.217*** |
| 9 | 0.260*** | -0.207*** | 0.307*** | -0.214*** |
| 10 | 0.256*** | -0.200*** | ——— | ——— |
| 11 | 0.254*** | -0.196*** | ——— | ——— |
| 12 | 0.250*** | -0.191*** | ——— | ——— |
| 13 | 0.251*** | -0.190*** | ——— | ——— |
| 14 | 0.252*** | -0.188*** | ——— | ——— |
| 15 | 0.252*** | -0.187*** | ——— | ——— |
| $\infty$ | 0.251*** | -0.182*** | ——— | ——— |

Note that in the case of using 10 papers as split point, the EP and ED are always the same when the number of papers in the look-back period is greater than or equal to 9 (there are not enough papers before the split point to allow for more papers in the look-back window), and hence the coefficients are the same as well. We list here only the coefficients of the EP and ED in each regression. ($*P < 0.1$; $**P < 0.05$; $***P < 0.01$.)

**Table A34. Regression coefficients of the EP and ED under different look-back years**

| look-back years | 10 career years as split point | | 10 papers as split point | |
|---|---|---|---|---|
| | EP | ED | EP | ED |
| 1 | 0.149*** | -0.163*** | 0.198*** | -0.211*** |
| 2 | 0.160*** | -0.202*** | 0.219*** | -0.244*** |
| 3 | 0.211*** | -0.243*** | 0.252*** | -0.260*** |
| 4 | 0.209*** | -0.224*** | 0.256*** | -0.248*** |
| 5 | 0.221*** | -0.223*** | 0.266*** | -0.245*** |
| 6 | 0.247*** | -0.230*** | 0.283*** | -0.247*** |
| 7 | 0.253*** | -0.224*** | 0.289*** | -0.243*** |
| 8 | 0.262*** | -0.215*** | 0.290*** | -0.239*** |
| 9 | 0.255*** | -0.194*** | 0.297*** | -0.239*** |
| 10 | 0.251*** | -0.182*** | 0.301*** | -0.239*** |
| 11 | ——— | ——— | 0.304*** | -0.239*** |
| 12 | ——— | ——— | 0.306*** | -0.235*** |
| 13 | ——— | ——— | 0.306*** | -0.232*** |
| 14 | ——— | ——— | 0.309*** | -0.231*** |
| 15 | ——— | ——— | 0.309*** | -0.228*** |
| $\infty$ | ——— | ——— | 0.307*** | -0.214*** |

Note that in the case of using 10 career years as split point, the EP and ED are always the same when the number of the look-back years is greater than or equal to 9 (there are not enough years before the split point to allow for more years in the look-back window), and hence the coefficients are the same as well. We list here only the coefficients of the EP and ED in each regression. (*$P < 0.1$; **$P < 0.05$; ***$P < 0.01$.)

**Table A35. PSM results with varying grouping quantiles and 10 papers as split points**

| Qtl | EP | | | | ED | | | | Combining EP and ED | | | |
|---|---|---|---|---|---|---|---|---|---|---|---|---|
| | Pairs | Treat | Control | P | Pairs | Treat | Control | P | Pairs | Treat | Control | P |
| 40 | 5838 | 1.718 | 1.657 | *** | 2238 | 1.595 | 1.659 | *** | 680 | 1.700 | 1.567 | *** |
| | | (0.008) | (0.008) | (***) | | (0.012) | (0.012) | (***) | | (0.023) | (0.022) | (***) |
| 30 | 4403 | 1.717 | 1.641 | *** | 1162 | 1.558 | 1.659 | *** | 11 | 2.026 | 1.413 | *** |
| | | (0.009) | (0.009) | (***) | | (0.017) | (0.017) | (***) | | (0.135) | (0.139) | (*) |
| 25 | 2630 | 1.746 | 1.662 | *** | 716 | 1.526 | 1.623 | *** | 16 | 1.603 | 1.266 | ** |
| | | (0.012) | (0.011) | (***) | | (0.022) | (0.021) | (***) | | (0.109) | (0.119) | (**) |
| 20 | 2630 | 1.746 | 1.662 | *** | 409 | 1.506 | 1.658 | *** | —— | —— | —— | —— |
| | | (0.012) | (0.011) | (***) | | (0.027) | (0.028) | (***) | | | | |
| 15 | 2630 | 1.746 | 1.662 | *** | 189 | 1.444 | 1.642 | *** | —— | —— | —— | —— |
| | | (0.012) | (0.011) | (***) | | (0.034) | (0.040) | (***) | | | | |

The column of pairs reports how many pairs are left in the groups after matchings. The Qtl column marks the quantile used in the grouping. The ED, EP and combining EP and ED columns correspond to the experiments where the ED, EP and combining both are the variable in question in PSMs, respectively. The Treat columns record the mean of the log-citations per paper of the treatment group, while the Control columns report those of the control group, where the exact meanings of treatment and control depend on the corresponding context. The P columns report whether the matched groups are statistically significantly different under $t$-tests—reported in rows without parentheses, and the Kruskal–Wallis tests—reported in rows with parentheses. (NS, not significant; $*P < 0.1$; $**P < 0.05$; $***P < 0.01$.)

**Table A36. PSM results with varying grouping quantiles and 10 career years as split points**

| Qtl | EP | | | | ED | | | | Combining EP and ED | | | |
|---|---|---|---|---|---|---|---|---|---|---|---|---|
| | Pairs | Treat | Control | P | Pairs | Treat | Control | P | Pairs | Treat | Control | P |
| 40 | 3549 | 1.611 | 1.557 | *** | 1941 | 1.498 | 1.572 | *** | 417 | 1.569 | 1.440 | *** |
| | | (0.010) | (0.010) | (***) | | (0.013) | (0.013) | (***) | | (0.027) | (0.027) | (***) |
| 30 | 2601 | 1.557 | 1.497 | *** | 1040 | 1.466 | 1.567 | *** | 72 | 1.550 | 1.380 | * |
| | | (0.011) | (0.011) | (***) | | (0.017) | (0.018) | (***) | | (0.072) | (0.065) | (*) |
| 25 | 2507 | 1.562 | 1.505 | *** | 678 | 1.484 | 1.573 | *** | 32 | 1.531 | 1.229 | ** |
| | | (0.012) | (0.012) | (***) | | (0.021) | (0.024) | (***) | | (0.101) | (0.081) | (**) |
| 20 | 1347 | 1.572 | 1.497 | *** | 414 | 1.421 | 1.553 | *** | —— | —— | —— | —— |
| | | (0.016) | (0.016) | (***) | | (0.027) | (0.030) | (***) | | | | |
| 15 | 949 | 1.573 | 1.441 | *** | 187 | 1.399 | 1.515 | ** | —— | —— | —— | —— |
| | | (0.019) | (0.019) | (***) | | (0.036) | (0.040) | (*) | | | | |

The table has the same notation as Table A35.

**Table A37. Results of the null model in PSM**

| Shuffling Level | Split Point | EP | ED | Combining EP and ED |
|---|---|---|---|---|
| Paper-level | 10 career year | 0 | 0 | 0 |
| | 10 paper | 0 | 0 | 0 |
| Author-level | 10 career year | 4 | 4 | 4 |
| | 10 paper | 0 | 1 | 0 |

The numbers in the table reports in how many times out of 1,000 repeated experiments the average differences within groups after matching in synthetic datasets are larger than the actually observed difference from the real data. The smaller the number, the better it is in supporting the robustness of our findings.

**Table A38. Results of the null model in PSW**

| Shuffling Level | Split Point | E(C-D) | E(B-D) | E(A-D) |
|---|---|---|---|---|
| Paper-level | 10th career year | 0 | 0 | 0 |
| | 10th paper | 0 | 0 | 0 |
| Author-level | 10th career year | 0 | 0 | 0 |
| | 10th paper | 0 | 7 | 0 |

The numbers in the table reports in how many times out of 1,000 repeated experiments ATEs in synthetic datasets are larger than the actually observed ATEs from the real data. The smaller the number, the better it is to support the robustness of our findings. The groups A, B, C and D are defined in Section A4.1.

**Table A39. Regression coefficients of the EP and ED after adding Gaussian noise to the independent variables**

| Added noise and variable | 10 career years as split point | | 10 papers as split point | |
|---|---|---|---|---|
| | EP | ED | EP | ED |
| $\epsilon \sim N(0, 0.1^2)$ to EP | 0.17*** | -0.16*** | 0.21*** | -0.17*** |
| $\epsilon \sim N(0, 0.3^2)$ to ED | 0.19*** | -0.04*** | 0.20*** | -0.17*** |
| $\epsilon \sim N(0, 10^2)$ to $\mathrm{LogCit}_{\mathrm{past}}$ | 0.34*** | -0.58*** | 0.37*** | -0.66*** |

We introduce Gaussian noise with different $\sigma$ to the independent variables EP, ED and $\mathrm{LogCit}_{\mathrm{past}}$ separately, and run the regression Eq. (A4). We increment the value of $\sigma$ starting from 0.1 until the regression results become statistically insignificant. We record the maximum $\sigma$ at which the results remain significant, as well as the coefficients of EP and ED at this perturbation level. Our experiments find that the maximum $\sigma$ of perturbation that can be added to EP while still maintaining significant results is 0.1. Similarly, for ED, the maximum $\sigma$ is 0.3. For $\mathrm{LogCit}_{\mathrm{past}}$, we observe that a perturbation level of more than 10 is unnecessary. Both EP and ED have a range of $[0, 1]$, so adding Gaussian noise with a standard deviation of 0.1 or 0.3 is considered a significant perturbation to either variable. The same applies to $\mathrm{LogCit}_{\mathrm{past}}$, which has a range of $[0, 4.3)$. (*$P < 0.1$; **$P < 0.05$; ***$P < 0.01$.)

**Table A40. The regression results of the future productivity on past switching behavior.**

| Split point | EP | ED |
|---|---|---|
| 10 papers | 0.954 | -7.79*** |
| 10 career years | -1.38* | -0.388 |

We set the dependent variables as future publication count, with the same regression setup as in the main text (see Eq. (A4)). Here, we list only the coefficients of the EP and ED in each regression. (*$P < 0.1$; **$P < 0.05$; ***$P < 0.01$.)